\documentclass[prd,aps,showpacs,preprintnumbers,amsmath,amssymb,nofootinbib,showkeys]{revtex4-1}

\pdfoutput=1
\usepackage{hyperref}
\usepackage{natbib}
\usepackage{booktabs}
\usepackage{subfiles}
\usepackage{aas_macros}
\usepackage{xcolor}
\usepackage{float}
\usepackage{graphicx}
\usepackage{dcolumn}
\usepackage{bm}
\usepackage{cases}
\usepackage[mathlines]{lineno}
\usepackage[lofdepth,lotdepth,caption=false]{subfig}

\newcommand{\vbar}{\bar{v}}
\newcommand{\vesc}{v_{esc}}
\newcommand{\Rstar}{R_{\star}}
\newcommand{\Mstar}{M_{\star}}
\newcommand{\be}{\begin{equation}}
\newcommand{\ee}{\end{equation}}
\newcommand{\pn}{p_N(\tau)}

\newcommand{\vn}{v_{N}}

\newcommand{\mess}{\frac{3(\vn^2-\vesc^2)}{2\vbar^2}}
\newcommand{\avg}[1]{\langle #1 \rangle}
\newcommand{\sqsixoverpi}{\left(\frac{6}{\pi}\right)^{1/2}}
\newcommand{\vevbsq}{\frac{\vesc^2}{\vbar^2}}
\newcommand{\nmax}{N_{max}}
\newcommand{\soneinf}{\sum_{N=1}^{\infty}}
\newcommand{\szeroinf}{\sum_{j=0}^{\infty}}
\newcommand{\unit}[1]{\mathrm{#1}}
\newcommand{\GeV}{\unit{GeV}}
\newcommand{\cc}{\unit{cm}^3}
\newcommand{\Msun}{M_{\odot}}
\newcommand{\Rsun}{R_{\odot}}
\newcommand{\tenovervbar}{\frac{10~\unit{km}\unit{s}^{-1}}{\vbar}}
\newcommand{\sigmaoversigma}{\left(\frac{\sigma}{1.26\times10^{-40}~\unit{cm}^2}\right)}

\newcommand{\sigmav}{\avg{\sigma v}}
\newcommand{\ccpers}{\cc\unit{s}^{-1}}
\newcommand{\percc}{\unit{cm}^{-3}}

\begin{document}
\title{Constraining Dark Matter properties with the first generation of stars}

\author{Cosmin Ilie}
\email[E-mail at: ]{cilie@colgate.edu}
 \altaffiliation[Additional Affiliation: ]{Department of Theoretical Physics, National Institute for Physics and Nuclear Engineering,  Magurele, P.O.Box M.G. 6, Romania}
\author{Caleb Levy}
\affiliation{ Department of Physics and Astronomy, Colgate University\\
13 Oak Dr., Hamilton, NY 13346, U.S.A.
}%
\author{Jacob Pilawa}%
\affiliation{Department of Astronomy, University of California, Berkeley\\
Berkeley, CA 94720 U.S.A.}
\author{Saiyang Zhang}
\affiliation{Theory Group, Department of Physics, University of Texas\\
Austin, TX 78712, U.S.A.}

\date{\today}%


\begin{abstract}
Dark Matter (DM) can be trapped by the gravitational field of any star, since collisions with nuclei in dense environments can slow down the DM particle below the escape velocity ($\vesc$) at the surface of the star. If captured, the DM particles can self-annihilate, and, therefore, provide a new source of energy for the star. We investigate this phenomenon for capture of DM particles by the first generation of stars [Population~III (Pop~III) stars], by using the multiscatter capture formalism. Pop~III stars are particularly good DM captors, since they form in DM-rich environments, at the center of$~\sim 10^6 \Msun$ DM minihalos, at redshifts $z\sim 15$. Assuming a DM-proton scattering cross section ($\sigma)$ at the current deepest exclusion limits provided by the XENON1T experiment, we find that captured DM annihilations at the core of Pop~III stars can lead, via the Eddington limit, to upper bounds in stellar masses that can be as low as a few $\Msun$ if the ambient DM density ($\rho_X$) at the location of the Pop~III star is sufficiently high. Conversely, when Pop~III stars are identified, one can use their observed mass ($\Mstar$) to place bounds on $\rho_X\sigma$. Using adiabatic contraction to estimate the ambient DM density in the environment surrounding Pop~III stars, we place projected upper limits on $\sigma$, for $\Mstar$ in the $100-1000~\Msun$ range, and find bounds that are competitive with, or deeper than, those provided by the most sensitive current direct detection experiments for both spin independent and spin dependent interactions, for a wide range of DM masses. Most intriguingly, we find that Pop~III stars with mass $\Mstar\gtrsim 300\Msun$ could be used to probe the SD proton-DM cross section below the ``neutrino floor,'' i.e. the region of parameter space where DM direct detection experiments will soon become overwhelmed by neutrino backgrounds.

\end{abstract}


\maketitle


\section{Introduction}\label{sec:Intro}

One of the most profound mysteries nature has presented us with is usually wrapped in two very descriptive, although sometimes misleading, words: Dark Matter (DM). It was Fritz Zwicky who, in 1933, coined the term \textit{dunkle Materie} (i.e.~Dark Matter) when describing the non-luminous mass that he inferred must have been present in abundance in the Coma Cluster of galaxies~\cite{Zwicky:1933,zwicky1937masses}. It took almost four decades until this idea re-emerged at the forefront of the literature. In 1970, Vera Rubin and Kent Ford showed that rotation curves of stars in galaxies are ``flat,'' a fact that can be interpreted as evidence of non-luminous matter at galactic scales~\cite{Rubin:1970}. Since then, a large body of evidence has emerged that supports the Dark Matter hypothesis. Only $20\%$  of the matter in the universe is made of regular, baryonic matter. The other $80\%$ is Dark Matter, whose existence is inferred via its gravitational effects, on all scales. DM leaves its imprint in the Cosmic Microwave Background radiation \cite{Komatsu:2009,Komatsu:2011,Ade:2015,Aghanim:2018}, since it provides the gravitational restoring force for the acoustic oscillations of the photon-baryon plasma before recombination. 

Under the influence of gravity, the primordial density fluctuations generated by cosmic inflation grow into over-dense regions dominated by dark matter in what is commonly referred to as hierarchical structure formation. DM forms minihalos that grow, via mergers, into larger and larger halos with a rich sub-structure. Numerical simulations show that those over-dense regions are connected by DM filaments, and separated by large, under-dense regions. As such, DM provides the scaffolding upon which regular, baryonic matter gravitationally collapses to form
galaxies and galaxy clusters. Using gravitational lensing, the Sloan Digital Sky Survey confirmed the predominance of dark matter in galaxies~\cite{AdelmanMcCarthy:2005}. Moreover, gravitational lensing has been used to map the structures DM forms at galaxy cluster~\cite{Natarajan:2017} and cosmological~\cite{Madhavacheril:2014,Vikram:2015,Hikage:2018} scales.

Today, the experimental hunt for Dark Matter has three prongs: particle production, direct detection, and indirect detection. So far, the Large Hadron Collider (LHC) has found no evidence of any physics outside of the standard model of particle physics, which, in turn, implies constraints on phenomenological models of DM. Indirect detection experiments seek to observe the products of annihilation (or decay) of DM that could emerge from nearby astrophysical sites where DM densities are high. Of those such places, the center of our galaxy and nearby dwarf spheroidal satellites of the Milky-Way are prime targets. Expected signals include, but are not limited to, gamma rays. An antiproton and a gamma-ray excess compared to known backgrounds have  been found in Alpha Magnetic Spectrometer and Fermi data, respectively. Both can be explained by the same DM particle model, a $\sim60~\unit{GeV}$ DM particle self annihilating~\cite{Goodenough:2009,hooper2011dark,Cholis:2019}. Alternatively, the gamma-ray signal could come from point sources, such as pulsars~\cite{Gordon:2013,YUAN:2014}, and the anti-proton excess could be due to collisions between cosmic-ray protons accelerated in the presence of a local supernova remnant (SNR) and the protons in the SNR cloud~\cite{Kohri:2015}.  Dwarf spheroidal satellite galaxies of the Milky Way are another prime target for detecting DM-DM annihilation signals. In lack thereof, the Fermi satellite data was used to place the most stringent bounds on the dark matter annihilation cross section to date~\cite{Ackermann:2015,Fermi-LAT:2016,Ahnen:2016}.

Direct detection experiments are extremely challenging. They are very sensitive, to the point of being able to detect the minute amount of energy a dark matter particle deposits inside the detector as it collides with an atomic nucleus~\cite{Goodman:1985,Drukier:1986}. Shielding from cosmic ray backgrounds means that these experiments have to be performed in deep, underground laboratories. Of the ten currently operational direct detection experiments, only the DAMA/LIBRA experiment in Gran Sasso, Italy produced a  detection signal~\cite{BERNABEI:1998,Bernabei:2014,Bernabei:2018}. Since 1998, the DAMA/LIBRA experiment finds an annual modulation in its signal that matches the modulation predicted by~\cite{Drukier:1986}. Although this is the cleanest hint of a dark matter detection yet, unfortunately, it has not been confirmed by other direct detection experiments exploring the same region of the parameter space, such as XENON1T. To settle this controversy, a new NaI experiment (the same detector material as DAMA/LIBRA) has been developed: COSINE~\cite{Adhikari:2018}. It will soon either refute or confirm the DAMA signal~\footnote{Recently, another experiment (ANAIS) has analysed their three year data and found no annual modulation~\cite{Amare:2021}.}. Another hint of DM detection came recently from XENON1T, the world's most sensitive DM direct detection experiment. An excess in the electronic recoil events could be explained by, among other things, solar axions~\cite{Aprile:2020}. While solar axions are not a dark matter candidate, their detection, if confirmed, would be the first discovery of a particle outside of the standard model of particle physics. This would provide insights into the production of axions in the early universe, which could serve as dark matter candidates. 

In lack of a clear, independently confirmed detection signal from direct detection experiments, we are left with exclusion limits on how strong DM and baryonic matter can interact. As experiments become more and more sensitive, they rule out larger and larger swaths of the possible DM-nucleon scattering cross section $\sigma$ vs DM particle mass ($m_X$) parameter space. However, an increase in sensitivity comes at a price. In the near future, it is expected that the XENON1T experiment will become sensitive to neutrinos. At that stage, any possible DM signal would be swamped by an overwhelming neutrino background, the so-called neutrino floor. As such, new detection strategies will have to be implemented. In this paper we discuss one such strategy, which relies on the capture of Dark Matter by the first generation of stars, the so called Population~III (Pop~III) stars. 

Astrophysical objects have a long history as DM probes in the literature. In the 80's, some of the seminal papers developing the mathematical formalism for capture of Dark Matter~\cite{Press:1985,Spergel:1985,Gould:1987,Gould:1987resonant} consider the potentially observable effects on the Sun from DM trapped inside it. All those works assumed that one collision with nuclei is sufficient to capture a DM particle inside a celestial object. This assumption can be bypassed by using the multiscatter capture formalism~\cite{Gould:1992ApJ, Bramante:2017,Dasgupta:2019juq,Ilie:2020Comment,Bell:2020}. As such, one can estimate capture rates in very dense environments, where, on average, a DM particle will collide multiple times per crossing with regular matter inside the astrophysical capturing object. The potential observable effects of captured DM have been used in the literature to constrain DM properties by using Pop~III stars~\cite{Freese:2008cap,Iocco:2008,Ilie:2019}, Neutron Stars~\cite{Baryakhtar:2017, Bramante:2017,Raj:2017wrv,Croon:2017zcu,Bell:2018pkk,Chen:2018ohx,Gresham:2018rqo,Acevedo:2019,Bell:2019pyc,Hamaguchi:2019oev,Leroy:2019,Leung:2019,Joglekar:2019,Bell:2020,Bell:2020b,Bell:2020NSSINS,Garani:2020,Genolini:2020,Joglekar:2020,Keung:2020,Kumar:2020,Perez-Garcia:2020}, White Dwarfs~\cite{Bertolami:2014wua,Bramante:2017,Dasgupta:2019juq,Horowitz:2020axx,Panotopoulos:2020kuo}, and exoplanets~\cite{Leane:2020wob}, to name a few.

In this work, we demonstrate how the observation of any Pop~III star can be used to place very stringent constraints on the strength of the proton-DM scattering cross section. Most importantly, if the ambient DM density ($\rho_X$) is sufficiently high, Pop~III stars can be used to probe below the neutrino floor, which will soon limit direct detection experiments on earth. The paper is organized as follows: in Sec.~\ref{sec:FirstStars} we review the main properties, as inferred from numerical simulations, of Pop~III stars, in Sec.~\ref{sec:obs} we briefly review the formalism used to calculate how efficiently DM is captured or evaporated by astrophysical objects and apply it to Pop.~III stars; in the process, we find that the heating from annihilations of captured DM inside the star leads to an upper bound on the stellar mass ($\Mstar$). In Sec.~\ref{sec:Constrain} we show how one can use the mere observation of a Pop.~III star of a given mass to place constraints on the product between the  DM-nucleon cross section ($\sigma$) and the ambient DM density ($\rho_X$). Assuming direct detection experiments will identify DM in the near future, and using upper bounds from XENON1T on $\sigma$, we then obtain projected bounds on $\rho_X$, for Pop~III stars of various masses. Conversely, using the adiabatic contraction formalism to estimate the possible range of $\rho_X$ at the location of Pop~III stars, and including the possible effects of DM annihilations on the ambient DM density, we calculate exclusion regions in the $\sigma$ vs $m_X$ parameter space corresponding to Pop~III stars of masses between $100-1000\Msun$. We find that Pop~III stars can be used to probe below the neutrino floor for SD experiments, such as PICO. For spin-independent (SI) experiments, at the higher end of $m_X$, i.e. $m_X\gtrsim 10^5~\GeV$, we find that Pop~III stars are placing constraints on $\sigma$ that are stronger than those placed by the most sensitive direct detection experiments currently available, such as XENON1T. Regarding sub-GeV DM, we considered the case of strongly interacting thermal DM models, such as SIMP/CoSIMP DM, as well as the standard thermal Weakly Interacting Massive Particles (WIMPs), and found exclusion regions in the $\sigma-m_X$ parameter space that are deeper than any current experiments.  Sec.~\ref{sec:discussion} is dedicated to a discussion of the implications and limitations of our approach. The paper ends with five appendices, in the following order: in Appendix~\ref{sec:MSCapture} we review the technical details of the multiscatter DM capture formalism and present derivations of analytic closed form formulae for the total capture rates in various limiting regimes of interest. This can be very useful in practice, for future research, since calculating the capture rates numerically can turn out to be computationally expensive. In Appendix~\ref{sec:CapDMTemp} we estimate the temperature of captured DM ($T_X$), which will be necessary when evaluating evaporation rates for DM when considering sub-GeV DM models. In Appendix~\ref{sec:Evaporation} we derive and validate a closed form analytic approximation of the evaporation rates of DM from Pop~III stars. In Appendix~\ref{sec:Equilibration} we discuss in more detail the DM models considered in this paper (thermal WIMPS, thermal sub-GeV Co/SIMP DM, and non-thermal superheavy DM) and the conditions necessary for the equilibrium between capture and annihilation/evaporation to be attained on timescales shorter than the lifetime of the star.  Finally, in Appendix~\ref{sec:DMHalos}, we apply the commonly used adiabatic compression formalism to estimate the ambient DM density relevant for the capture of DM by Pop~III stars. Additionally, we estimate the role of DM annihilations in the ambient medium, and find the so called annihilation plateau for each of the DM models considered. 

%

\section{The First Stars in the Universe}\label{sec:FirstStars}
Below we give a brief summary of the status of the literature regarding the formation of the first stars, also called Pop~III stars, our candidate targets as DM probes. They formed at the center of DM mini-halos ($M_{halo}\sim 10^6\Msun$), when the universe was roughly $400$~Myrs old, corresponding to redshifts $z\sim15$. At that epoch, pristine, zero metallicity gas from Big Bang Nucleosynthesis is cool enough to start its gravitational infall into the potential well provided by the high DM density regions at the center of the halo.  As the gas collapses it will form one, or sometimes a few clumps, separated by distances as large as a few parsecs. Those gas clumps are as massive as $20,000~\Msun$ each, with the most massive one located close to the center of the DM halo (see Fig.~14 of~\cite{Barkana:2000}). The balance between heating and cooling, which for pristine, zero metallicity clouds is quite poor, determines the stage when this collapse stops. If fragmentation during this phase plays an important role, the outcome would be that each of those gas clumps forms several Pop~III stars. Conversely, if fragmentation is suppressed, the formation of Pop~III stars is monolithic. Currently the consensus is that: ``At the end of the initial collapse, a small protostellar core has formed at the center of the minihalo.''~\citet{Bromm:2013}. This protostellar core is surrounded by an accretion disk, roughly 10 A.\ U.\ in size, which can sometimes fragment. As shown in~\citet{Greif:2012}, for example, the most massive of those stellar fragments remains close to the center; in addition, there could be other, smaller fragments fragments in highly complex orbits, ``most of which migrate towards the center of the cloud'' (see Fig.~5 of~\cite{Greif:2012}). This picture is confirmed by most hydrodynamical simulations, that demonstrate that typically one or just a few Pop~III stars form per mini-halo, with masses up to $\sim 1000\Msun$ and within the inner $10$~A.\ U.\ of the DM halo. More explicitly, simulations have shown that the most massive protostars remain close to the center of the cloud, which, itself is aligned with the center of the DM halo~\cite{Barkana:2000,Abel:2001,Bromm:2003,Yoshida:2006,Yoshida:2008,Loeb:2010,Bromm:2013,Machida:2013,Klessen:2018}, with some of the smaller fragments being dynamically ejected from the central region of the halo, while others  will move inwards and get accreted by the central, most massive object~\cite{Clark:2011,Greif:2012,Smith:2012,Stacy:2016}. 

Fragmentation of the collapsing gas and/or of the accretion disk is the primary mechanism that controls the multiplicity of Pop~III stars per micro DM halo, and prior to Pop~III host halo mergers is the only mechanism that determines this important parameter. In the first decade of the twentieth century a preliminary standard model has emerged in the literature, in view of the consensus in the results from the vast majority of simulations~\cite{Abel:2001,Barkana:2000,Bromm:2003,Yoshida:2006,OShea:2007,Yoshida:2008,Bromm:2009}. It was believed that Pop~III stars generally form in isolation, and `` at most one massive ($M\gg \Msun$) metal free star forms per pre-galactic halo.''(from abstract of Ref.~\citet{Abel:2001}). This picture has recently undergone some scrutiny, since more recent simulations started to indicate that protostellar disks around primoridial stars can become gravitationally unstable and fragment to build up binary or higher-order multiple stellar systems~\cite{Clark:2008,Clark:2011,Stacy:2010}. However the following consensus emerges from most simulations: the first stars have a top-heavy initial mass function, and they form in relatively small numbers per DM halo. The first to simulate fully three-dimensional Pop~III star formation including both fragmentation (up to the resolution of the simulation) and radiative feedback found an average of 3 stars forming in each of the cosmologically simulated DM halos of Ref.~\cite{Susa:2014}. Additionally, each subsequent generation of stars that are formed in a given halo will be, on average, smaller mass. Thus more massive Pop~III stars are going to be found in more solitary environments, allowing a single, massive Pop~III star to dominate the baryonic mass of a halo. More recent simulations find similar halo statistics. Figure 7 of \cite{Skinner:2020} demonstrates that halos form a median of 4 stars and a maximum of 16 stars, with a sharp decrease in the number of halos forming greater than 6 massive Pop~III stars. The centrality of these objects is also of importance. Susa \textit{et al.}~\cite{Susa:2014} find that the most massive Pop~III stars form, on average, more centrally than less massive stars with a significant fraction of stars larger than $100 M_\odot$ forming within $100$A.\ U.\ of the center of the mini-halo. The same simulations found no haloes with more than one star if the first star is formed over around $150 M_\odot$ (Fig. 11 of ~\cite{Susa:2014}). We note that the primary targets of our method of constraining DM will be Pop~III stars more massive than $\sim150\Msun$,  and, as such, those are expected to form in isolation, one per DM halo, and with locations closely aligned with the center of the DM halo. 

One of the key effects that needs to be taken into account when estimating the typical mass of a Pop~III stars is radiative feedback, which has the potential to shut off accretion, and, as such, limit the stellar mass. However, regarding this issue, a recent review by prominent authors of this field states~\cite{Haemmerle:2020}: ``any firm conclusions about the resulting mass spectrum of Pop. III stars in the
presence of radiative feedback seems premature at this stage.'' However, note that cosmological 3D hydrodynamical simulations regularly find final masses in excess of $100~\Msun$~\cite{Susa:2014}, with some finding Pop~III stars as massive as $1000~\Msun$~\cite{Hirano:2014}.  

There are two main mechanisms that suppress fragmentation in the cloud: magnetic field interactions in the gas cloud; and, dark matter self-annihilation. Several mechanisms have been studied which generate magnetic fields in these pristine gas clouds, however a generic feature is that these magnetic fields are relatively weak~\cite{Haemmerle:2020}. In the presence of these fields, angular momentum transfer occurs by both protostellar jets and magnetic breaking. This causes gas to fall directly onto the protostar instead of forming a disc. Without a disc, fragmentation cannot occur, and the result is a single, massive, central star \cite{Machida:2013}. A weaker suppression can come from considering dark matter annihilation during the collapse of mini-halos. Without dark matter annihilation, fragmentation is seen to occur in collapsing Pop~III halos~\cite{Clark:2008,Turk:2009,Clark:2011,Stacy:2010}. Adiabatic contraction can bring dark matter with the collapsing baryons, causing density and thus the annihilation rate to increase several orders of magnitude (since annihilation rate scales with the square of the number density). It is unclear if dark matter annihilation is energetic enough to overcome H$_2$ cooling, however simulations show that dark matter annihilation can impact the dynamics of the accretion disc and thus reduce the level of fragmentation~\cite{Smith:2012,Stacy:2012,Stacy:2014}. In light of these (radiative feedback, magnetic fields, dark matter annihilation, and other) effects, any absolute statements regarding the mass spectrum of Pop~III stars are premature beyond what we have seen in simulations to this point: a low number of massive, central Pop~III stars forming in each halo. Consequently, these very massive, central stars are hot and emit a lot of photo-ionizing radiation. As such, Pop~III stars usher the epoch of re-ionization, when the baryonic gas in the universe becomes fully ionized. This transition is complete by redshift $z\sim 7$. 

We end this section with a brief discussion regarding the role of the heating from DM annihilation on the formation of the first stars. This has been initially investigated by Ref.~\cite{Freese:2008ds}, who found that under certain conditions, dark matter heating can overcome the dominant cooling mechanisms. This would subsequently halt the collapse of the protostellar gas cloud when the baryon number density is roughly $n\sim 10^{17}~\unit{cm}^{-3}$, well below the typical $n\sim 10^{22}~\unit{cm}^{-3}$ when DM heating is not included. As such, DM heating could lead to the formation of a new phase in the stellar evolution, a Dark Star. These puffy objects are powered by dark matter annihilations. Dark stars can grow to be supermassive~\cite{Freese:2010smds}, and could be observed with the upcoming James Webb Space Telescope (JWST)~\cite{Ilie:2012}. Their observation would indirectly confirm the existence of Dark Matter. In contrast, if Dark Matter heating plays little role in the formation of the first stars, a proto Pop~III star is born when the baryons have collapsed up to $n\sim 10^{22}~\unit{cm}^{-3}$. Pop~III stars and Dark Stars have very different photometric signatures~\cite{Zackrisson:2011,Ilie:2012}, and as such, JWST could be used to disambiguate between those two. For the reminder of this paper we will assume that at least some of the first stars will be Pop~III stars, and that those objects will be found with an upcoming telescope, such as JWST. In fact, we want to point out that~\citet{Vanzella:2020} has already found a candidate Pop~III stellar system at $z\sim 7$ in the MUSE deep lensed Hubble Space Telescope field.   

\section{Capture and evaporation of DM by Pop~III stars and their observational effects}\label{sec:obs}

Via collisions with nuclei inside any compact, astrophysical object, such as stars, neutron stars (NSs), or white dwarfs, a DM particle can be slowed below the escape velocity at the surface of the object, and thus become trapped by its gravitational field. Subsequent collisions lead to further slowing down, and eventually the captured DM sinks towards the center of the star, forming a self-gravitating DM core. This is, in essence, what is commonly referred to as DM capture. This phenomenon was studied initially for Weakly Interacting Dark Matter (WIMPs) in the 1980s, when the single scattering capture formalism of~\cite{Press:1985,Gould:1987,Gould:1987resonant} was developed. In practice, the formalism is limited to the case when DM particles are, on average, experiencing at most one collision with nuclei inside the star as they traverse it, hence ``single scattering.'' This is a valid approximation when the capturing object is not too compact, and/or when the cross section of interaction between DM and baryons is not too high, which is a direct consequence of the average number of collisions per crossing of an object of radius $\Rstar$, with target nuclei number density $n_T$, also called the optical depth, given by: $\tau=2\Rstar~\sigma~n_T$. Whenever $\tau\ll 1$, one can safely apply the single scatter formalism of DM capture. Conversely, when $\tau\gg 1$, one should use the more general, multi-scatter formalism, developed by~\cite{Gould:1992ApJ}~\footnote{See also~\cite{Bramante:2017,Dasgupta:2019juq,Ilie:2020Comment,Dasgupta:2020dik}}. In the next few paragraphs, we give a brief review of the multiscatter formalism, and the closed form analytical approximations we derived for the total capture rates, in various limiting regimes. The interested reader should consult  Appendix~\ref{sec:MSCapture} for technical details. As DM particles, coming from a reservoir with number density $n_X$, cross an astrophysical object with $n_T$ number density of scattering nuclei, they will be captured at a rate given by~\cite{Bramante:2017}:

\be
\label{eq:Ctot}
C_{tot} = \sum_{N=1}^{\infty} C_{N} = \sum_{N=1}^{\infty} \underbrace{\pi \Rstar^2}_\textrm{capture area}\times \,\underbrace{n_X \int_0^{\infty} \dfrac{f(u)du}{u}\,(u^2+v_{ esc}^2)}_\textrm{DM flux}\times \, \underbrace{p_{ N}(\tau)}_\textrm{prob. for $N$ collisions}\times \, \underbrace{g_{ N}(u)}_\textrm{prob. of capture}.
\ee
Throughout, we will denote by $C_N$ the capture rate after exactly $N$ collisions with nuclei inside the star. Note that for non-relativistic DM, such as is the case in our work, $n_X=\frac{\rho_X}{m_X}$. Therefore, since Pop~III stars form in DM-rich environments (see Appendix~\ref{sec:DMHalos} for estimates of $\rho_X$), they are particularly efficient at capturing DM. In Ref.~\cite{Ilie:2019}, we showed that the probability of $N$ collisions between DM and nuclei inside the star has the following closed form:
\be\label{eq:pN}
\pn=\frac{2}{\tau^2}\left(N+1-\frac{\Gamma(N+2,\tau)}{N!}\right), 
\ee
where $\Gamma(a,b)$ is the incomplete gamma function. For the probability that a DM particle is slowed down below $\vesc$ by exactly N collisions we assume, following~\cite{Bramante:2017}:   
\be\label{eq:gN}
g_N(u)=\Theta(u_{max;N}-u),
\ee
where $\Theta(x)$ is the Heaviside step function. Throughout, we denote by $u_{max;N}=\vesc\left[(1-\beta_+/2)^{-N}-1\right]^{1/2}$, the maximum value of the velocity a DM particle can have, far from the star, such that it will be slowed down below the escape velocity after $N$ collisions. Here $\beta_+\equiv 4mm_X/(m+m_X)^2$, with $m$ being the mass of the target nucleus. In our work, we assume a Maxwell-Boltzmann distribution $f_{MB}(u)$ for the velocities of DM particles surrounding the star. There is only one unique parameter describing such a distribution, the velocity dispersion ($\vbar$). In Appendix~\ref{sec:DMHalos}, we estimate that the dispersion velocity for DM in $10^6\Msun$ minihalos where Pop~III stars form is, to within factors of order unity, $\vbar=10\unit{km}\unit{s}^{-1}$. For more details on how we implement the calculation of $C_{tot}$ from Eq.~(\ref{eq:Ctot}) numerically, and for useful analytical approximations and their derivation, see Appendix~\ref{sec:MSCapture}. To facilitate the understanding of the main body of the paper, without the need to refer to appendices, we summarize below the main results regarding the behavior of the capture rates.  

For $m_X\gg m$, and in the multiscatter regime ($\tau\gg 1$) we find (see Eqns.~(\ref{eq:CtotApprox}) and~(\ref{eq:SumRegI})) that the total capture rate has the following scaling:
\be\label{eq:CtotMS}
C_{tot}\sim \frac{\rho_X\sigma}{m_X^2\vbar^3}\frac{\Mstar^3}{\Rstar^2}, 
\ee
whereas for single scattering capture ($\tau\ll1$) we find two distinct scaling relations:
\begin{subnumcases}
{C_{tot}=C_1\sim} 
\frac{\rho_X\sigma}{m_X^2\vbar^3}\frac{\Mstar^3}{\Rstar^2}, & \text{if  }$m_X\gg 3m\left(\frac{\vesc}{\vbar}\right)^2$ \label{eq:CtotSSHM}\\
\frac{\rho_X\sigma}{m_X\vbar}\frac{\Mstar^2}{\Rstar}, & \text{if  } $\frac{m}{3}\left(\frac{\vbar}{\vesc}\right)^{2}\ll m_X\ll 3m\left(\frac{\vesc}{\vbar}\right)^2$ \label{eq:CtotSSLM}
\end{subnumcases}
It  is noteworthy, and perhaps somewhat unexpected, that at the higher end of the DM particle mass where the single scattering approximation holds, we recover the same scaling with relevant parameters as one has in the multiscatter capture regime, as one can see from Eqns.~(\ref{eq:CtotMS}) and~(\ref{eq:CtotSSHM}).

For sub-GeV DM particles, one needs to include the effects of ``evaporation,'' i.e. the loss of captured DM, as they may be up-scattered to velocities above the escape velocity ($v_{esc}$) via collisions with nuclei, especially near the center of the star, where nuclei are the most energetic. In Appendix~\ref{sec:Evaporation} we obtain and validate the following approximation for the evaporation rate of DM particles from a star:
\be\label{eq:EvapRate}
E\approx\frac{3V_{\star}\bar{n}_pu_c\sigma}{2V_1\sqrt{\pi}}e^{-\frac{v_{esc}^2\mu}{u_c^2\Theta}(1+\xi_1/2)}.
\ee
We assumed that the internal structure of the star is well modeled by a $n=3$ polytrope, such as is the case for the radiation pressure dominated Pop~III stars considered here. Throughout, $V_{\star}$ represents the volume of the star, $\bar{n}_p$ is the average proton number density, $u_c\equiv\sqrt{\frac{2T_c}{m_p}}$, i.e. the average thermal velocity of protons at the center of the star, $\mu\equiv m_X/m$, $\Theta\equiv T_X/T_c$, and $\xi_1\approx6.89$ is the first node of the Lane-Emden function for $n=3$. Additionally, $V_i\equiv\int_{\star}dVe^{-im_X\Phi/T_X}$, with $\Phi(r)$ being the gravitational potential inside the star. Note how the exponential term suppresses the evaporation rates for $m_X\gtrsim 1\GeV$.

The interplay of capture, annihilation, and evaporation of DM inside the star can be modeled by the differential equation:
\begin{equation}\label{eq:Ndot}
    \dot{N}_{X} = C_{tot} - \Gamma_{A}-EN_{X},
\end{equation}
where $C_{tot}$ is the total capture rate and $\Gamma_{A}$ is the annihilation rate, which can be recast as: $\Gamma_A=C_AN_X^j$, with $C_A$ being an $N_X$-independent annihilation coefficient, and $j$ being the number of DM particles entering each annihilation event.
In this paper we consider four different scenarios for the annihilation events: p/s-wave annihilations ($j=2$; DM+DM$\to$ SM+SM), SIMP DM~\cite{Hochberg:2014} ($j=3$; DM+DM+DM$\to$DM+DM), or Co-SIMP~\cite{Smirnov:2020}($j=2$; DM+DM+SM$\to$DM+SM). 

For $j=2$ Eq.~(\ref{eq:Ndot}) has the following analytic solution:
\be\label{eq:NXt}
N_{X}(t)=\sqrt{\frac{C_{tot}}{C_A}}\frac{\tanh\left(\frac{\kappa t}{\tau_{eq}}\right)}{\kappa+\frac{1}{2}E\tau_{eq}\tanh\left(\frac{\kappa t}{\tau_{eq}}\right)},
\ee
with $\tau_{eq}\equiv1/\sqrt{C_{tot}C_A}$, and  $\kappa\equiv\sqrt{1+E^2\tau_{eq}^2/4}$. Whenever $t\gg\tau_{eq}/\kappa$, the number of DM particles inside the star ($N_X$) attains a constant, limiting value. Previous work on Pop~III stars showed that, for WIMP-like dark matter, equilibrium ($\dot{N}_X=0$) is quickly reached within the lifetime of the star~\cite{Freese:2008cap}. The same holds true for superheavy dark matter ($m_X\gtrsim 10^8\unit{GeV}$), assuming an annihilation cross section at the unitarity limit, as shown in~\cite{Ilie:2019}. In Appendix~\ref{sec:Equilibration} we revisit and generalize those investigations, including the role of evaporation for light DM. For non-thermal DM, where the annihilation cross section is not constrained by the thermal relic abundance, we obtain the minimum annihilation cross section such that equilibrium between capture/evaporation/annihilation can be reached within a fraction of the lifetime of the star. This can potentially have important repercussions on our ability to constrain DM-proton interaction cross sections, since DM will annihilate outside of the star as well, at a rate controlled by the anninhilation cross section. If this process operates for sufficiently long times, the DM ambient densities are reduced to a time-dependent plateau value (the annihilation plateau) from their initial, adiabatically contracted assumed profile. In Appendices~\ref{sec:Equilibration} and~\ref{sec:DMHalos} we discuss in detail the interplay between the requirement of efficient equilibration of the DM capture and annihilation/evaporation processes inside the star and the annihilation of DM in the vicinity of the star. In all of the bounds on $\sigma$ we present we will include, whenever significant, the effects of this annihilation plateau. 

\begin{figure} [!thb]
    \centering
    \includegraphics[width=.8\linewidth]{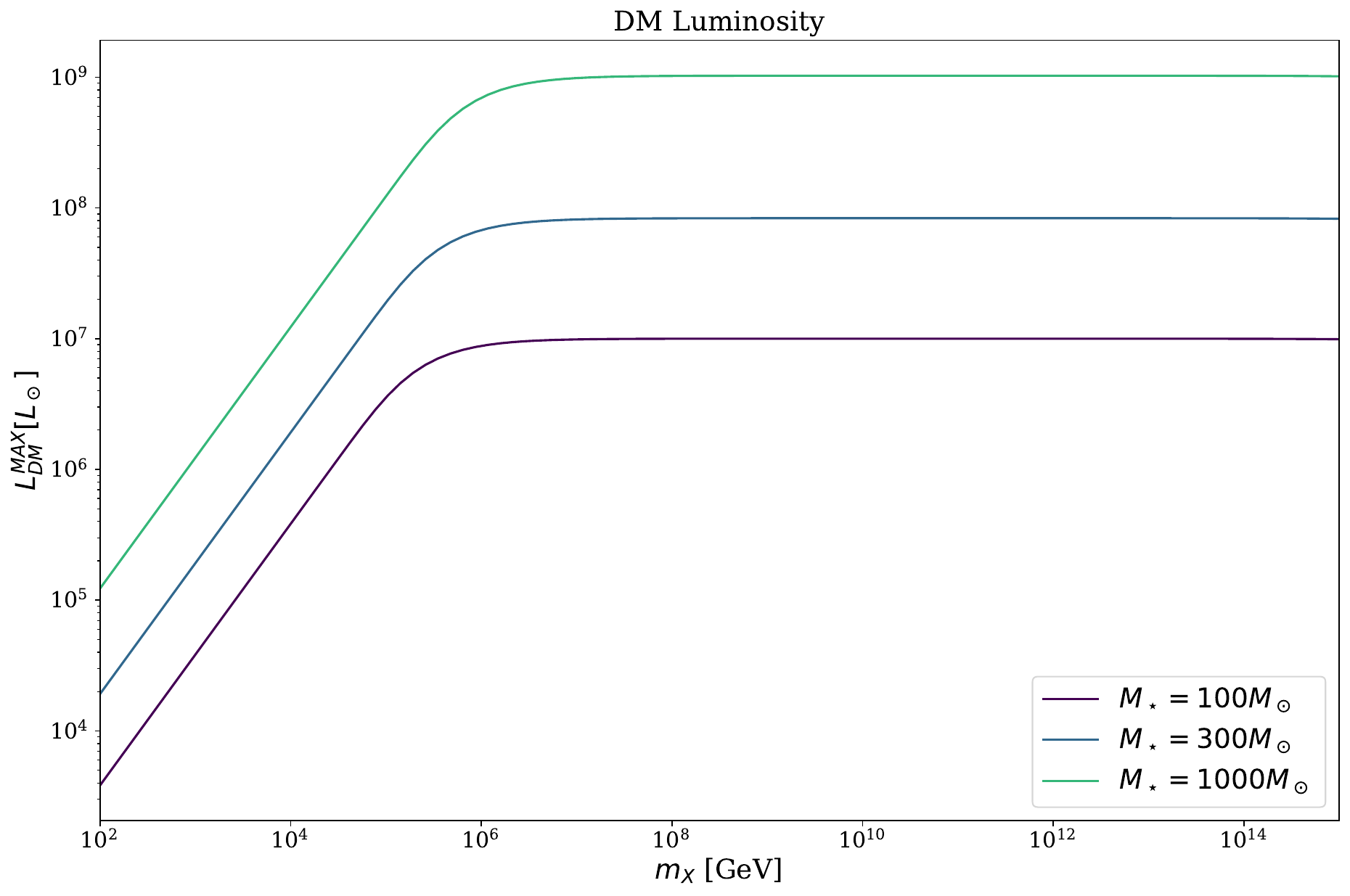}
    \caption{Upper bounds on the luminosity from captured dark matter annihilations for Pop~III stars of various masses in the mass range of $10^{2} - 10^{15}  \text{GeV}$. We assume here $\vbar = 10^{6} \unit{cm} \unit{ s}^{-1}$ and $\rho_X = 10^{16} \GeV \unit{ cm}^{-3}$. For the proton-DM cross section we used XENON1T SI bounds.}
    \label{fig:LDM}
\end{figure}

After this equilibrium is reached, the rate of change of DM particles in the stellar core becomes zero, and one obtains a stable energy source from dark matter annihilations with luminosity. At $m_X\gtrsim 1\GeV$, when we can neglect evaporation, this becomes:

\begin{equation}\label{eq:LDMeqCtotNoEvap}
    L_{DM} = f \Gamma_{A} m_X = fC_{tot}m_X,
\end{equation}
with $f$ being the model-dependent, order unity, fraction of the rest mass energy ($m_X$) that is deposited inside the star as a result of DM annihilations. For simplicity we assume $f=1$, i.e. all the energy from DM annihilations gets deposited inside the star. Our results scale linearly with $f$, and as such it would be straightforward to adjust them for any arbitrary $f$. An example calculation of dark matter luminosity from annihilations is presented in Fig.~\ref{fig:LDM}, for $m_X\gtrsim 10^2\GeV$. Assuming XENON1T~\cite{Aprile:2018} Spin Independent (SI) upper bounds on $\sigma$:
\begin{equation}\label{Eq:X1Tbounds}
    \sigma\lesssim 8\times 10^{-41}~\unit{cm}^2\left(\frac{m_X}{10^8~\unit{GeV}}\right),
\end{equation}
we obtain the maximum possible luminosity due to captured DM ($L_{DM}$) at a given DM particle mass ($m_X$), for Pop~III stars of mass $100$, $300$, and $1000\Msun$, respectively.

From Fig.~\ref{fig:LDM}, we note the two distinct trends for the upper bounds that any DM direct detection experiment places on $L_{DM}$: a constant value at high $m_X$, and $L^{max}_{DM}\propto m_X$ for low $m_X$. This transition at $m_X=3m(\vesc/\vbar)^2$ is to be expected, in view of our results for the scaling of $C_{tot}$ (Eqns.~(\ref{eq:CtotSSHM})~-~(\ref{eq:CtotSSLM})). Moreover, since $L_{DM}\propto\sigma/m_X$ for $m_X\gtrsim 3m(\vesc/\vbar)^2$, and in view of the upper bound from XENON1T $\sigma\propto m_X$, we can understand the trend from Fig.~\ref{fig:LDM}, where at the high-mass end $L^{max}_{DM}\propto m_X^0$. Conversely, for $m_X\lesssim 3m(\vesc/\vbar)^2$, a consequence of the capture rate being proportional to $\sigma/m_X$ is that $L^{max}_{DM}\propto m_X$, as found numerically in Fig.~\ref{fig:LDM}. 

As alluded to before, at $m_X\lesssim 1\GeV$ we include the effects of DM evaporation. In the case of $j=2$ (p/s-wave annihilation or Co-SIMP DM), in view of Eq.~(\ref{eq:NXt}), one can include analytically the role of the evaporation in the DM luminosity after equilibrium has been reached in the following way:
\be\label{eq:LDMeqCtot}
L_{DM} = f \Gamma_{A} m_X = \frac{fC_{tot}m_X}{(\kappa+\frac{1}{2}E\tau_{eq})^2},
\ee
As expected, the effect of evaporation is to reduce the amount of the rest mass energy from captured DM particles that is deposited inside the star. Additionally, note that even if $E$ is significant, its effects on the DM luminosity can be irrelevant if $E\tau_{eq}\ll1$. 

Using these calculations for the luminosity from dark matter self-annihilations, we can estimate an upper bound on the mass of Pop~III stars shining at the Eddington Limit. For stars that are radiation pressure dominated, the mass and luminosity become linearly proportional. Additionally, any further accretion or additional luminosity for a star is disrupted by the radiation pressure and is not allowed. We can write the Eddington Luminosity as:
\begin{equation}
    L_{Edd} = \frac{4\pi c G M_\star}{\kappa_\rho},
\end{equation}
where $G$ is the Universal gravitational constant, $c$ is the speed of light, $M_\star$ is the mass of the star in question, and $\kappa_\rho$ is the stellar atmospheric opacity. The dominant opacity source in metal-free, hot atmospheres is Thompson electron scattering, which is a function of the hydrogen fraction ($X$) of the star: $\kappa_\rho = \kappa_{es} = 0.2(1+X) \unit{cm}^2 \unit{s}^{-1}$. As the star ages, the hydrogen fraction decreases while the fraction of other elements increases, making $\kappa_\rho$ a function of the age/metallicity of the star. In this work we assume a big bang nucleosynthesis (BBN) composition of Pop~III stars, resulting in an Eddington luminosity of:
\begin{equation}
    L_{Edd} = 3.7142 \times 10^{4} (M_\star/M_\odot)L_\odot
\end{equation}

Re-interpreting the Eddington Luminosity not as a maximum luminosity, but as a maximum mass bound, we can calculate what the maximum mass of Pop~III stars would be, via  the following criterion:
\begin{equation}
    L_{nuc}(\Mstar) + L_{DM}(\Mstar)\leq L_{Edd}(\Mstar),
    \label{eq:LeddMaxMass}
\end{equation}
 with the bound being saturated for a star of $\Mstar=M_{max}$. We include contributions to the luminosity from both DM-DM annihilations and hydrogen burning in the core. For the contribution from nuclear fusion, we find an interpolating function that fits well the Zero Age Main Sequence (ZAMS) Pop~III models as tabulated in Table \ref{table:params}. We therefore assume that, to a good approximation, the rate of hydrogen fusion, and therefore $L_{nuc}$, will not be affected by dark matter annihilations taking place inside the stellar core. A full hydrodynamic simulation, which is beyond the scope of this paper, would be required to account for the possible effect of the DM annihilations on the nuclear luminosity. However, based on the stellar thermostat effect, we expect that the core temperature, and therefore the nuclear luminosity, will not change significantly even if there is an additional source of energy from DM annihilations. If anything, DM annihilations would lead to an increase in core temperature, and therefore to an increase in the nuclear luminosity. As such, our bounds should be viewed as conservative.  

In Table~\ref{table:params} we list the relevant parameters, such as mass, radius, escape velocity, and luminosity due to nuclear fusion, for Pop~III stellar models from~\cite{Iocco:2008,Windhorst:2019,Ohkubo:2009}.

\begin{table*}[!htb]
\centering
\begin{tabular}{llll}
\hline
$M_\star [M_{\odot}]$  & $R_\star [R_{\odot}]$    & $v_{esc} [v_{esc,\odot}]$ & $L_{nuc} [L_{\odot}]$ \\ \hline 
1         & 0.875 & 1.072      & $1.91 \times 10^0$  \\
1.5       & 0.954 & 1.257      & $1.05 \times 10^1$  \\
2         & 1.025 & 1.401      & $3.29 \times 10^1$  \\
3         & 1.119 & 1.642      & $1.46 \times 10^2$  \\
5         & 1.233 & 2.019      & $8.46 \times 10^2$  \\
10        & 1.400 & 2.680      & $7.27 \times 10^3$  \\
15        & 1.515 & 3.156      & $2.34 \times 10^4$  \\
20        & 1.653 & 3.488      & $5.11 \times 10^4$  \\
30        & 2.123 & 3.769      & $1.45 \times 10^5$  \\
50        & 2.864 & 4.190      & $4.25 \times 10^5$  \\ 
100       & 4.118 & 4.942      & $1.40 \times 10^6$  \\        
200       & 6.140 & 5.723      & $3.97 \times 10^6$  \\        
300       & 7.408 & 6.382      & $6.57 \times 10^6$  \\
400       & 9.030 & 6.674      & $9.89 \times 10^6$  \\
600       & 11.24 & 7.326      & $1.61 \times 10^7$  \\        
1000      & 12.85 & 8.845      & $2.02 \times 10^7$  \\ \hline
\end{tabular}
\caption{Stellar mass, radius, and luminosity in solar units for the Zero Age Main Sequence (ZAMS) Pop~III models of \cite{Iocco:2008,Windhorst:2019,Ohkubo:2009} we consider in this paper.}
\label{table:params}
\end{table*}

\begin{figure} [!thb]
    \centering
    \includegraphics[width=.8\linewidth]{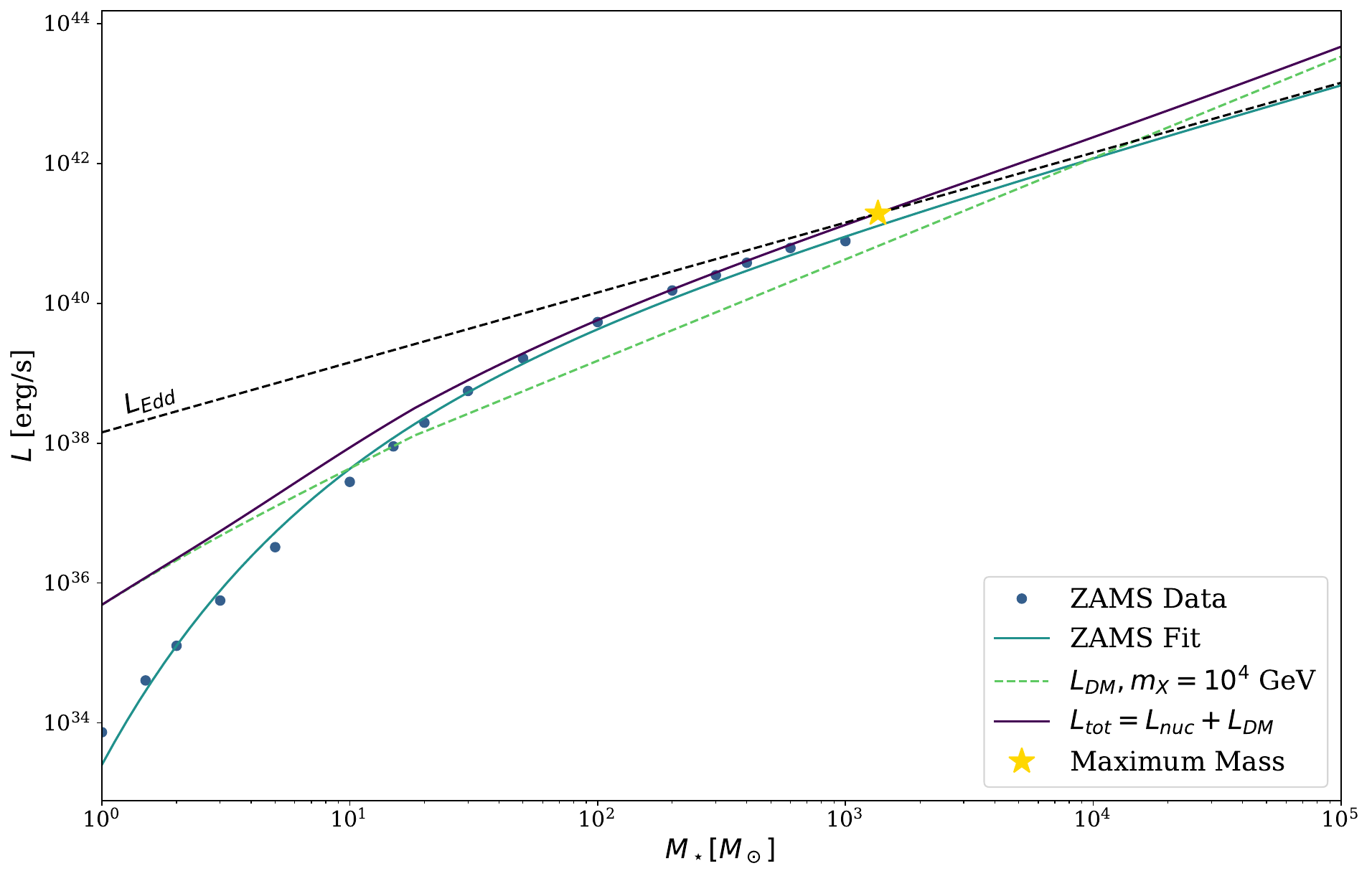}
    \caption{Luminosity as a function of stellar mass for $10^{4}$ GeV dark matter and a DM density of $\rho_X = 10^{16}$ GeV cm$^{-3}$. Two things to note are: (a) the nuclear fusion luminosity approaches the Eddington limit for large stellar masses and (b) the sum of $L_{DM}$ and $L_{nuc}$ and its intersection with $L_{Edd}$ defines the maximum mass for a given density and mass of Dark Matter. In this specific calculation, we find a maximum mass of $M_{max} \sim 1100 M_\odot$. We can see the same result in Fig.~\ref{fig:mmax} -- the maximum mass of a Pop III star considering $10^4$ GeV dark matter at an ambient density of $10^{16}$ GeV cm$^{-3}$ is of order $10^3 M_\odot$.}
    \label{fig:zamsfit}
\end{figure}

As mentioned previously, in order to apply our Eddington limit criterion (Eq.~(\ref{eq:LeddMaxMass})), and therefore find the maximum mass a Pop~III can have if the effects of captured DM annihilations are taken into account, we need a fitting formula for $L_{nuc}(\Mstar)$. We find:
\begin{equation} \label{eq:Lnuc}
L_{n u c} \simeq 10^{\frac{\log \left(3.71 \times 10^{4} L_{\odot} \mathrm{s} / \mathrm{erg}\right)}{1+\exp (-0.85 \log(x)-1.95)}} \cdot x^{\frac{2.01}{x^{0.48}+1}} \operatorname{erg} / \mathrm{s},
\end{equation}
where $x \equiv \frac{M_\star}{M_\odot}$ and $L_\odot \equiv 3.846 \times 10^{33} \operatorname{erg} / \mathrm{s}$. This formula interpolates between the lower mass regime ($L_{nuc} \propto M_\star^3$) and the Eddington limited regime ($L_{nuc} \propto M_\star$). An example calculation of the Eddington limit, DM luminosity, and nuclear luminosity is presented in Fig.~\ref{fig:zamsfit}. 

In Fig.~\ref{fig:zamsfit}, the maximum mass corresponds to the intersection of the sum $L_{DM} + L_{nuc}$ with the Eddington limit. We note a break in the power law for the Dark Matter luminosity around $M_\star \sim 20 M_\odot$. This power law comes from the dependence on radius with mass that we derived for our Pop~III star data. The piecewise expression we adopt for the rest of this paper is:

\be\label{eq:HomoRels}
  \frac{\Rstar}{R_\odot}\approx
  \begin{cases}
                                   0.88\left(\frac{M_\star}{\Msun}\right)^{0.20} & \text{if $M_\star \lesssim 20M_\odot$} \\
                                   0.32\left(\frac{M_\star}{\Msun}\right)^{0.55} & \text{if $M_\star \gtrsim 20M_\odot$.} \\
  \end{cases}
\ee
These homology relations were found by fitting two distinct power laws to the data in Table \ref{table:params}. Since in Fig.~\ref{fig:zamsfit} we consider the case of a $10^{4}\GeV$ DM particle, in view of Eq.~(\ref{eq:CtotSSLM}), $L_{DM}\propto \Mstar^2/\Rstar$, and in view of Eq.~(\ref{eq:HomoRels}) we predict, and confirmed numerically, that $L_{DM}\propto\Mstar^{1.8}$ (for $\Mstar\lesssim 20\Msun$), and $L_{DM}\propto\Mstar^{1.45}$ (for $\Mstar\gtrsim 20\Msun$). We note that for $m_X\gtrsim3m(\vesc/\vbar)^2$, we expect a different scaling of $L_{DM}$ with $\Mstar$, in view of Eqns.~(\ref{eq:CtotMS})-~(\ref{eq:CtotSSHM}). Namely, $L_{DM}\propto\Mstar^{2.6}$ (for $\Mstar\lesssim 20\Msun$), and $L_{DM}\propto\Mstar^{1.9}$ (for $\Mstar\gtrsim 20\Msun$), respectively. The main point is that {\it{all}} of those indicate an increase with stellar mass faster than $\Mstar$. Therefore, for a sufficiently large $\Mstar$, one is guaranteed to find that the sum $L_{nuc}+L_{DM}$ reaches the Eddington limit, which directly implies a maximum stellar mass. 

\begin{figure} [!thb]
    \centering
    \includegraphics[width=.8\linewidth]{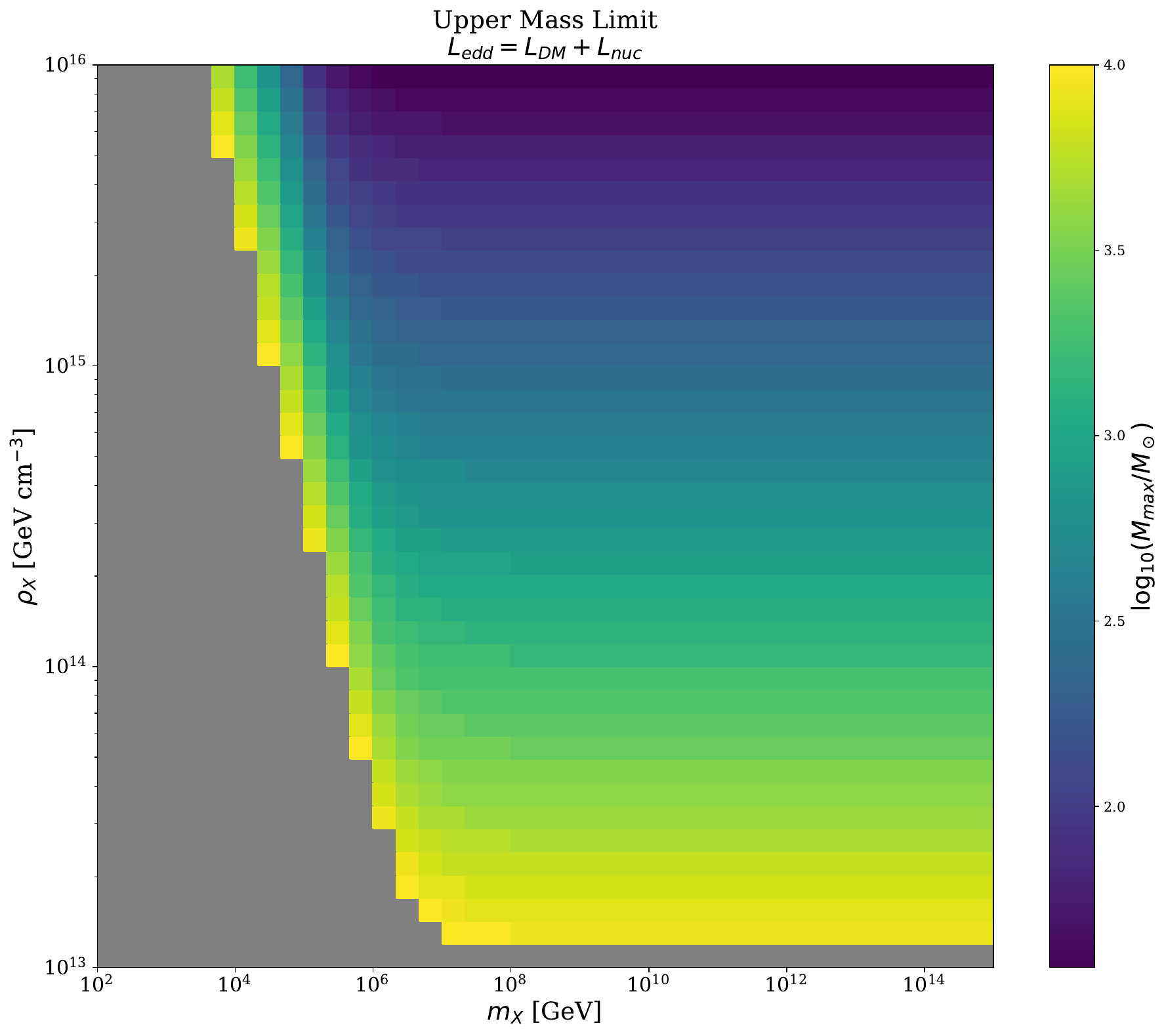}
    \caption{Maximum stellar mass as a function of $m_X$ and $\rho_X$ assuming X1T SI DM-proton cross section bounds and $\vbar = 10^{6} \unit{cm} \unit{s}^{-1}$ when including the effects of annihilation of captured dark matter by Pop~III stars. The gray area corresponds to bounds weaker than $10^4\Msun$, where other mechanisms, such as fragmentation of the gas cloud or radiative feedback, would be dominant in determining the maximum stellar mass~\cite{Stacy:2016}. }
    \label{fig:mmax}
\end{figure}

Upper bounds on Pop~III stellar masses obtained by imposing the sub-Eddington condition (Eq.~(\ref{eq:LeddMaxMass})) and assuming XENON1T SI limits on $\sigma$, are shown in Fig.~\ref{fig:mmax}. For $\vbar$ we have assumed a fiducial value of $10~\unit{km/s}$, representative of the $10^6\Msun$ minihalos hosting Pop~III stars~\cite{Freese:2008cap} (see also Appendix~\ref{sec:DMHalos}). Reading the plot vertically, we note that for a given $m_X$, an increase in the ambient DM density, $\rho_X$, leads to tighter bounds, as evidenced by the darkening of the colors in the heatmap as we progress upward, towards higher $\rho_X$, in bins of fixed $m_X$. This is to be expected, since $L_{DM}\propto\rho_X$, and therefore $M_{max}$ is inversely proportional to the ambient DM density. We now move to discussing the trends in the heatmap if we read it horizontally, keeping $\rho_X$ fixed. Remember, $L^{max}_{DM}\propto m_X$, whenever $m_X\lesssim3m(\vesc/\vbar)^2$, and  $L^{max}_{DM}\propto m_X^0$, when $m_X\gtrsim3m(\vesc/\vbar)^2$, as evidenced by the two distinct trends of $L^{max}_{DM}$ in Fig.~\ref{fig:LDM}. This implies upper bounds on Pop~III stellar mass that are insensitive with $m_X$ at the higher end of the DM particle mass range, and bounds that become weaker as we decrease $m_X$, whenever $m_X\lesssim3m(\vesc/\vbar)^2$. Both of those trends can be seen in Fig.~\ref{fig:mmax}.

In the next section we demonstrate that the mere observation of a Pop~III star of any mass can be, in principle, used to place constraints on the DM-proton scattering cross section.


\section{Constraining DM properties using Pop~III stars}\label{sec:Constrain}

In this section we demonstrate a method for placing bounds on Dark Matter properties through the observation of Pop III stars. As discussed in the previous section, the capture and annihilation of Dark Matter particles by Pop III stars in dense DM environments provides an additional source of stellar luminosity. This extra power source places limits on the maximum mass it can attain via the Eddington luminosity, as demonstrated in Sec.~\ref{sec:obs} (Fig.~\ref{fig:mmax}). Here, instead, we pose the following question: what information about Dark Matter can we ascertain if we were to observe any Pop III star, of a given mass? The mere existence of the star already implies something about the luminosity due to captured DM: $L_{DM}\leq L_{Edd}(\Mstar)-L_{nuc}(\Mstar)$, which is just the sub-Eddington condition of Eq.~(\ref{eq:LeddMaxMass}), re-arranged in order to demonstrate the idea of constraining DM properties. Whenever $\vbar\ll\vesc$ (which is the case for Pop~III stars), and $m_X\gg m$~\footnote{For the case of $m_X\ll m$ all we need to do is to replace $m_X\leftrightarrow m$ in $A_N$.}, it turns out that this condition can be recast as:

\be
f m_X \underbrace{\sqrt{24 \pi} G M_\star R_\star\frac{\rho_X}{m_X\bar{v}} \sum_{N=1}^{\infty} p_N (\tau) \left(1-\left(1+\frac{2 A_{N}^{2} \bar{v}^{2}}{3v_{esc}^{2}}\right) e^{-A_{N}^{2}}\right)}_\textrm{Total Capture Rate ($C_{tot}\equiv\sum_{N=1}^{\infty}C_N$)} \leq L_{Edd}(M_\star) - L_{nuc}(M_\star),
\label{eq:ConstrainDM}
\ee
with $A_N^2=(3Nm\vesc^2)/(m_X\vbar^2)$, when $m_X\gg m$, and $A_N^2=(3Nm_X\vesc^2)/(m\vbar^2)$, when $m\gg m_X$. We have used the fact that $L_{DM}=fm_XC_{tot}$ (Eq.~(\ref{eq:LDMeqCtotNoEvap}))~\footnote{At low $m_X$ we include the effects of DM Evaporation as per Eq.~(\ref{eq:LDMeqCtot}). }, and for the total capture rate $C_{tot}=\sum_{N=1}^{N=\infty} C_N$, with $C_N$ given by Eq.~(\ref{eq:CNapprox}). In all of our numerical results we used the full, non-approximated, $C_N$ from Eq.~(\ref{eq:CN}). However, in order to gain physical insight and understand the behavior of our bounds, it is easiest if we use the approximated $C_N$ of Eq.~(\ref{eq:CNapprox}), as done in Eq.~(\ref{eq:ConstrainDM}). Since the sum on the lhs is directly proportional to the DM-proton scattering cross section~\footnote{This subtle point can be most easily understood if we look at the limiting behaviours from Eq.~(\ref{eq:CtotMS}) and Eqns.~(\ref{eq:CtotSSHM})-(\ref{eq:CtotSSLM}). For more details see Appendix~\ref{sec:MSCapture}.}, we can use Eq.~(\ref{eq:ConstrainDM}) to place upper bounds on $\sigma\times\rho_X$. Using constraints on $\sigma$ from direct detection experiments, we can break this degeneracy, and use our method to place upper bounds on $\rho_X$ at the center of minihalos hosting Pop~III stars. Additionally, one can estimate the ambient DM density at the location of the star. In Appendix~\ref{sec:DMHalos}, we apply the well-established adiabatic contraction formalism to do just that. This leads to the exciting possibility of constraining the DM-proton scattering cross section ($\sigma$) via Eq.~(\ref{eq:ConstrainDM}) by finding numerically the value of $\sigma$ which saturates the inequality.

\begin{figure} [!htb]
    \centering
    \includegraphics[width=.8\linewidth]{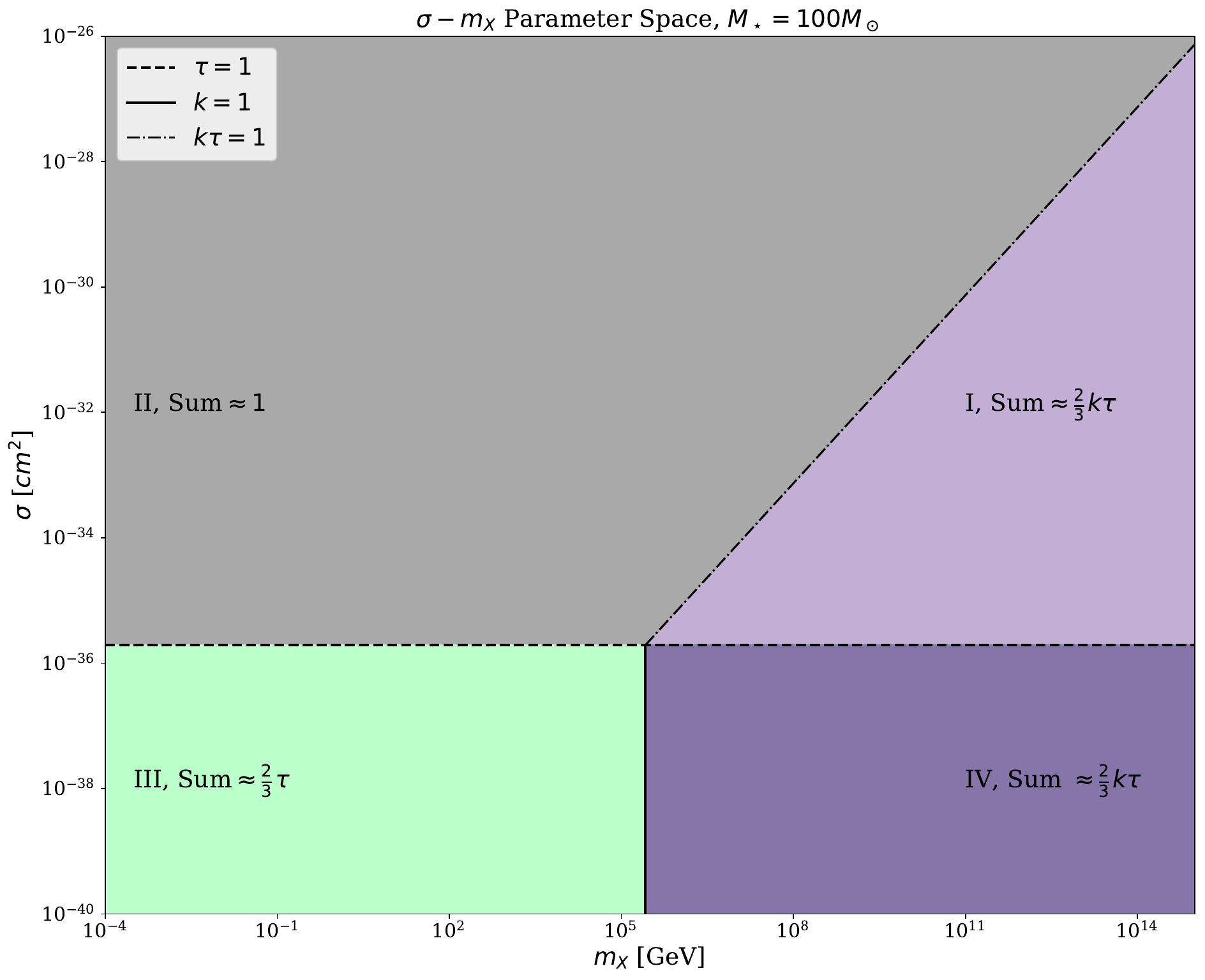}
    \caption{Leading order approximate values of $\sum_{N=1}^{\infty} p_N (\tau) \left(1-\left(1+\frac{2 A_{N}^{2} \bar{v}^{2}}{3v_{esc}^{2}}\right) e^{-A_{N}^{2}}\right)$ in various regions of the $\sigma-m_X$ parameter space are given in each corresponding region. The line of $\tau=1$ separates the single scatter ($\tau\lesssim1$) from multiscatter regime ($\tau\gtrsim1$). The multiscatter region can be further subdivided into two regions: Region~I ($\tau\gtrsim 1$ and $k\tau\lesssim1$) and Region~II ($\tau\gtrsim 1$ and $k\tau\gtrsim1$), where the sum takes the value: $2/3k\tau$, and $1$, respectively. We define $k$ in the following way: $k\equiv A_1^2=\frac{3\vesc^2}{\vbar^2}\frac{\min(m_X;m)}{\max(m_X;m)}$. Furthermore, note that the line $k=1$ separates the single scatter capture in two two distinct regions: $k\gg1$, where the sum is $2/3\tau$ (Region III), and $k\ll1$, where the sum is $2/3k\tau$ (Region IV). Most remarkably, we find that in region IV (single scatter, and $k\tau\ll1$) and region I (multiscatter, and $k\tau\ll1$) the sum, and therefore the capture rates, have the exact same parametric scaling.}
    \label{fig:sigma_mx_ps}
\end{figure}

We start with our projected bounds on $\rho_X\times\sigma$, inferred from assuming the possible identification of of Pop~III stars of various mass. To this aim, we recast Eq.~(\ref{eq:ConstrainDM}) by isolating on the lhs all the unconstrained parameters (in this case $\rho_X$ and $\tau\propto\sigma$):
\be\label{eq:ConstrRhoSigma}
\rho_X\sum_{N=1}^{\infty} p_N (\tau) \left(1-\left(1+\frac{2 A_{N}^{2} \bar{v}^{2}}{3v_{esc}^{2}}\right) e^{-A_{N}^{2}}\right)\leq\frac{L_{Edd}(M_\star) - L_{nuc}(M_\star)}{\sqrt{24\pi}f}\frac{\vbar}{G\Mstar\Rstar}
\ee
Next, we approximate the sum in Eqns.~(\ref{eq:ConstrainDM})-(\ref{eq:ConstrRhoSigma}), and  find that it takes three possible values: $1$ (Region II, i.e. $\tau\gg1$ and $k\tau\gg1$), $2/3\tau$ (Region III, i.e. $\tau\ll1$, and $k\gg1$), and $2/3k\tau$ (Region IV, i.e. $k\ll1$ and $\tau\ll1$, and Region I, i.e. $k\tau\gg1$ and $\tau\gg1)$ . This allows us to explicitly express $L_{DM}\propto\rho_X\sigma$, as expected. See Appendix~\ref{sec:MSCapture} for details on how we obtained the three approximate values mentioned above. In Fig.~\ref{fig:sigma_mx_ps}, we plot the various regions of validity for those three approximations in the $\sigma$-$m_X$ parameter space.
Note that the location of the $k=1$, $\tau=1$, and $k\tau=1$ lines that separate the $\sigma-m_X$ parameter space into the four regions (labeled I-IV in Fig.~\ref{fig:sigma_mx_ps}) will be different for different mass stars, which can be most easily understood from the following scaling relations for $\tau=2\Rstar\sigma n_T$ (with $n_T$ the number density of target nuclei) and $k\equiv3\frac{\min(m;m_X)}{\max(m;m_X)}\frac{\vesc^2}{\vbar^2}$ (with $m$ being the mass of the target nuclei):
\begin{eqnarray}
k & \approx & 10^{4}\frac{\Mstar}{\Msun}\frac{\Rsun}{\Rstar}\left(\tenovervbar\right)^2\frac{\min(m_X;m)}{\max(m_X;m)}\label{eq:k}\\
\tau & \approx & 10^{-5}\sigmaoversigma\frac{\Mstar}{\Msun}\left(\frac{\Rsun}{\Rstar}\right)^2\label{eq:tau}.
\end{eqnarray}
Throughout our work, we assume that collisions with the more abundant $H$ nuclei (i.e. protons) dominates the capture. This simplification leads to an underestimate of the total capture rates, and therefore all of our bounds would become more stringent if the effects of collisions with $He$ nuclei were taken into account. 
In obtaining the result of Eq.~(\ref{eq:tau}), we assumed the fraction of $H$ in a Pop~III star to be given by BBN, i.e. $X\approx 0.75$. 

Having found approximations for $\sum_{N=1}^{\infty} p_N (\tau) \left(1-\left(1+\frac{2 A_{N}^{2} \bar{v}^{2}}{3v_{esc}^{2}}\right) e^{-A_{N}^{2}}\right)$, we can use them to calculate the capture rates in each of the four regions identified (see the underbraced part of Eq.~(\ref{eq:ConstrainDM}) or Eq.~(\ref{eq:CtotApprox})). Perhaps the most intriguing, and somewhat unexpected region of the $\sigma-m_X$ parameter space is what we called Region II, in which the sum attains its maximum value, $1$. Physically, in that region, the scattering cross section is sufficiently high to efficiently lead to the capture of {\it{all}} DM particles crossing the star. This leads to a particularly simple form for the total capture rate, which now becomes just the number of DM particles crossing the star per unit time, i.e. flux $\times$ area:

\be\label{eq:CtotScalII}
C_{tot}^{II}\approx 8\times 10^{43}~\unit{s}^{-1}\left(\frac{\rho_X}{10^{14}~\GeV~\unit{cm}^{-3}}\right)\left(\frac{10^2~\GeV}{m_X}\right)\left(\tenovervbar\right)\frac{\Mstar}{\Msun}\frac{\Rstar}{\Rsun}.
\ee
Continuing to Region III (single scatter and $k\gg1$) we find:
\be\label{eq:CtotScalIII}
C_{tot}^{III}\approx 5.4\times 10^{38}~\unit{s}^{-1}\left(\frac{\rho_X}{10^{14}~\GeV~\unit{cm}^{-3}}\right)\sigmaoversigma\left(\frac{10^2~\GeV}{m_X}\right)\left(\tenovervbar\right)\left(\frac{\Mstar}{\Msun}\right)^2\left(\frac{\Rstar}{\Rsun}\right)^{-1}.
\ee
This is just the scaling from Eq.~(\ref{eq:CtotSSLM}), with numerical factors explicitly shown here. Moving to regions IV (single scatter and $k\ll 1$) and I (multi scatter and $k\tau\ll1)$, we find, remarkably, that the capture rates have the exact same form:
\be\label{eq:CtotScalIIV}
C_{tot}^{I}=C_{tot}^{IV}\approx6.26\times10^{28}~\unit{s}^{-1}\left(\frac{\rho_X}{10^{14}~\GeV~\unit{cm}^{-3}}\right)\sigmaoversigma\left(\frac{10^8~\GeV}{m_X}\right)^2\left(\tenovervbar\right)^3\left(\frac{\Mstar}{\Msun}\right)^3\left(\frac{\Rstar}{\Rsun}\right)^{-2}.
\ee
This is a highly counter-intuitive result, since in Region IV the single scatter approximation holds, whereas Region I is where the multiscatter approach is necessary. The fact that there is a smooth continuity between those two is not unexpected. What is surprising is the large swath of parameter space for which both the single scatter approximation and the multiscatter yield exactly the same result, even if in one case the controlling parameter $\tau$ is much larger than one (Region I) vs. much less than unity (Region IV). 

\begin{figure} [!thb]
    \centering
    \includegraphics[width=.8\linewidth]{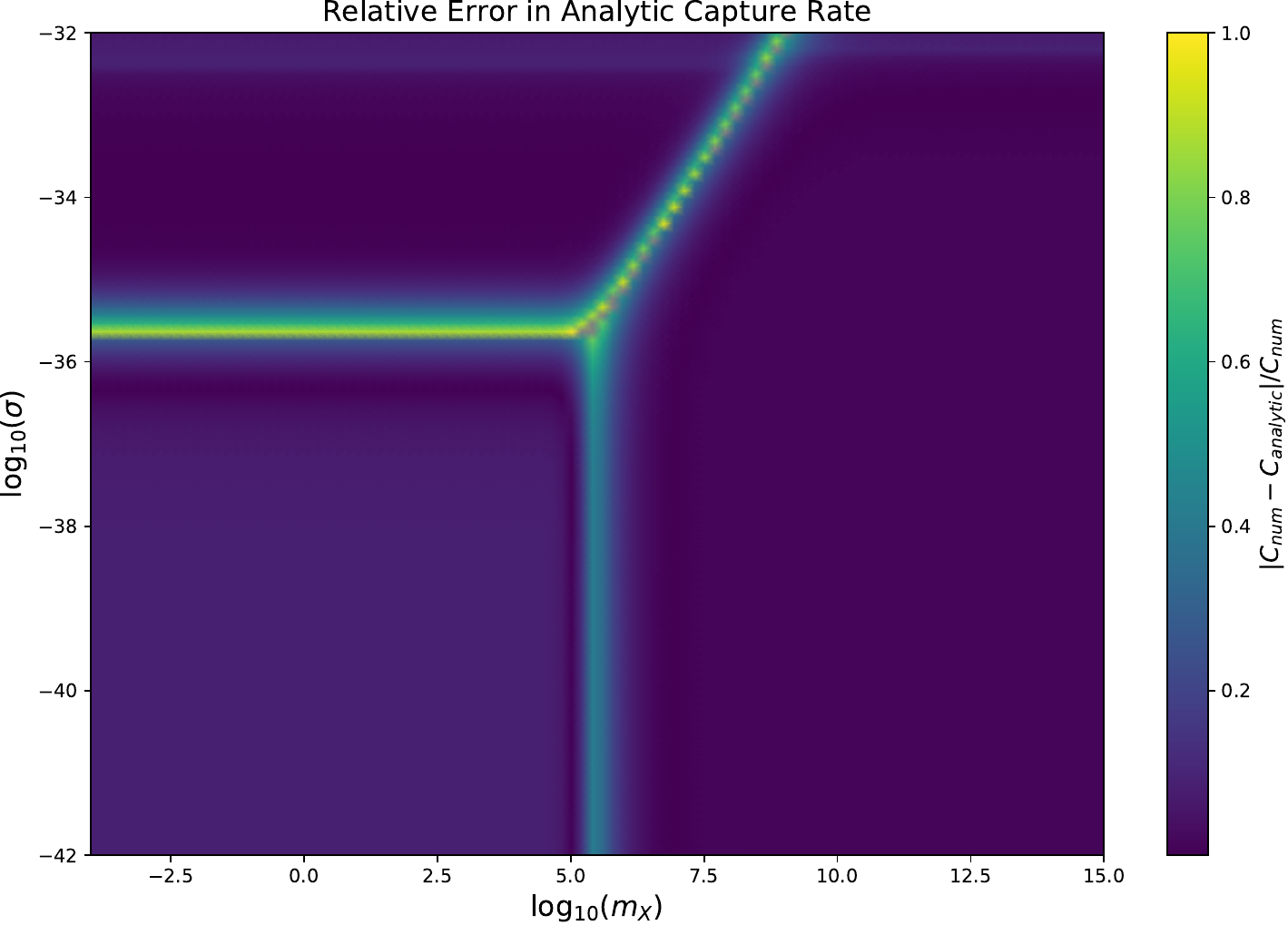}
    \caption{The relative error in the analytic approximations of the capture rate in the four regions of interest (I-IV), when compared to a full numerical calculation. Note how, apart from the naturally emerging boundary lines defined by $\tau=1$, $k=1$, and $k\tau=1$, our approximations hold very well, with a relative error less than $10\%$ throughout.}
    \label{fig:CApproxValidation}
\end{figure}

 In Fig.~\ref{fig:CApproxValidation} we present a numerical validation of our analytic approximations of the total capture rates of Eqns.~(\ref{eq:CtotScalII})-(\ref{eq:CtotScalIIV}). Note the excellent agreement between the full numeric result and our approximations, which only breaks down at boundaries of regions II and III. 

\begin{figure} [!thb]
    \centering
    \includegraphics[width=.8\linewidth]{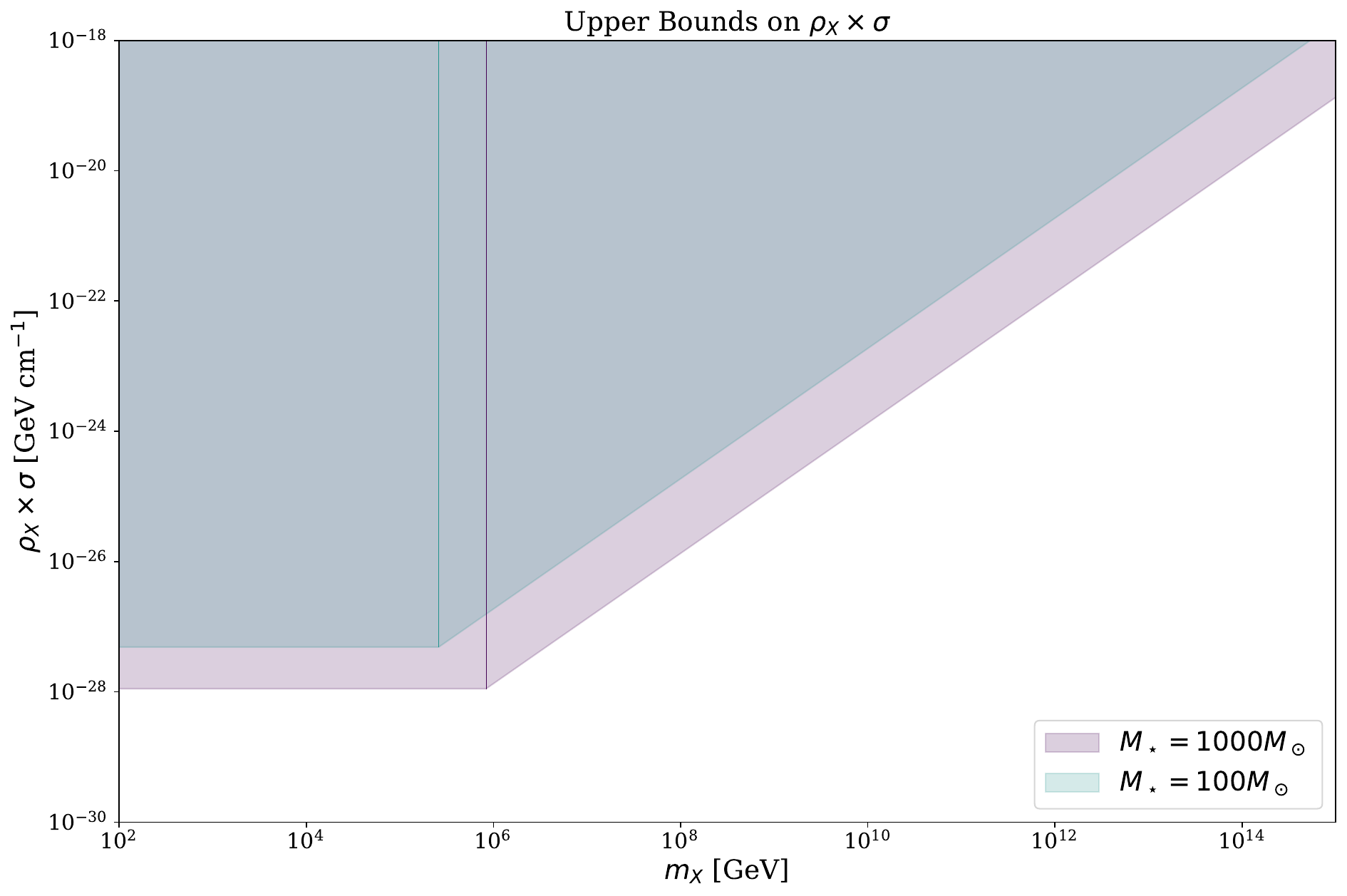}
    \caption{Projected bounds on $(\rho_X\times\sigma)$ vs. $m_X$ imposed by the potential observation of Pop~III stars. In obtaining these limits we only assume the observation of a hypothetical Pop~III star, of a given mass. The thin vertical lines correspond to $k\equiv 3\frac{m}{m_X}\frac{\vesc^2}{\vbar^2}=1$, for each star.}
    \label{fig:RhoSig-Mchi_Bounds}
\end{figure}

 From Eq.~(\ref{eq:ConstrainDM}), and using the three different forms of $C_{tot}$ from Eqns.~(\ref{eq:CtotScalII})-(\ref{eq:CtotScalIIV}), we can place numerical bounds on $\rho_X\sigma$. Note that from the independence of $\sigma$  in $C_{tot}^{II}$, the total capture rate in Region~II, we could directly constrain the DM density at the location of the star, without any knowledge of $\sigma$. For  compact objects, such as Neutron Stars, it turns out that Region~II is in parameter space that is not yet ruled out by direct detection experiments, for both SD and SI $\sigma$. This means that, in principle, if the effects of DM heating could be observed in Neutron Stars,  besides acting as probes of DM, NS could also be used to constrain the DM density in their environment, if capture becomes so efficient such that the entire DM flux crossing the NS is trapped. Returning to the focus of our paper, Pop~III stars, we get the following constraints on $\rho_X\sigma$:
\be\label{eq:ConstRhoSigmaRegions}
\rho_X\sigma\lesssim
\begin{cases}
\left(\frac{\pi}{6}\right)^{1/2}\frac{\vbar}{\vesc^2}\frac{m}{X\Mstar}\frac{L_{Edd}(\Mstar)-L_{nuc}(\Mstar)}{f}, & \text{for Region~III}\\
\left(\frac{\pi}{54}\right)^{1/2}\frac{\vbar^3}{\vesc^4}\frac{m_X}{X\Mstar}\frac{L_{Edd}(\Mstar)-L_{nuc}(\Mstar)}{f}, & \text{for Regions~IV and I.}
\end{cases}
\ee
The above equation comes from Eq.~(\ref{eq:ConstrRhoSigma}) by using the appropriate approximations for the sum on the lhs: $2/3\tau$ (Region~III) and $2/3k\tau$ (Regions~IV and I). As usual, by $X$ we denote the hydrogen mass-fraction of the star, and $f$ the fraction of the annihilation energy deposited in the star. In obtaining the bounds for Region~IV ($\tau\ll1$ and $k\ll1$) and Region~I ($\tau\gg1$ and $k\tau\ll1)$, we have explicitly replaced $k$ with its definition: $k\equiv 3\frac{\min(m_X;m)}{\max(m_X;m)}\frac{\vesc^2}{\vbar^2}$. From Eq.~(\ref{eq:ConstRhoSigmaRegions}), we expect that our bounds on $\rho_X\sigma$ vs. $m_X$ will be constant for lower $m_X$ (i.e. Region~III, where $k\gg 1$) and will scale linearly with $m_X$ at larger DM particle mass, corresponding to $k\ll 1$ (Regions~I and~IV), a trend that can be seen explicitly in Fig.~\ref{fig:RhoSig-Mchi_Bounds}.
 As expected, the tightest bounds in Fig.~\ref{fig:RhoSig-Mchi_Bounds} are placed through the observation of the most massive Pop III stars, since more massive stars lead to more efficient capture rates, and therefore a larger $L_{DM}$. This is a major benefit of our method, as more massive stars are easier to detect than their less massive counterparts due to their greater luminosity. 

\begin{figure} [!thb]
    \centering
    \includegraphics[width=.8\linewidth]{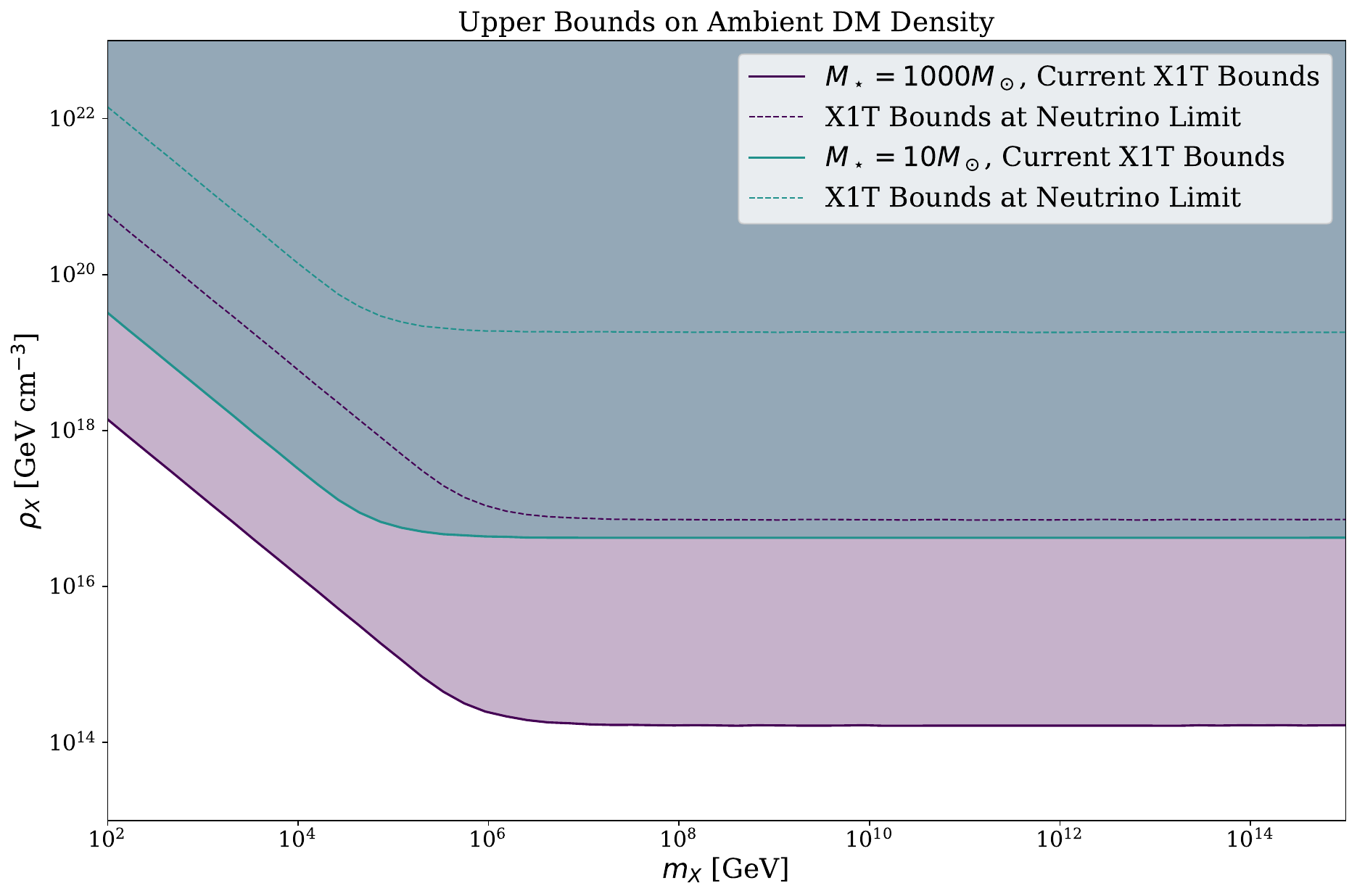}
    \caption{Projected constraints on ambient DM density at the center of Pop~III host DM mini-halos as a function of DM mass, assuming $\sigma$ have been positively identified by SD direct detection experiments. For the solid/dashed exclusion limit lines we assume the DM-proton scattering cross section at the current XENON1T limit/XENON neutrino floor given by~\cite{Aprile:2018, Kavanagh:2018}. 
    The shaded regions represent the regions in the $\rho_X - m_X$ parameter space ruled out by the detection of Pop~III star with mass $M_\star$. The purple lines/regions represent the bounds when $M_\star = 1000 M_\odot$ while the blue lines/regions represent the bounds when $M_\star = 10 M_\odot$.}
    \label{fig:Rho-Mchi_Bounds}
\end{figure}

As mentioned before, the DM luminosity is sensitive to the product $\rho_X\times\sigma$, both in the single and multiscatter capture regimes. As such, without any other information aside from the mass of a hypothetically observed Pop~III star, we can only constrain this product. We now proceed to break down the degeneracy between $\rho_X$ and $\sigma$, and place exclusion limits on each of those two independent parameters. If direct detection experiments are to find the DM particle, both $\sigma$ and $m_X$ are going to be in a relatively narrow swath of the  $\sigma-m_X$ parameter space, between current bounds and the neutrino floor. In Fig.~\ref{fig:Rho-Mchi_Bounds}, we calculate projected bounds on the ambient DM density at the location of Pop~III stars, implied by the observation of a Pop~III star, and assuming the DM-proton scattering cross section is anywhere in the band of parameter space where SI direct detection experiment could identify it. This represents a method for constraining the central DM density in halos, a parameter that is beyond the reach of current numerical simulations.  
In order to place the constraints on $\rho_X$ presented in Fig.~\ref{fig:Rho-Mchi_Bounds}, we numerically solve Eq.~(\ref{eq:ConstrRhoSigma}) for the DM density that saturates the sub-Eddington bound for a variety of proton-DM cross sections and for hypothetical Pop~III stars with mass ranging between $10\Msun$ and $1000\Msun$. For $\sigma$, we assume values that are still allowed by direct detection experiments, but above the neutrino floor. Note that our projected bounds will actually become weaker as direct detection experiments further constrain $\sigma$ to lower values. This is to be expected, since pushing $\sigma$ to lower values implies higher $\rho_X$ in order to maintain the capture rate, and $L_{DM}$, constant. We note a broken power law behavior for our projected upper bounds on $\rho_X$ with $m_X$: at high $m_X$ the bounds are constant, whereas at lower $m_X$ they scale like $m_X^{-1}$. Both of those are a consequence of the two different scaling relations valid for the total capture rate: $C_{tot}\sim\rho_X\sigma/m_X$ (Region~III, where $k\gg1$) and $C_{tot}\sim\rho_X\sigma/m_X^2$ (Regions~IV and~I, where $k\ll1$). Since we use bounds from direct detection: $\sigma\propto m_X$, leading to $L^{max}_{DM}\sim\rho_X m_X$ (Region~III) and $L^{max}_{DM}\sim\rho_X$ (Regions~IV and~I). This, in turn, leads to the observed broken power law trend in the $\rho_X$ projected upper bounds. Additionally, for higher mass stars, the bounds are stronger, which is a consequence of higher capture rates for the case of more massive Pop~III stars.  For the most massive star we consider here, $M_\star = 1000 M_\odot$, using the current best bounds on $\sigma$ places a limit on $\rho_X$ as low as $10^{14}$ GeV cm$^{-3}$ for DM masses $\gtrsim 10^{6}$ GeV. This value ranges from $\sim 10^{14} - 10^{17}$ GeV cm$^{-3}$ for $\sigma$ between the current best bounds and the neutrino floor. For the lowest DM mass ($m_X = 10^2$ GeV), our bounds range from $\sim 10^{18} - 10^{21}$ GeV cm$^{-3}$. A similar analysis on the $M_\star = 10 M_\odot$ case shows a limit of $\sim 10^{17} - 10^{20}$ GeV cm$^{-3}$ for DM masses $\gtrsim 10^{5}$ GeV and $\sim 10^{19} - 10^{22}$ GeV cm$^{-3}$ for $m_X = 10^2$ GeV. We want to emphasise once more that our constraints on $\rho_X$ are forecast bounds, assuming identification of DM from direct detection experiments and the observation of a Pop~III star of a given mass.   

We discuss below the implications of our results regarding the possibility of constraining the DM density  at the center of DM halos hosting Pop~III stars. Fig.~\ref{fig:Rho-Mchi_Bounds} outlines the main findings: a way to constrain the DM ambient density towards the center of halos with maximum values as low as $\rho_X \sim 10^{14}$ GeV cm$^{-3}$ for the most massive stars. Analytically, Dark Matter halo profiles can be well understood by the Navarro-Frenk-White (NFW) profile, as outlined in \cite{Navarro:1997}. These profiles become altered due to the infall of baryonic matter to the center of the halo, which pulls the dark matter closer towards the center in a process known as adiabatic contraction. Previous work has been done to study DM capture and annihilation in the first stars using these adiabatically contracted profiles \cite{Freese:2008cap}. However, although analytical methods can be used to estimate the ambient DM density at the edge of the baryonic core, as done in \cite{Freese:2008dmdens} and this paper (See Appendix \ref{sec:DMHalos}), numerical simulations, such as those done in \cite{Abel:2001}, are unable to resolve the density towards the edge of the baryonic core. Hence, we provide a novel method for constraining this property through the observation of Pop~III stars, in conjunction with the possible upcoming identification of $\sigma$ and $m_X$ by direct detection experiments.
Our findings, outlined in Fig.~\ref{fig:Rho-Mchi_Bounds}, demonstrate that realistic bounds on the DM density can be placed across all DM masses. 

We conclude this section with the most exciting application of our method: using Pop~III stars to place upper bounds on the DM-proton scattering cross section. In~\cite{Ilie:2020BNF} we apply this method to the candidate Pop~III complex at $z \sim 7$ identified in the MUSE Hubble deep lensed field by~\cite{Vanzella:2020} and find exclusion limits on $\sigma$ that are competitive, or deeper than, those obtained by the most sensitive direct detection experiments to date: XENON1T(SI), and PICO60(SD). Additionally, for SD constraints, our bounds probe well below the neutrino floor. Moreover, for sub-GeV DM, we placed bounds on the DM-proton interaction cross section for WIMP DM and the theoretically motivated Co-SIMP model~\cite{Smirnov:2020}. 

In this paper, we will focus on the projected upper limits resulting from the potential detection of Pop III stars at redshifts of $z\sim 10 - 20$, which is where JWST is most likely to find Pop~III stars. For constraining $\sigma$, we assume a central DM density corresponding to adiabatically contracted NFW profiles with enhanced densities of $\rho_X = 10^{13} - 10^{16}$ GeV cm$^{-3}$ for Pop III stars formed at $z \sim 10 - 20$ (See Appendix \ref{sec:DMHalos} for details on DM densities at the center of Pop III forming halos). The range in $\rho_X$ corresponds to different assumptions on the number density of the collapsing baryonic cloud when compression of the DM densities due to infall of baryons will cease to be efficient. We represent our uncertainty in the central density by placing a range of constraints on $\sigma$, corresponding to the possible range of DM densities. Additionally, we take into consideration the possible effects DM annihilation would have on the ambient DM density. For $2\to2$ processes, such as p/s wave annihilations, one finds: $\rho^{-1}_X(t)=\rho^{-1}_{X0}+\rho^{-1}_{AP}(t)$, with $\rho^{-1}_{X0}$ being the initial DM density, and the annihilation plateau (value reached at later times) given by: $\rho_{AP}(t)=m_X/(\avg{\sigma v} t)$. Regarding $\sigmav$, for WIMPs we use the value that leads to freezeout of the observed thermal relic abundance, appropriate for each case: the standard $\sigmav\sim 10^{-26}~\unit{cm}^3\unit{s}^{-1}$(s-wave) and $\sigmav\sim 10^{-24}/x~\ccpers$~(p-wave)\cite{Lopes:2016}. Unless otherwise specified, $x=m_X/T_X$, with $T_X$ being the captured DM temperature, which we calculate in Appendix~\ref{sec:CapDMTemp}. For non-thermal DM, when $\sigmav$ is not fixed by the relic abundance, in Appendix~\ref{sec:Equilibration} we calculate the lower bound on $\sigmav$ that leads to an equilibration of the capture and annihilation/evaporation processes in a timescale much shorter than the lifetime of the star. We find that this is in both cases much lower than the unitarity limit, therefore equilibration is physically possible. For the case of thermal DM we have explicitly checked in the same Appendix that the freezeout $\sigmav$ is sufficiently high to ensure rapid equilibration. 

\begin{figure} [!htb]
    \centering
    \includegraphics[width=0.9\linewidth]{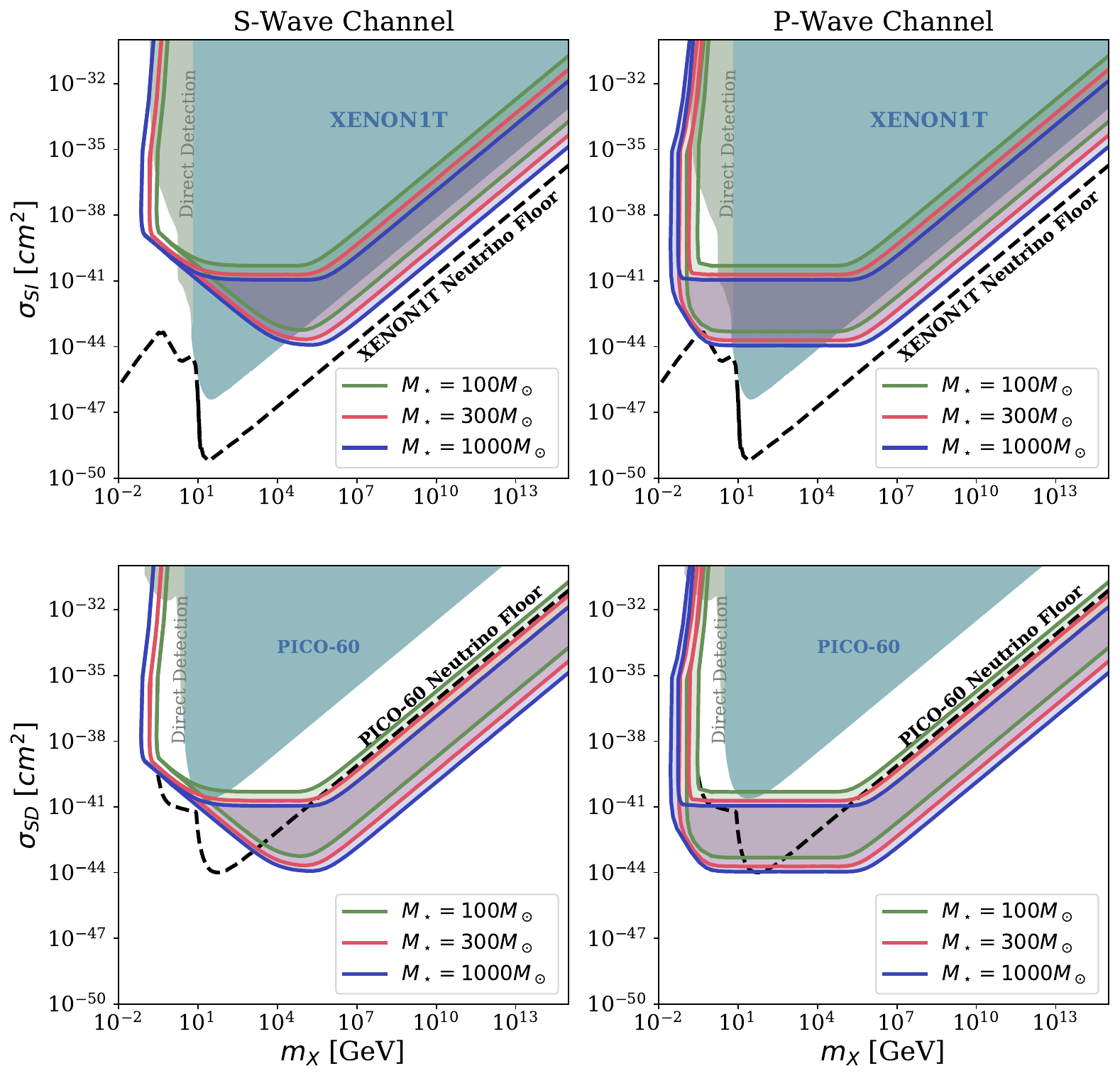}
    \caption{Projected bounds on DM-nucleon scattering cross section as a function of DM mass placed by the potential observation of Pop~III stars with masses ranging from $100\Msun$ (green) to $1000\Msun$ (purple). Ambient DM densities are found from adiabatically contracted NFW profiles. We represent the inherent uncertainty of this procedure by colored shaded regions. They each represent the range of upper bounds on $\sigma$ which can be placed for a given Pop~III stellar mass, when $\rho_X$ takes the following range of possible values:  $10^{13}~\GeV~\unit{cm}^{-3}\lesssim\rho_X(0)\lesssim 10^{16}~\GeV~\unit{cm}^{-3}$. For each star, we consider the effects of annihilations in the region surrounding the star on the ambient density for $t = 1$ My. The left panels represent the bounds placed for the s-wave annihilation channel, with the top being spin independent bounds and the bottom spin dependent bounds. The right panels are bounds placed on WIMP DM annihilating through the p-wave channel, again with the top panel being spin independent and the bottom spin dependent. For the spin-independent bounds, the blue region is the excluded region from the XENON1T experiment, the grey region from the most stringent bounds below 6 GeV \cite{Aprile:2018,Bringmann:2018cvk,Aprile:2019ldm,Abdelhameed:2019}, and the solid black line represents the neutrino floor for the XENON1T experiment. For the SD parameter space, the blue region is the excluded region from the PICO-60 experiment, the grey region from the most stringent bounds below 6 GeV \cite{Bringmann:2018cvk,Amole:2019fdf,Aprile:2019ldm,Aprile:2019} and the solid black line represents the neutrino floor for this experiment. Note that the detection of all Pop~III masses considered here can be used to rule out previously unexplored parameter spaces for sufficiently high DM densities.}
    \label{fig:Sigma-Mchi_Bounds}
\end{figure}

 Fig.~\ref{fig:Sigma-Mchi_Bounds} shows our main results: competitive bounds can be placed on the $\sigma-m_X$ parameter space through the detection of Pop~III stars in sufficiently high density DM regions detected long after they enter the zero age main sequence. For comparison, we have included the current best bounds on this parameter space available from the XENON1T one-year direct detection experiment for SI interactions and the PICO-60 experiments for SD interactions. We have also included the deepest bounds which could be placed by each experiment (black lines). Direct detection experiments on Earth are fast-approaching limits on their ability to constrain DM parameter space due to the flood of atmospheric neutrinos \cite{Kavanagh:2018}. Below the so-called ``Neutrino Floor," these experiments will be unable to discern DM signals from the background flux of neutrinos and will thus lose constraining power. Our results suggest that we can compete with the current bounds placed by the XENON1T one-year experiment for SI interactions. When considering SD interactions, for all DM densities and stellar masses considered, we predict that Pop~III stars, if observed and confirmed, would rule out a large swath of parameter space currently untouched by the best bounds given by direct detection. Perhaps the most exciting finding of this work is that we are able to probe below the neutrino floor region limiting SD direct detection experiments. 
 
Above $m_X\sim 10^6$ GeV, the linear relationship for our projected limits on $\sigma$ from Fig.~\ref{fig:Sigma-Mchi_Bounds} can be easily understood. When $k\ll1$, i.e. for higher $m_X$, the luminosity due to captured DM annihilations scales like $L_{DM}\propto\sigma/m_X$, and therefore the bounds on $\sigma$ scale linearly with $m_X$. Of course, this assumes $\rho_X$ is independent of $m_X$, i.e. given by the initial, adiabatically contracted profile. For low-mass WIMP DM, annihilations are much more efficient in the ambient medium surrounding the star and so at late times the DM density becomes dependent on $m_X$, as per the ``annihilation plateau'' discussed in detail in Appendix~\ref{sec:DMHalos}. The bounds in Fig.~\ref{fig:Sigma-Mchi_Bounds} are those for a star detected at around $t\sim 10^6$ years after entering the main sequence, when it is most likely to be detected. The annihilation plateau effect is evident in the bounds on s-wave DM in Fig.~\ref{fig:Sigma-Mchi_Bounds} for $1~\GeV \lesssim m_X\lesssim 10^6 ~\GeV$. For the higher densities, in this mass region the bounds scale like $\sigma \propto m_X^{-1}$. Without the annihilation plateau, when $m_X$ is in this range, $k\gg 1$, and so $L_{DM}\propto \frac{\sigma\rho_X}{m_X^0}$. However, at late times, for s-wave annihilation the ambient DM density scales like $\rho_X\propto m_X$, and so the bounds become inversely related to $m_X$. For the p-wave channel, the annihilation plateau becomes evident at lower DM masses than the s-wave channel due to its lower annihilation cross-section. In the right panels of Fig.~\ref{fig:Sigma-Mchi_Bounds}, the annihilation plateau effect can be seen for the $M_\star=1000~M_\odot$ star in the mass region $10^{-1}~\GeV \lesssim m_X\lesssim 1~\GeV$. Here, the bounds have the relationship $\sigma\propto m_X^{-2}$. This is because the p-wave annihilation cross section scales like $\sigmav\propto m_X^{-1}$ and the annihilation plateau like $\rho_{AP}\propto \frac{m_X}{\sigmav}$. Thus, at late times, $\rho_X\propto m_X^2$. In both the p-wave and s-wave cases, below $m_X\sim 1~\GeV$ the effects of evaporation become prominent and thus our bounds become asymptotic as seen in both sides of Fig.~\ref{fig:Sigma-Mchi_Bounds}. As the bounds cross the boundary of region II and region III of the $\sigma-m_X$ parameter space, they begin curving towards the right, forming a small section of lower bounds. As noted in the discussion of Fig.~\ref{fig:sigma_mx_ps}, in region II the DM capture rate is independent of $\sigma$, and so the bounds we find in this region have $\sigma$-dependence from evaporation only and are from solving Eq.~(\ref{eq:LeddMaxMass}) with the DM luminosity given by Eq.~(\ref{eq:LDMeqCtot}).

 Figure~\ref{fig:Sigma-Mchi_Bounds} shows that in the highest density environments predicted in adiabatically contracted Pop~III star forming DM halos, the observation of Pop~III stars places tighter bounds on $\sigma$ than possible with direct detection experiments. For the lowest densities we consider here  ($\rho_X \sim 10^{13} \text{ GeV cm}^{-3}$), the bounds we place are deeper than current bounds on SD-interactions across all DM masses. Referring to Appendix~\ref{sec:DMHalos}, we can see that this density is approximately that of the DM density at the edge of the baryonic core for baryonic densities of $n_{B} = 10^{13}$ cm$^{-3}$ for a potential $z \sim 15$ system. These bounds are quite conservative as the baryonic cloud continues to collapse up to the formation of a proto-stellar core, at  $n_{B} \sim 10^{22}$ cm$^{-3}$, which would correspond to an adiabatically contracted value for $\rho_X$ of $\sim 10^{19}~\GeV~\unit{cm}^{-3}$. Realistically, we expect the typical density for a Pop~III host DM minihalo to be somewhere between $10^{13}-10^{19}~\GeV~\unit{cm}^{-3}$. To be conservative, we will consider an upper limit of $\rho_X\sim 10^{16}~\GeV\percc$.
 Increasing the DM density from $10^{13}$ GeV cm$^{-3}$ has the effect of placing tighter constraints on $\sigma$. For our highest DM density considered ($\rho_X = 10^{16}$ GeV cm$^{-3}$), the bounds placed by all stellar masses are deeper than the XENON1T one-year bounds for SI interactions, once DM masses are $\gtrsim 10^5$ GeV. Since $L_{DM}\propto \rho_X$, our projected bounds are deeper for higher $\rho_X$, as evidenced in Fig.~\ref{fig:Sigma-Mchi_Bounds}. Also, higher mass Pop~III stars lead to more stringent bounds, since more massive stars are more efficient DM captors. We assumed that the Pop~III stars are within 10~A.\ U.\ of the center of the DM halo, as demonstrated by numerous hydrodynamical simulations~\cite{Barkana:2000,Abel:2001,Bromm:2003,Yoshida:2006,Yoshida:2008,Loeb:2010,Greif:2012,Bromm:2013,Klessen:2018} which show that Pop~III stars form either in isolation, or a few per DM mini-halo, with most of them within the central 10~A.\ U.\ and the most massive ones closest to the center. 
 
\begin{figure} [!htb]
    \centering
    \includegraphics[width=0.9\linewidth]{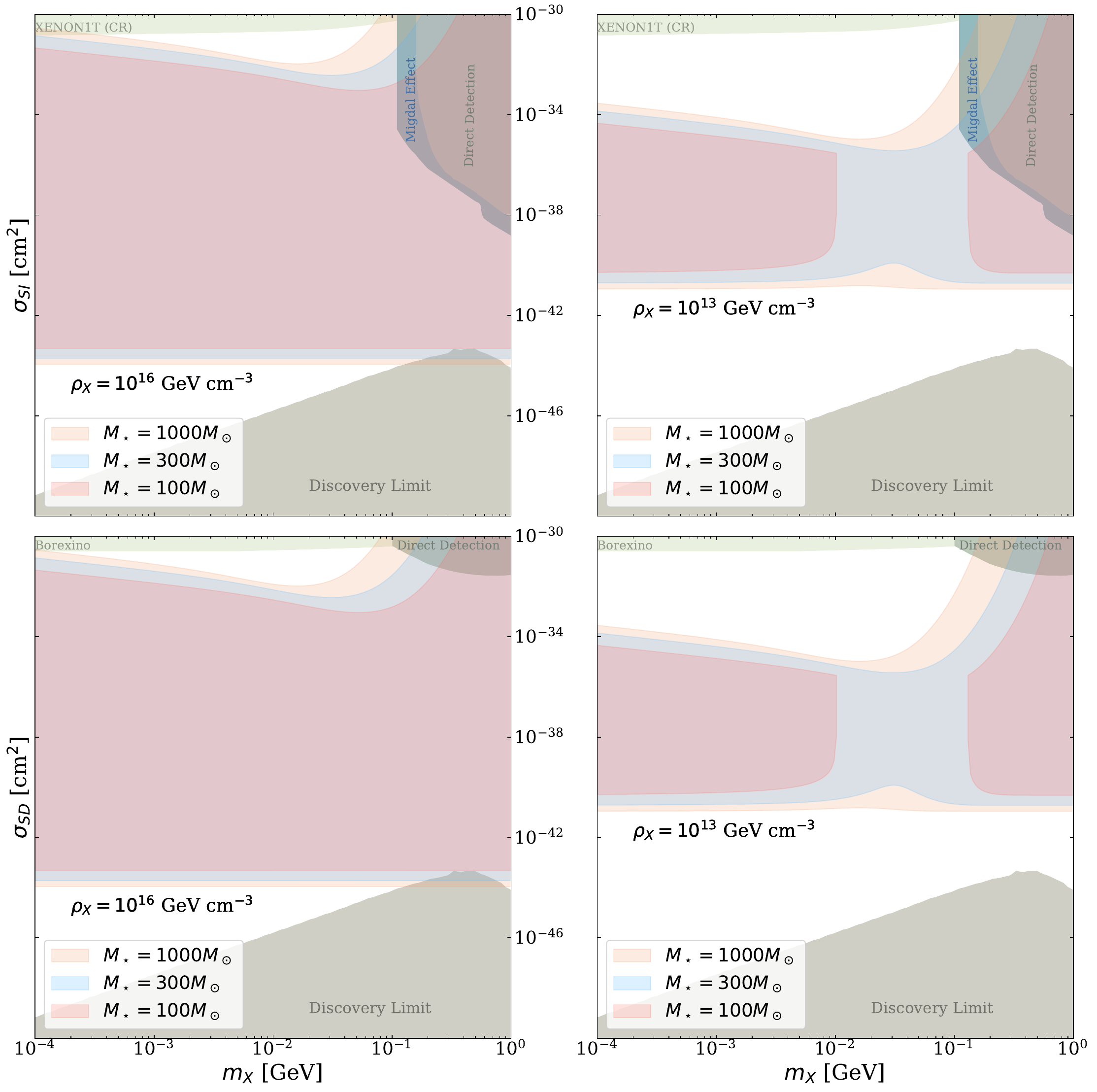}
    \caption{Projected bounds in the $\sigma-m_X$ parameter space for low mass ($10^{-4}~\GeV \lesssim m_X\lesssim 1~\GeV$) WIMP DM which annihilate via s-wave processes. The top and bottom panels compare our bounds to the most recent exclusion limits for both the SI~\cite{Aprile:2018,Bringmann:2018cvk,Aprile:2019ldm,Abdelhameed:2019} and SD~\cite{Bringmann:2018cvk,Amole:2019fdf,Aprile:2019ldm,Aprile:2019} interactions, as well as the ``discovery limit'' of direct detection experiments~\cite{Billard:2013,Battaglieri:2017}. We assume an adiabatically contracted halo with initial densities $\rho_X(0)\sim 10^{16}=\GeV~\unit{cm}^{-3}$ (left) and $\rho_X(0)= 10^{13}~\GeV~\unit{cm}^{-3}$ (right). The bounds are placed at $t=0$ and thus do not include the ``annihilation plateau.'' In all observations of a Pop.~III star, large portions of previous unexplored parameter space are ruled out. The precise shape of these regions is described in the text.  }
    \label{fig:A2_PRD_BOUNDS_sigmaSI_mx_lowmass_WIMP_s-wave}
\end{figure}

\begin{figure} [!htb]
    \centering
    \includegraphics[width=0.9\linewidth]{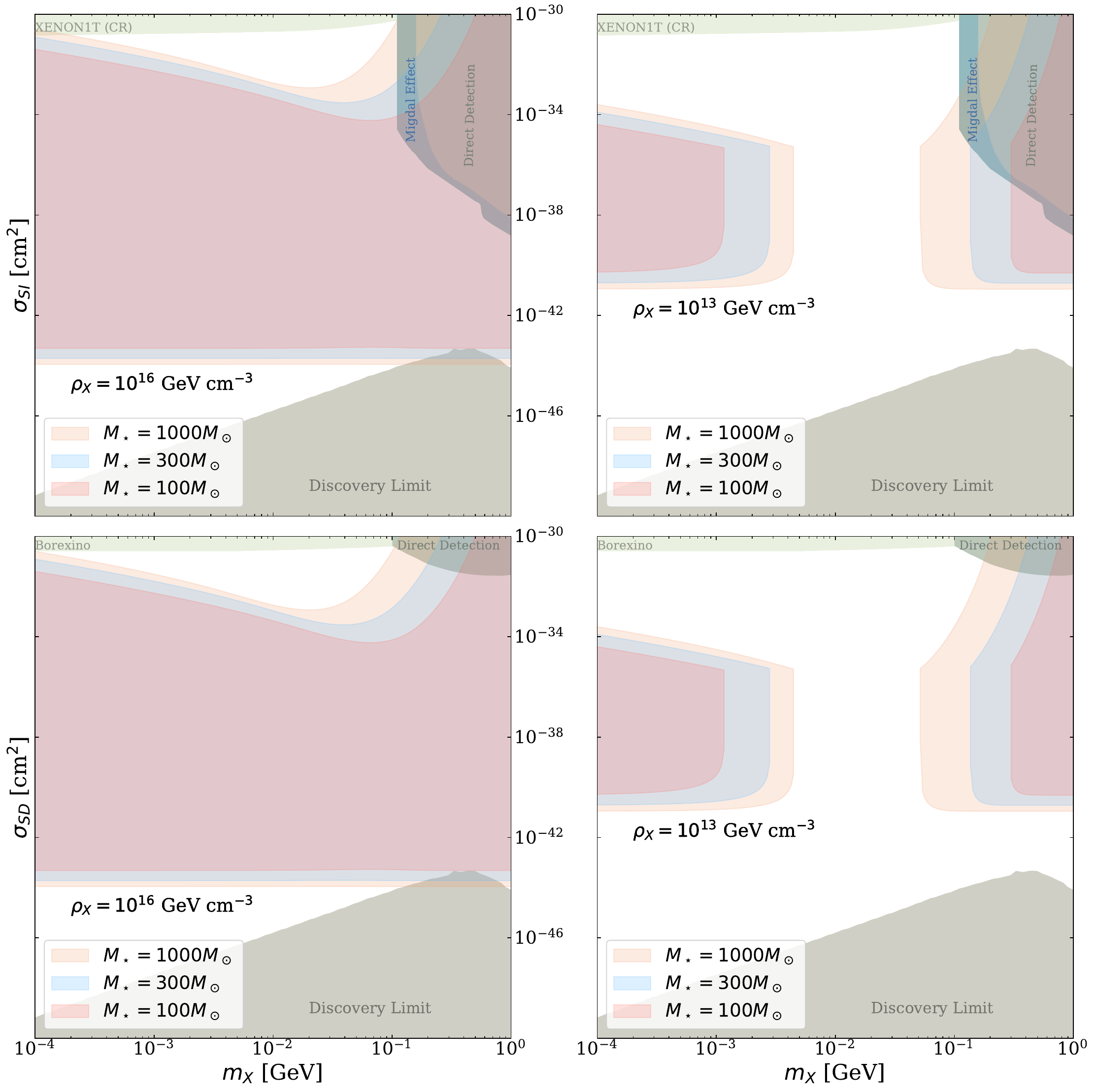}
    \caption{Projected bounds in the $\sigma-m_X$ parameter space for low mass ($10^{-4}~\GeV \lesssim m_X\lesssim 1~\GeV$) WIMP DM which annihilate via p-wave processes. The top and bottom panels compare our bounds to the most recent exclusion limits for both the SI~\cite{Aprile:2018,Bringmann:2018cvk,Aprile:2019ldm,Abdelhameed:2019} and SD~\cite{Bringmann:2018cvk,Amole:2019fdf,Aprile:2019ldm,Aprile:2019} interactions, as well as the ``discovery limit'' of direct detection experiments~\cite{Billard:2013,Battaglieri:2017}. We assume an adiabatically contracted halo with initial densities $\rho_X(0)\sim 10^{16}=\GeV~\unit{cm}^{-3}$ (left) and $\rho_X(0)= 10^{13}~\GeV~\unit{cm}^{-3}$ (right). The bounds are placed at $t=0$ and thus do not include the ``annihilation plateau.'' In all observations of a Pop.~III star, large portions of previous unexplored parameter space are ruled out. The precise shape of these regions is described in the text.  }
    \label{fig:A2_PRD_BOUNDS_sigmaSI_mx_lowmass_WIMP_p-wave}
\end{figure}

 Next, for illustrative purposes, we will show the possibility of constraining $\sigma$ through the detection of a young Pop~III star, such that the annihilation plateau is not relevant. Fig.~\ref{fig:A2_PRD_BOUNDS_sigmaSI_mx_lowmass_WIMP_s-wave} and Fig.~\ref{fig:A2_PRD_BOUNDS_sigmaSI_mx_lowmass_WIMP_p-wave} show projected bounds in the low-mass WIMP DM regime, for both spin-dependent (bottom panels) and spin-independent (top panels) DM, when considering s-wave and p-wave annihilation processes, respectively. BBN places the a stringent limit on the lowest value of $m_X$ for WIMPs at roughly $10~\unit{MeV}$~\citep{Sabti:2019}, which is a value we will adopt here. More information on these models can be found in Appendix D. In each case we consider three stars of mass $M=100M_\odot$, $300M_\odot$, and $1000M_\odot$. We assume an adiabatically contracted halo with ambient DM densities of $\rho_X \sim 10^{16}$ GeV cm$^{-3}$ for the left panels and $\rho_X \sim 10^{13}$ GeV cm$^{-3}$ for the right panels of the figures. Our bounds are placed using a hypothetical star that just formed, and, as such do not include the effects of the annihilation plateau. We include those effects in Fig~.\ref{fig:Sigma-Mchi_Bounds}, in which case the effects of DM annihilations in the ambient medium are considered for roughly 1 Myrs, i.e. the expected lifetime on the Zero Age Main Sequence of such massive Pop~III stars.  For both s-wave and p-wave annihilations, for the (conservative) densities considered, our method rules our large portions of parameter space for both the spin-independent and spin-dependent DM models. These excluded regions include portions of parameter space currently inaccessible to ground based direct detection experiments. Interestingly, our method results not in a strict upper bound on $sigma$, but rather excluded regions of parameter space defined by an upper and lower bound. The flat, lower limit (upper bound on $sigma$) of our excluded regions arise via the same mechanism as in Fig.~\ref{fig:Sigma-Mchi_Bounds}: because we are not considering the annihilation plateau, $L_{DM} \sim \sigma m_X$ for low mass DM, giving insensitivity to DM mass. We also see a region of parameter space in which we lose constraining power and instead are left with an open ``funnel'' region, joining the upper and lower limits. In this region, evaporation dominates capture, since evaporation is independent of the ambient DM density, whereas capture rate scales linearly with density $\rho_X$. Thus, we lose the ability constrain regions of parameter space as we move to lower mass stars and lower ambient DM densities due to the lower capture rates associated with these systems. This effect happens for a wider range of DM densities and stellar parameters in the p-wave case due to its lower annihilation cross section. We note that the $M_\star = 300M_\odot$ case when considering s-wave annihilations shows part of the ``transition'' regime, where evaporation starts to become more dominant over capture rate. The upper limits (lower bounds on $\sigma$) of our excluded region arise from solving Eq~.~(\ref{eq:ConstrainDM}) in region II of the $\sigma-m_X$ parameter space (Fig.~4) and depends on both the DM capture rate and the DM evaporation rate. 
 
\begin{figure} [!htb]
    \centering
    \includegraphics[width=.9\linewidth]{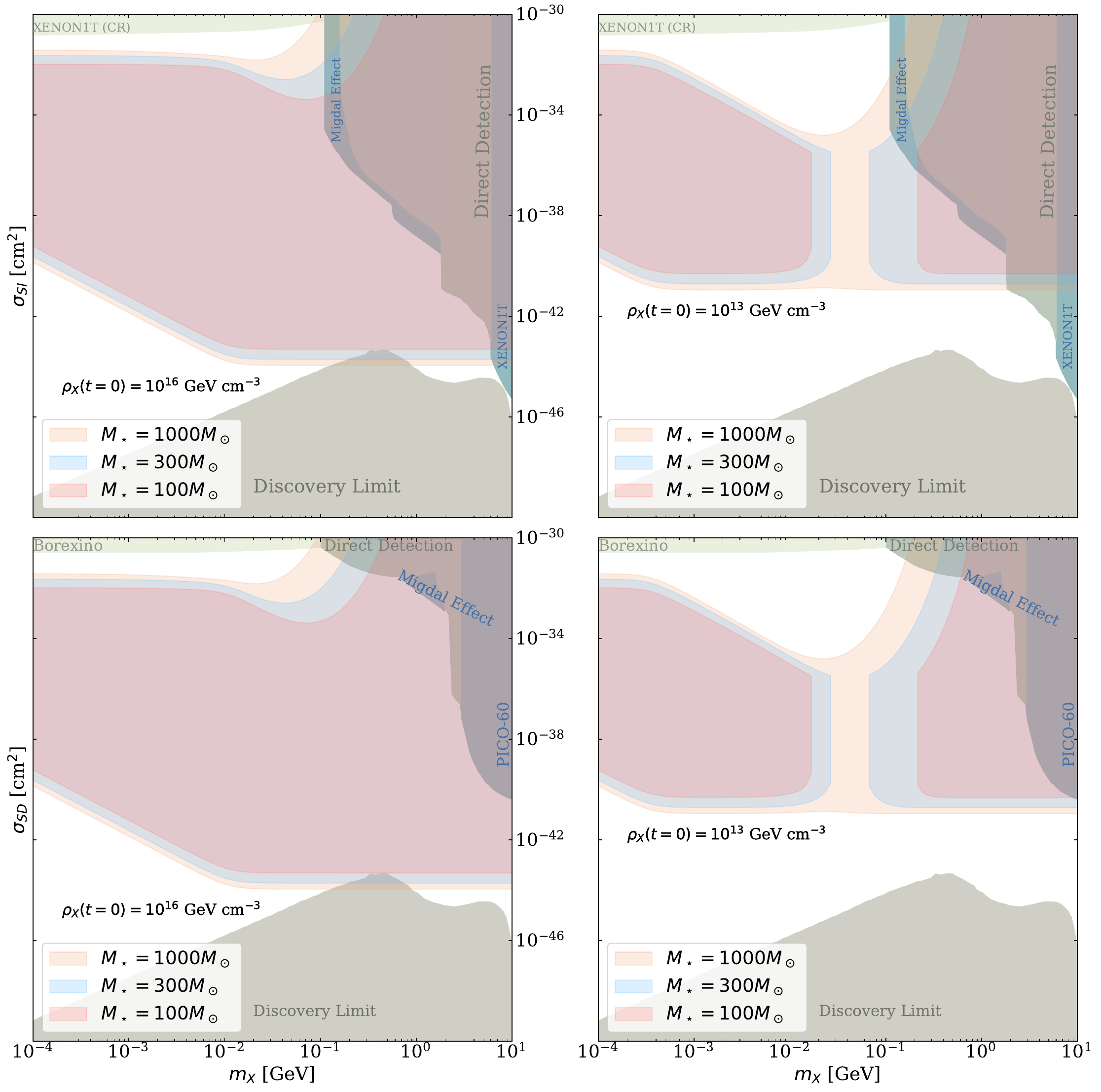}
    \caption{Top/bottom panels contrast our bounds to most recent exclusion limits for SI~\cite{Aprile:2018,Bringmann:2018cvk,Aprile:2019ldm,Abdelhameed:2019} and SD~\cite{Bringmann:2018cvk,Amole:2019fdf,Aprile:2019ldm,Aprile:2019} interactions from various experiments, each with the name listed inside the corresponding region. Additionally, we plot the limiting region, inaccessible to direct detection experiments, labeled ``Discovery Limit'', i.e. the neutrino floor~\cite{Billard:2013,Battaglieri:2017}.Right/left panels correspond to the two ends of the $\rho_X$ interval considered: $10^{13}-10^{16}\GeV\percc$.The initial ambient DM densities ($\rho_X(t=0)$)used when placing these constraints are those given by adiabtically contracted NFW profiles. The densities considered here are at $t = 1$ My. Projected bounds on $\sigma - m_X$ parameter space from the potential observation of Pop III stars with masses ranging from $M_\star= 100 M_\odot - 1000 M_\odot$. The DM particle models considered when placing these bounds are the SIMP/CoSIMP models when the effects of the annihilation plateau due to SIMP dark matter are most prominent and thus the projected bounds most conservative (See Appendix \ref{sec:DMHalos}).}
    \label{fig:Sigma-Mchi_Bounds_lowmass}
\end{figure}

 We next move our focus to to non-WIMP sub-GeV DM modes. Fig.~\ref{fig:Sigma-Mchi_Bounds_lowmass} shows the projected bounds in the low-mass region for strongly interacting thermal DM. We focus our attention on two such models: the Strongly Interacting Massive Particles (SIMP)~\cite{Hochberg:2014} that can annihilate via the following $3\to 2$ process: $DM+DM+DM\to DM+DM$, and the Co-SIMP model of~\cite{Smirnov:2020}, in which the following process is responsible for thermal production: $DM+DM+SM\to DM+SM$. For more details on those models see Appendix~\ref{sec:Equilibration}. If we allow for both of these processes to happen simultaneously, the CoSIMP channel is dominant inside the star, in view of a high density of baryons, whereas outside of the star the SIMP annihilations would be dominant, with CoSIMP annihilations being essentially negligible. Moreover, we point out that out of those two models, only the Co-SIMP DM interactions can lead to a transfer of energy to baryons inside a star, which is one of the fundamental assumptions we make in this work. We take an initial DM density in the $\rho_X \sim 10^{13} - 10^{16}$ GeV cm$^{-3}$ range,  and, assuming SIMP annihilations outside of the star, we time-evolved $\rho_X$ to $t = 1$ My to include the effects of the so called ``annihilation plateau.'' For more details on the initial, and the time evolved $\rho_X$ see Appendix~\ref{sec:DMHalos}. Note that, for all densities and stars considered, we rule out large swaths of previously unexplored parameter space. For reference, we have placed current direct detection limits from SI and SD searches. We also include a projected SI neutrino floor for future He-based experiments \cite{Ruppin:2014} and show that, in high density environments, we place projected constraints below this neutrino floor for DM masses in the range $m_X \sim 0.1 - 1$ GeV. Intriguingly, the projected bounds encapsulate a region defined by an upper and lower bound. The upper bound arises from solving the inequality $L_{DM} \leq L_{Edd} - L_{nuc}$ in region III of $\sigma-m_X$ parameter space (See Fig.~\ref{fig:sigma_mx_ps}) and results from the sensitivity of both capture and evaporation on $\sigma$. The lower bound, on the other hand, results from solutions in region II and relies solely on the sensitivity of evaporation on $\sigma$. This is evident when considering the independence of the capture rate on $\sigma$ in region II in contrast to the universal dependence of the evaporation rate on $\sigma$. The result is a bounded region ruling out swaths of parameter space that are not currently constrained by direct detection experiments. An important feature of these bounds is the funnel region of unconstrained parameter space for $\rho_X(t = 0) = 10^{13}$ GeV cm$^{-3}$ and $M_\star = 100 M_\odot$ and $M_\star = 300 M_\odot$. This effect emerges from the loss of constraining power due to the dominance of evaporation over capture for these parameters. Note that $C_{tot} \sim \rho_X$, irrespective of $m_X$ and $\sigma$, while $E \sim \rho_X^0$. Thus, a lower DM density implies a lower rate of capture but an unaffected evaporation rate. This leads to a diminishing effect on DM luminosity and effectively a loss of constraining power in the funnel region. 

The broken power law of the upper bounds we place in Fig.~\ref{fig:Sigma-Mchi_Bounds_lowmass} can be understood in the following way: at the higher end of $m_X$, the DM density is not affected by the ``annihilation plateau,'' and, as such, the upper bounds on $\sigma$ are insensitive to $m_X$, a consequence of $L_{DM}\sim \sigma m_X^0$ in Region~III of parameter space. At lower $m_X$, we note an inverse relationship between $\sigma$ and $m_X$ in the upper bounds. Specifically, for  $m_X \sim 10^{-4}-10^{-2}$ GeV for $\rho_X(0) = 10^{16}$ GeV cm$^{-3}$ and $m_X \sim 10^{-4} - 10^{-3}$ GeV for $\rho_X(0) = 10^{13}$ GeV cm$^{-3}$. This inverse relationship is easily understood by considering Eq. (\ref{eq:ConstrRhoSigma}) for Region III and Eq. (\ref{eq:rhot_simp}) in the limit that $\rho_{AP} \ll \rho_0$. Since $\sigma \sim 1/\rho_X$ and $\rho_X \approx \rho_{AP}^{3\rightarrow 2}\ \sim m_X$ at $t = 1$ My for lower DM masses, the following relationship emerges $\sigma \sim 1/m_X$, as demonstrated in the plot. A similar analysis for the lower bound would show that the flattening of this bound (in the same mass ranges we see an inverse relationship for the upper bound) is also a result of the annihilation plateau.

An intriguing question to ask is: for a given stellar mass, what minimum ambient DM density is necessary for constraining below the neutrino floor or the current XENON1T bounds? We will focus here only on the case of $m_X\gtrsim 100~\GeV$. To answer this question, we note that at large $m_X$, both our method and the direct detection experiments predict bounds that scale linearly with $m_X$. For the neutrino floor bounds we take:
\be
\sigma_{NF, X1T} \approx 10^{-50.7}\frac{\text{cm}^{2}}{\text{GeV}} m_X, 
\ee
\be
\sigma_{NF, C_3 F_8} \approx 10^{-46.1}\frac{\text{cm}^{2}}{\text{GeV}} m_X,
\ee
while the current XENON1T limits are given by Eq.~(\ref{Eq:X1Tbounds}). Under the $\tau \gg 1$ limit, we can approximate the sum in Eq.~(\ref{eq:ConstrRhoSigma}) with $2/3k\tau$, and solve for $\rho_X$:

\be\label{eq:RhoNF}
\rho_{X;NF, X1T} \equiv 10^{50.7} \frac{\text{GeV}}{\text{cm}^{2}} \sqrt{\frac{8\pi}{243}} \frac{1}{M_\star} \frac{\bar{v}^3}{v_{esc}^4}  \frac{L_{Edd}(M_\star)-L_{nuc}(M_\star)}{f},
\ee
\be\label{eq:RhoX1T}
\rho_{X;X1T} \equiv 10^{47.9} \frac{\text{GeV}}{\text{cm}^{2}} \sqrt{\frac{8\pi}{243}} \frac{1}{M_\star} \frac{\bar{v}^3}{v_{esc}^4}  \frac{L_{Edd}(M_\star)-L_{nuc}(M_\star)}{f}.
\ee
This provides an analytical method to estimate the DM density necessary for our method to predict upper bounds on $\sigma$ that are at the XENON neutrino floor and the current XENON1T one year limits. For densities higher than $\rho_{X;NF}$ ($\rho_{X;X1T}$), the observation of a Pop~III star of mass $\Mstar$ will place bounds deeper than the neutrino floor (XENON1T 1-year bounds). In Table \ref{table:Rhos}, we give $\rho_{X;NF,X1T}$ and $\rho_{X;X1T}$ for Pop III stars in the mass range $M_\star = 100-1000~M_\odot$. Note that the values for $\rho_{X;NF}$ range from $\sim 10^{17}$ GeV cm$^{-3}$ up to $\sim 10^{18}$ GeV cm$^{-3}$. For $\rho_{X;X1T}$, the values range from $\sim 10^{14}$ GeV cm$^{-3}$ to $\sim 10^{15}$ GeV cm$^{-3}$. This means that placing bounds tighter than the current best bounds from the XENON1T 1-year experiment is very plausible through the observation of Pop III stars.

\begin{table*}[!htb] 
\centering
\begin{tabular}{|c| c| c|}
\hline
$M_\star [M_\odot]$    & \text{Log}$_{10}(\rho_{X;NF, X1T}/ \text{ GeV cm}^{-3})$ & \text{Log}$_{10}(\rho_{X;X1T}/ \text{ GeV cm}^{-3})$ \\ \hline
100  & 18.0 & 15.2 \\
200  & 17.6 & 14.8 \\
300  & 17.3 & 14.5 \\
400  & 17.2 & 14.4 \\
600  & 16.9 & 14.1 \\
1000 & 16.8 & 14.0 \\ \hline
\end{tabular}
\caption{Table showing the DM densities that would imply bounds on $\sigma - m_X$ parameter space competitive with the neutrino floor ($\rho_{X;NF}$, Eq.~(\ref{eq:RhoNF})) or the current XENON1T 1-year experiment ($\rho_{X;X1T}$, Eq.~(\ref{eq:RhoX1T})) for a given stellar mass. DM densities higher than the values quoted here would lead to bounds deeper than the neutrino floor/XENON1T experiment for a given Pop~III mass.}
\label{table:Rhos}
\end{table*}

Regarding the SD proton-DM interactions, we point out that all of our bounds, even for the smallest $\Mstar$ and lowest $\rho_X$ considered are many orders of magnitude deeper than those placed by the PICO-60 experiment, for $m_X\gtrsim 10^5~\GeV$. Moreover, Pop~III stars more massive than $\sim300~\Msun$ probe below the C3F8 neutrino floor, even for the lowest $\rho_X\sim 10^{13}~\GeV\percc$.

We conclude this paper with Sec.~\ref{sec:discussion}, where we summarize our main results and discuss their implications and potential limitations. 


\section{Summary and Discussion}\label{sec:discussion}

 In this paper, we study the observable effects of DM capture on Pop~III stars. In Sec.~\ref{sec:obs}, we find that the additional heat source due to captured DM annihilations can lead to upper limits on Pop~III stellar masses. Assuming the DM-proton scattering cross section ($\sigma$) at the upper bound given by XENON1T for SI scattering, and for sufficiently high ambient DM densities at the location of the star  ($\rho_X$), we find that this maximum Pop~III stellar mass can be as low as $~\sim 10\Msun$ (see Fig.~\ref{fig:mmax}). In Sec.~\ref{sec:Constrain}, we provide a novel way to place competitive bounds on the product of two very important DM parameters: the DM density at the center of mini-halos hosting Pop~III stars ($\rho_X$), and the DM-proton scattering cross section ($\sigma$) (see Fig.~\ref{fig:RhoSig-Mchi_Bounds}). In practice, our projected bounds are obtained by assuming the upcoming, potential identification of Pop~III stars and their corresponding masses, and by imposing the Eddington luminosity limit. Having constrained $\sigma\times\rho_X$, we can break this degeneracy if we know either of those two parameters. In Fig.~\ref{fig:Rho-Mchi_Bounds}, we forecast  limits on the DM density at the center of Pop~III star hosting minihalos by assuming direct detection experiments will identify DM somewhere in the allowed region of the $\sigma-m_X$ parameter space, between the current XENON1T bounds, and the neutrino floor. If, conversely, SD experiments such as PICO-60, identify DM first, then our projected bounds on $\rho_X$ will be even deeper, since our method is insensitive to the SI/SD distinction, and direct detection experiments can only find $\sigma_{SD}$ with values larger any possible $\sigma_{SI}$, that is not yet ruled out. In Fig.~\ref{fig:Sigma-Mchi_Bounds}, we present upper limits on  $\sigma$ vs. $m_X$, assuming adiabatically contracted DM densities in the Pop~III star host minihalo. Most intriguingly, we show that with our method, Pop~III stars can be used to probe below the neutrino floor.  We note here that a major benefit of our method is that higher mass stars allow us to place tighter bounds due to their enhancement of DM luminosity. This is beneficial because the future detection of Pop~III stars is more likely to occur for more massive stars. Lastly, in Fig.~\ref{fig:Sigma-Mchi_Bounds_lowmass}, we present our SI/SD bounds on $\sigma$, for thermal sub-GeV DM, assuming CoSIMP/SIMP DM. We point out that if we assume only CoSIMP DM, our bounds will not be affected by the ``annihilation plateau,'' and therefore rule out even a larger swath of parameter space. In a future publication we plan to extend our sub-GeV analysis to other DM models. 

We also recognize that this method makes assumptions about DM properties, such as its ability to self-annihilate, and so it is somewhat limited in that regard. However, note that for thermal DM, annihilations are a key ingredient in the DM production mechanism. Therefore, this is not an assumption of the model, but rather a necessity to explain the observed relic abundance. However, the unitarity limit places an upper bound on the mass of a thermal relic, of roughly $300~\unit{TeV}$~\cite{Griest:1990}. Mechanisms for thermal DM to bypass the unitarity limit have been identified in the literature. For example, see~\cite{Harigaya:2016nlg} for a thermal DM model, with $m_X$ up to $\sim\unit{PeV}$. For higher mass DM, self annihilations are not a requirement, but rather an assumption we make. It is, however, a natural one, as in most models the DM particle is its own antipartner. Secondly, we make the assumption that Pop~III stars can reach masses in excess of $100\Msun$, and that those objects usually are found within the inner $10$~A.\ U.\ of the host DM microhalo. Regarding the mass spectrum of Pop~III stars, simulations are not yet conclusive. However, once found, the mass will be the primary observable that we use, so no assumption needs to be made there. Regarding the centralicity of the first stars, this assumption is supported by N-body simulations~\cite{Barkana:2000,Abel:2001,Bromm:2003,Yoshida:2006,Yoshida:2008,Loeb:2010,Bromm:2013,Klessen:2018} that find that even when the gas cloud fragments, and forms multiple stars, the most massive one is usually closest to the center, and most of those stars are within the central $10~$ A.\ U.\ of the center of the DM halo. Most importantly for our work, Pop~III stars more massive than $\sim150\Msun$ almost exclusively form in isolation, one per microhalo~\cite{Susa:2014}. For more details on this point see Sec.~\ref{sec:FirstStars}. When we use our formalism to place bounds on the DM-proton cross section, a potential limitation comes from the uncertainty in the ambient DM density at the center of the DM halos. For this work we used the well established adiabatic contraction formalism (see Fig.~\ref{fig:DMDensityProfiles} in Appendix~\ref{sec:DMHalos}), which is supported by numerical simulations of high redshift DM microhalos~\cite{Abel:2001,Sellwood:2005,Gendin:2011,Davis:2013mha} (also see Fig.~\ref{fig:DM_AC}). At lower redshifts, well after the first stars have formed, baryonic feedback effects are expected to be important, and, as such, adiabatic compression should be suppressed. Even so, for the Milky Way DM halo, rotation curves Gaia DR2 data offer the first experimental evidence of DM density compression in presence of baryons~\cite{Cautun:2020}. However, one should point out that the current resolution of hydrodynamic N-body simulations is not sufficiently high to probe the inner parsec regions of the DM minihalos deep enough, and therefore one needs to resort to analytical approximations, such as adiabatic contraction, when estimating $\rho_X$ in the ambient environment of Pop~III stars, near the center of their host halos. In the near future, more sophisticated simulations should be able to verify our estimates on $\rho_X$, and as such, narrow down the uncertainty bands in our $\sigma$ vs $m_X$ exclusion limits. We point out once more the complementarity of our method with direct detection experiments. If the proton-DM cross section interaction will be identified by such experiments, then we can use our method to place bounds on the DM density at the location of the first stars, once those are observed with JWST and/or the Roman telescopes. Direct, dynamical measurements of the DM density in those extremely distant microhalos would be nearly impossible, and as such our method could be used to bypass this limitation.  

\section{Acknowledgements} CI would like to thank Katherine Freese and Paolo Gondolo for sharing the code we used in Appendix~\ref{sec:DMHalos} to calculate the adiabatically contracted NFW profiles. This is the same code used in ~\citet*{Spolyar:2008dark}, where the conditions for the formation of Dark Stars were first identified. CL thanks the financial support from Colgate University, via the Research Council student wage grant, and the Justus ’43 and Jayne Schlichting Student Research Funds.

\appendix


\section{Multi Scatter Capture of Dark Matter} \label{sec:MSCapture}

We start this Appendix with a brief review of the formalism we used to calculate the rates of DM capture by Pop~III stars. Then we proceed to calculate closed form, analytic approximations for the total capture rates, that can be very useful from both a practical standpoint, and for explicitly displaying the dependence of the capture rates in the multiscatter regime with physical parameters of interest. We first introduced those closed form analytic approximations for the total capture rates in~\cite{Ilie:2020Comment}; however, in view of the word count limitations for that comment paper, we couldn't present derivations there. In this Appendix we fill in those details.  

For any astrophysical object, the main parameter that controls the capture is the optical depth $\tau\equiv 2\Rstar\sigma n_T$, with $\Rstar$ being the radius of the star, $\sigma$ being the DM-target nucleus scattering cross section, and $n_T$ being the number density of target nuclei inside the stars. Whenever $\tau\ll1$ one can use the single scattering formalism introduced by Gould~\cite{Gould:1987,Gould:1987resonant} in the late 1980s. Whenever $\tau\gtrsim 1$, one has to use the more general multiscatter formalism~\cite{Bramante:2017,Ilie:2020Comment}. In our work, we will use exclusively use the latter, since, in the limit of $\tau\ll1$, it naturally covers the single scattering regime. 

DM particles in the vicinity of any massive object are attracted by its gravitational field. As a DM particle crosses a star, it interacts with the nuclei inside, and after each collision it loses an energy of $\Delta E_{i} = -\beta_{+}E_{i}$. Here, $E_i$ represents the energy of the DM particle before the $i$th collision, and $\beta_{+}$ is related to the mass of the DM particle ($m_{X}$) and mass of the target nuclei ($m$) in the following way: $\beta_+\equiv4mm_X/(m+m_X)^2$. If collisions are efficient enough to slow the DM particle below the escape velocity at the surface of the star, the DM particle becomes trapped. 

The capture rates after exactly $N$ collisions ($C_N)$ depend on two distinct quantities: the flux of dark matter particles entering the surface of the star and the probability of capture after exactly $N$  collisions with the nuclei inside the star ($g_N$). Therefore, the total capture rate can be written as, in Eq.~(\ref{eq:Ctot}), which we reproduce here, for convenience:   
\be
\label{Eq:CtotAppendix}
C_{tot} = \sum_{N=1}^{\infty} C_{N} = \sum_{N=1}^{\infty} \underbrace{\pi \Rstar^2}_\textrm{capture area}\times \,\underbrace{n_X \int_0^{\infty} \dfrac{f(u)du}{u}\,(u^2+v_{ esc}^2)}_\textrm{DM flux}\times \, \underbrace{p_{ N}(\tau)}_\textrm{probability for $N$ collisions}\times \, \underbrace{g_{ N}(u)}_\textrm{probability of capture},
\ee
where $u$ represents the DM velocity far from the gravitational potential well of the star, $p_N(\tau)$ is the probability that a DM with optical depth $\tau$ experiences exactly $N$ collisions (given by Eq.~(\ref{eq:pN})), and $g_N(u)$ is the probability of capture after exactly $N$ collisions. The latter has the following approximate form~\cite{Bramante:2017}: $g_N(u)=\Theta(u_{max;N}-u)$,
where $\Theta(x)$ is the Heaviside step function, $u_{max;N}=\vesc\left[(1-\beta_+/2)^{-N}-1\right]^{1/2}$ is the maximum value of the velocity a DM particle can have, far from the star, such that it will be slowed down below the escape velocity after $N$ collisions.

Assuming a Maxwellian velocity distribution, the general formula of capture after $N$ scatters is~\cite{Bramante:2017,Ilie:2020Comment}:
\be\label{eq:CN}
C_{N}=\frac{1}{3}\pi R_{\star}^{2} p_{N}(\tau) \frac{\sqrt{6} n_{X}}{\sqrt{\pi} \bar{v}}\left(\left(2 \bar{v}^{2}+3 v_{e s c}^{2}\right)-\left(2 \bar{v}^{2}+3 v_{N}^{2}\right) \exp \left(-\frac{3\left(v_{N}^{2}-v_{e s c}^{2}\right)}{2 \bar{v}^{2}}\right)\right),
\ee
where $\vbar$ represents the dispersion velocity of DM particles inside the halo,  $v_N=\vesc(1-\avg{z}\beta_+)^{-N/2}$ is the velocity of DM after $N$ scatters, where $\langle z\rangle$ accounts for the scatter angle and has an average value of $\frac{1}{2}$~\cite{Bramante:2017}.

We note that the probability of exactly $N$ scatters, $\pn$, can be approximated as follows:
\begin{subnumcases}{\pn\approx}
\frac{2\tau^{N}}{N!(N+2)}+\mathcal{O}(\tau^{N+1}), & \text{if } $\tau\ll 1$ \label{eq:pnlt} \\
\frac{2}{\tau^{2}}(N+1)\Theta(\tau-N), & \text{if } $\tau\gg 1$\label{eq:pnht}.
\end{subnumcases}

 We verified numerically that the sums defining the total capture rates from Eq.~(\ref{Eq:CtotAppendix}) will generally converge if $N_{cut}\approx\tau$. In our work, we perform the sums numerically until they have converged. However, it is very useful for future work to investigate if a closed form can be found for $C_{tot}$, given the form of $C_N$ from Eq.~(\ref{eq:CN}). In Ref.~\cite{Ilie:2020Comment}, we presented such a closed form; in view of the word count limitations, we were not able to provide a derivation. We sketch it below.   
First, we define the exponential factor $R_v$ in Eq.~(\ref{eq:CN}), as $R_v\equiv\mess$.  Under different mass limits, $R_v$ behaves differently because of $\vn$. When $m_N~\simeq~m_X$, $\beta_+ \simeq 1$ and $\vn \sim \vesc (2^N-1)^{1/2}$. In the other limiting case where $m \ll m_X$, $\beta_+ \simeq 4m/m_X$ and $\vn \simeq \vesc (1+Nm/m_X)^{1/2}$. $R_v$ then becomes:
 \begin{subnumcases}{R_v\approx} 
\frac{3}{2}(2^N-1)\frac{\vesc^2}{\vbar^2},   & \text{ \!\!\!\!if } $m~\sim~m_X$ \label{eq:Rlowmx} \\
\frac{3}{2}N\frac{m_N}{m_X}\vevbsq, & \text{\!\!if } $m\ll m_X$.\label{eq:Rhighmx}
\end{subnumcases}

Eq.~(\ref{eq:CN}) can then be expanded under the limit $R_v \gg 1 $ and $R_v \ll 1 $ as:
\begin{subnumcases}{C_N\approx}
\frac{1}{3}\sqsixoverpi\pi R^2\pn n_X\frac{3\vesc^2+2\vbar^2}{\vbar},& \label{eq:CNbigR} \\
\frac{3}{2}\sqsixoverpi\pi R^2\pn\frac{n_X \vesc^4}{\vbar^3}\beta_+\avg{z}\left(N+N^2\beta_+\avg{z}\right)\label{eq:CNsmallR}&.
\end{subnumcases}

Using the $\tau\gg1$ approximation of $\pn$ from Eq.~(\ref{eq:pnht}), we get the total capture rate up to $\nmax$ number of scatters: 
\begin{subnumcases}{C_{tot,\nmax}\approx}
\left(\frac{2}{3\pi}\right)^{1/2}\frac{\pi R^{2}}{\tau^{2}}n_{X}\frac{3\vesc^{2}+2\vbar^{2}}{\vbar}\nmax(\nmax+3), &\label{eq:CtotNmaxbigR} \\
\sqsixoverpi\frac{\pi R^{2}}{\tau^{2}}n_{X}\frac{\vesc^{4}}{\vbar^3}\beta_{+}\avg{z}\nmax(\nmax+1)(\nmax+2)\left(1+\frac{\beta_{+}\avg{z}}{4}(1+3\nmax)\right).&
\label{eq:CtotoNmaxsmallR}
\end{subnumcases}
The above equations would hold only when $\tau \gg 1 $ and $\nmax < \tau$. When $\nmax \gtrsim\tau$, in view of the $\Theta(\tau-N)$ factor in the approximate form of $\pn$ (Eq.~(\ref{eq:pnht})), the sum converges around $\nmax \sim \tau$ and $C_{tot,\nmax} $ reduces to $C_{tot,\tau}\approx C_{tot}$. 

For Pop~III stars, the escape velocity is much larger than the thermal velocity of dark matter($\vesc \gg \vbar $). Assuming there is a definite hierarchy between $m_X$ and $m$, i.e. if $m_X\gg m$ or $m_X\ll m$ Eq.~(\ref{eq:CN}) could be simplified as: 
\be\label{eq:CNapprox}
C_{N} = \sqrt{24\pi}n_{X}GM_{\star}R_{\star}\frac{1}{\bar{v}} p_{N}(\tau)\left(1-\left(1+\frac{2 A_{N}^{2} \bar{v}^{2}}{3v_{esc}^{2}}\right) e^{-A_{N}^{2}}\right), \text{where} \  A_{N}^{2}=\frac{3N\vesc^2 }{\vbar^2}\frac{\min(m_X;m)}{\max(m_X;m)}.
\ee
We point out that the above equation is slightly different from the corresponding one in~\cite{Bramante:2017}, where the sign in front of the term $\frac{2 A_{N}^{2} \bar{v}^{2}}{3v_{esc}^{2}}$ appears as a $-$. This is one of the typos we found in~\cite{Bramante:2017}, which are explained in~\cite{Ilie:2020Comment}. Using Eq.~(\ref{eq:CNapprox}), the total capture rate becomes:
\be\label{eq:CtotApprox}
C_{tot}(m_{X})=\left(\text{const.}\right) \times n_{X}\sum_{N=1}^{\infty} p_{N}(\tau)\left(1-\left(1+\frac{2 A_{N}^{2} \bar{v}^{2}}{3v_{esc}^{2}}\right) e^{-A_{N}^{2}}\right),
\ee
where, for simplicity, we introduced the following notation: $const=\sqrt{24\pi}G\Mstar\Rstar/\vbar$. To further simplify this expression, and extract some useful information, we divide the analytical derivation into two cases: single and multi scatter. For the former ($\tau \ll 1$), $C_{tot}=C_{1}$ and $A_{N}^{2}=A_{1}^{2}\equiv k$. Since $k$ appears in the exponent, we have two distinct cases. Therefore, when $\tau\ll1$ and $k\gg 1$ (Region~III of Fig.~\ref{fig:sigma_mx_ps}), the total capture rate can be approximated as:
\be\label{eq:Ctotcase1}
C_{tot}(m_{X}) \simeq \left(\text{const.}\right) \times n_{X} p_{1}(\tau) \simeq \left(\text{const.}\right) \times n_{X} \frac{2\tau}{3}.
\ee
Conversely, when $\tau\ll1$ and $k \ll 1$ (Region~IV of Fig.~\ref{fig:sigma_mx_ps}), using $\frac{2k\vbar^2}{3\vesc^2}=2\frac{m}{m_X}\ll1$, the total capture rate becomes:
\be\label{eq:Ctotcase2}
C_{tot}(m_{X}) \simeq \left(\text{const.}\right) \times n_{X} p_{1}(\tau)\left(1-\left(1+2\frac{m}{m_X}\right) \left(1-k\right)\right)\simeq \left(\text{const.}\right) \times n_{X} \frac{2k\tau}{3}.
\ee
For the multi-scatter case($\tau \gg 1$), let us introduce the following notation:
\be\label{eq:sumT1T2T3}
\sum_{N=1}^{\infty} p_{N}(\tau)\left(1-\left(1+\frac{2 A_{N}^{2} \bar{v}^{2}}{3v_{esc}^{2}}\right) e^{-A_{N}^{2}}\right)=T_1-T_2-T_3,
\ee
where $T_1\equiv\soneinf\pn$, $T_2\equiv\soneinf\pn e^{-A_N^2}$, and $T_3\equiv\soneinf\pn\frac{2A_N^2\vbar^2}{3\vesc^2}e^{-A_N^2}$. We can simplify $A_{N}^{2}$ as $A_{N}^{2}=NA_{1}^{2}=Nk=N\frac{3 m\vesc^2 }{m_{X}\vbar^2}$. For $T_{1}$, we can directly get $T_1=1-p_{0}(\tau)\simeq 1$. For $T_{2}$ and $T_{3}$, we first need to expand the exponential terms into the sum of a series: $e^{-A_N^2} = e^{-Nk} = \szeroinf \frac{(-Nk)
^j}{j!}= \szeroinf \frac{(-k)
^j}{j!}N^j $. Then, for $\tau \gg 1$, and using the approximate form of $\pn$ from Eq.~(\ref{eq:pnht}) times $N$ to some power $j$, and summing from $N=1$ to $\infty$, is approximately equal to doing the integration over the same range. By keeping the leading order term of the integration, we get the following 
\be \label{eq:sumpn}
\soneinf \pn N \approx \int_{1}^{\tau}\frac{2}{\tau^2}(N+1)NdN \approx \frac{2}{3}\tau, \  \soneinf \pn N^2 \approx \int_{1}^{\tau}\frac{2}{\tau^2}(N+1)N^2dN \approx \frac{1}{2}\tau^2,\   \soneinf \pn N^3 \approx \frac{2}{5}\tau^3 \  ...
\ee
This leads to a more general format:
\be \label{eq:sumpngen}
\soneinf \pn N^j \approx \frac{2}{j+2}\tau^{j}.
\ee

Finally, by substituting Eq.~(\ref{eq:sumpngen}) into the definitions of $T_2$ and $T_3$, we obtained the following closed form:
\be\label{eq:T2}
T_{2}\equiv \soneinf\pn e^{-A_N^2}\approx \szeroinf  \frac{2(-k\tau)
^j}{j!(j+2)} =\frac{2e^{-k\tau}(-1+e^{k\tau}-k\tau)}{(k\tau)^2},
\ee
\be\label{eq:T3}
T_{3}\equiv \soneinf\pn\frac{2A_N^2\vbar^2}{3\vesc^2}e^{-A_N^2}=\frac{2m}{m_X}\soneinf\pn N e^{-A_N^2}\approx \szeroinf \frac{2\tau(-k\tau)
^j}{j!(j+3)} =\frac{4m}{m_X}\frac{e^{-k\tau}(-2+2e^{k\tau}-2k\tau-k^2\tau^2)}{k^3\tau^2}.
\ee

We can further approximate the expansions of $T_2$ and $T_3$ in Eqns.~(\ref{eq:T2})-~(\ref{eq:T3}), depending on the value of $k\tau$. By combining the $T_1$, $T_2$, and $T_3$ approximate values, and keeping only leading order terms, we get:

\begin{subnumcases}
{\sum_{N=1}^{\infty}p_{N}(\tau)\left(1-\left(1+\frac{2 A_{N}^{2} \bar{v}^{2}}{3v_{esc}^{2}}\right) e^{-A_{N}^{2}}\right)=}
\frac{2}{3}k\tau & \mbox{if } $k\tau\ll 1$ \mbox{and } $\tau\gg1$ \label{eq:SumRegI}\\
1 & \mbox{if } $k\tau\gg 1$ \mbox{and } $\tau\gg1$ \label{eq:SumRegII}
\end{subnumcases}

The two results above represent the values that $\sum_{N=1}^{\infty}p_{N}(\tau)\left(1-\left(1+\frac{2 A_{N}^{2} \bar{v}^{2}}{3v_{esc}^{2}}\right) e^{-A_{N}^{2}}\right)$ take in Regions~I and~II, respectively, of Fig.~\ref{fig:sigma_mx_ps}. The values in Regions~III and IV can be inferred from Eq.~(\ref{eq:Ctotcase1}) and Eq.~(\ref{eq:Ctotcase2}), respectively.  


\section{Temperature of Captured Dark Matter}\label{sec:CapDMTemp}

At any spatial point inside of a star, stellar material is in approximate local thermodynamic equilibrium at some temperature $T_X(\boldsymbol{r})$. Captured dark matter inside the star scatters off of those baryons, bringing the distribution of dark matter particles to a Maxwellian form:

\be
f_X(\boldsymbol{v}_X, \boldsymbol{r}) \sim \exp{\left(\frac{-E}{kT_X}\right)},
\ee
where $E = \frac12 m_X v_X^2 + m_X \Phi(r)$ is the total energy of the dark matter particle, $k$ is Boltzmann's constant, and $T_X$ is the dark matter kinetic temperature. This temperature is not a well defined quantity; dark matter particles traverse through a range of radii throughout their orbits, and thus experience a range of interactions at different local kinetic temperatures. In fact, there is no single value $T_X$ for which the above expression is exactly true since these dark matter particles are undergoing processes that will equilibrate the dark matter to different local temperatures as it traverses star. Following \cite{Spergel:1985}, we assume that the dark matter particles distribution is described by a single, orbit-averaged temperature $T_X$ which satisfies not the collisional Boltzmann equation, but rather it's first energy moment. For time-independent distributions, requiring that the first moment is satisfied is equivalent to there being no \textit{net} flow of energy into the dark matter distribution from the solar material. The effects of heatflow have been included by~\cite{Garani:2017}. Comparing their results with those of~\cite{Spergel:1985}, we note that, to leading order, the effects of heatflow from evaporated DM are subdominant. In what follows, for simplicity, we neglect those sub-leading effects, and follow the approach presented in~\cite{Spergel:1985}.  

Letting $\sigma_X(\theta)$ be the differential scattering cross section, and letting $\Delta E\left(\boldsymbol{v}_X,\boldsymbol{v}_p,\theta\right)$ be the energy transfer to a dark matter particle from a collision, the energy-moment equation is: 

\be\label{eq:noTransfer}
\int d^{3} r \int d^{3} v_{X} f_{X}\left(\boldsymbol{v}_{X}, \boldsymbol{r}\right) \int d^{3} v_{p} f_{p}\left(\boldsymbol{v}_{p}, \boldsymbol{r}\right) \int d \cos \theta \sigma_{X}(\theta)\left|\boldsymbol{v}_{X}-\boldsymbol{v}_{p}\right| \Delta E\left(\boldsymbol{v}_{X}, \boldsymbol{v}_{p}, \theta\right)=0
\ee

To proceed in a more general situation, we introduce the Knudsen number of a weakly interacting mixture of two Maxwell gases, defined as:

\be
\text{Kn} = \left(n_p \sigma_X L \right)^{-1},
\ee
where $n_p$ is the number density of background particles (protons), $\sigma_X$ is the interaction dark matter-proton interaction cross section, and $L$ is the length scale of the system over which interactions can occur. When $\text{Kn}\gg1$, dark matter undergoes many interactions over the length scale $L$. For dark matter-proton interactions inside the star, we adopt the length scale of \cite{Spergel:1985} as the radius at which the dark matter is in approximate thermodynamic equilibrium with the core:

\be
\frac32 k T_c = m_X \Phi(r_X) \rightarrow r_{X}=L=\left(\frac{9}{4 \pi} \frac{k T_{c}}{G \rho_{c} m_{p}}\right)^{1 / 2} \sqrt{\frac{m_{p}}{m_{X}}}
\ee

In the large Kn limit, we can treat the system with statistical mechanics. Such is the case for all of our systems. The lowest Knudsen number we encounter in our analysis is for a $100 M_\odot$ Pop.~III star in an ambient DM density of $\log_{10}\left(\rho_X [\GeV \text{ cm}^{-3}] \right) = 13$ which gives a value of $\text{Kn} \sim 10^2$ at both $m_X \sim 10^{-4} \GeV$ and $m_X \sim 10^{15} \GeV$ with a maximum value of $\text{Kn}$ at $m_X \sim 10^5 \GeV$. To estimate $\rho_{c,p}$, we use the polytropic approximation of Eq.~(\ref{eq:rhoCPoly}), and confirmed this approximation to hold from \textsc{MESA} simulations of ZAMS Pop~III stars to order-of-magnitude. We take $\sigma_X$ from our bounds without evaporation (see Fig.~\ref{fig:Sigma-Mchi_Bounds}).

With such high $\text{Kn}$, the processes governing the velocity distribution of baryons and dark matter allow us to treat both the protons and the dark matter as Maxwellian gases. Plugging in Maxwellian distributions for both the protons and the dark matter, one obtains, from Eq.~(\ref{eq:noTransfer}):

\be\label{eq:transcend}
\int d^{3} r \, n_{p}(r) \int d^{3} v_{X} \exp \left(\frac{-E}{k T_{X}}\right) \int d^{3} v_{p} \exp \left[\frac{-m_{p} v_{p}^{2}}{2 k T(r)}\right]\left|\boldsymbol{v}_{X}-\boldsymbol{v}_{p}\right|\langle\Delta E\rangle=0,
\ee
where $\langle\rangle$ denotes the average over scattering angle $\theta$. This is, in principle, a simple transcendental equation for $T_X$ for a given energy transfer $\langle \Delta E \rangle$, relative velocity $|\boldsymbol{v}_{X}-\boldsymbol{v}_{p}|$, and $T(r)$. As outlined in \S4 and Appendix A of \cite{Spergel:1985}, we can re-write \ref{eq:transcend} as:

\be\label{eq:transcendental}
\int_{0}^{R_{\star}} n_{p}(r)\left[\frac{m_{p} T_{X}+m_{X} T(r)}{m_{X} m_{p}}\right]^{1 / 2}\left[T(r)-T_{X}\right] \exp \left[\frac{-m_{X} \Phi(r)}{k T_{X}}\right] r^{2} d r=0,
\ee
where $\Phi(r)$ is the gravitational potential defined by: $\Phi(r)\equiv\int_0^r \,dr'G M(r')/r'^2$. Next we use the $n=3$ polytropic model approximation in order to calculate $\Phi$. This assumption is always valid whenever the ratio between the radiation pressure and the gas pressure is a constant throughout the star, and this is the case for the radiation pressure dominated $\Mstar\gtrsim100\Msun$ Pop~III stars on the Zero Age Main Sequence (ZAMS). We have also checked this assumption by using the \textsc{MESA} stellar evolution code.

For a polytrope of an arbitrary index $n$, the following relationship holds: $\rho(\xi)=\rho_c\theta^n(\xi)$. We denoted by $\rho_c$ the central density, $\xi\equiv (r/\Rstar)\xi_1$ being the dimensionless radial variable, and $\xi_1$ being the first node of the Lane-Emden function $\theta$, which corresponds to the surface of the star ($\Rstar$). The Lane-Emden function obeys the following differential equation:
\be\label{eq:LE}
\frac{1}{\xi^2}\frac{d}{d\xi}\left(\xi^2\frac{d\theta(\xi)}{d\xi}\right)=-\theta(\xi)^n.
\ee
For $n=3$ one can show numerically that $\xi_1\approx6.89$. Moreover, the Lane-Emden function obeys the following boundary conditions at $\xi=0$ (the center of the star): $\theta(0)=1$, and $d\theta/d\xi=0$. One can show that for a polytrope the amount of baryonic mass enclosed by a radius $r$, corresponding to a dimensionless radial variable $\xi$, is: $$M(r)=-4\pi\rho_c\left(\Rstar/\xi_1\right)^3\xi^2d\theta/d\xi.$$
Using the definition of the gravitational potential from above, we find that the integral can be performed analytically, with the following result:
\be\label{eq:PhiPoly}
\Phi(\xi)=4\pi G \rho_c \left(\frac{\Rstar}{\xi_1}\right)^2\left[1-\theta(\xi)\right].
\ee
Moreover, we find that the central density for a polytropic star can be expressed as:
\be\label{eq:rhoCPoly}
\rho_c=\frac{\Mstar}{-4\pi\left(\frac{\Rstar}{\xi_1}\right)^3\xi_1^2\left(\frac{d\theta}{d\xi}\right)_1}.
\ee
Combining Eqns.~(\ref{eq:PhiPoly})-(\ref{eq:rhoCPoly}), and the fact that for $n=3$ we can use the following approximation:
$\xi_1^2\left(\frac{d\theta}{d\xi}\right)_1\approx2$,
we get he following, simpler form of the gravitational potential for a $n=3$ polyropic star:
\be\label{eq:Phin3}
\Phi(\xi)\approx v_{esc}^2\frac{\xi_1}{4}(1-\theta(\xi)),
\ee
with $v_{esc}^2=2G\Mstar/\Rstar$, being the escape velocity at the surface of the star.

Using the $n=3$ polytrope, and the assumption of $P\sim P_{gas}\sim P_{rad}$ (i.e. a constant ratio between the gas and radiation pressure throughout the star) we can show that: $n(\xi)=n_c\theta^3(\xi)$ (i.e. $n=3$ polytrope) implies that $T(\xi)=T_c\theta(\xi)$. We adopt $T_c\sim10^{8}$~K, a value verified by simulations of Pop~III stars with \textsc{MESA}. Introducing the following dimensionless variables:
\be
\xi \equiv \frac{r}{R_\star}\xi_1 \text{ , } \mu \equiv \frac{m_X}{m} \text{ , } \tilde{\Phi}(r) \equiv \frac{m \Phi(r)}{k T_c} \text{ and } \Theta \equiv \frac{T_X}{T_c}
\ee
we can rewrite Eq.~(\ref{eq:transcendental}) as:

\be\label{eq:numsol}
\int_{0}^{\xi_{1}} \theta(\xi)^{3} \exp \left(\frac{- \mu}{\Theta}\Phi(\xi)\right)\left(\frac{\Theta+\mu \theta(\xi)}{\mu}\right)^{1 / 2}[\Theta-\theta(\xi)] \xi^{2} d \xi=0
\ee

\begin{figure}[H]
    \centering
    \includegraphics[width=.8\linewidth]{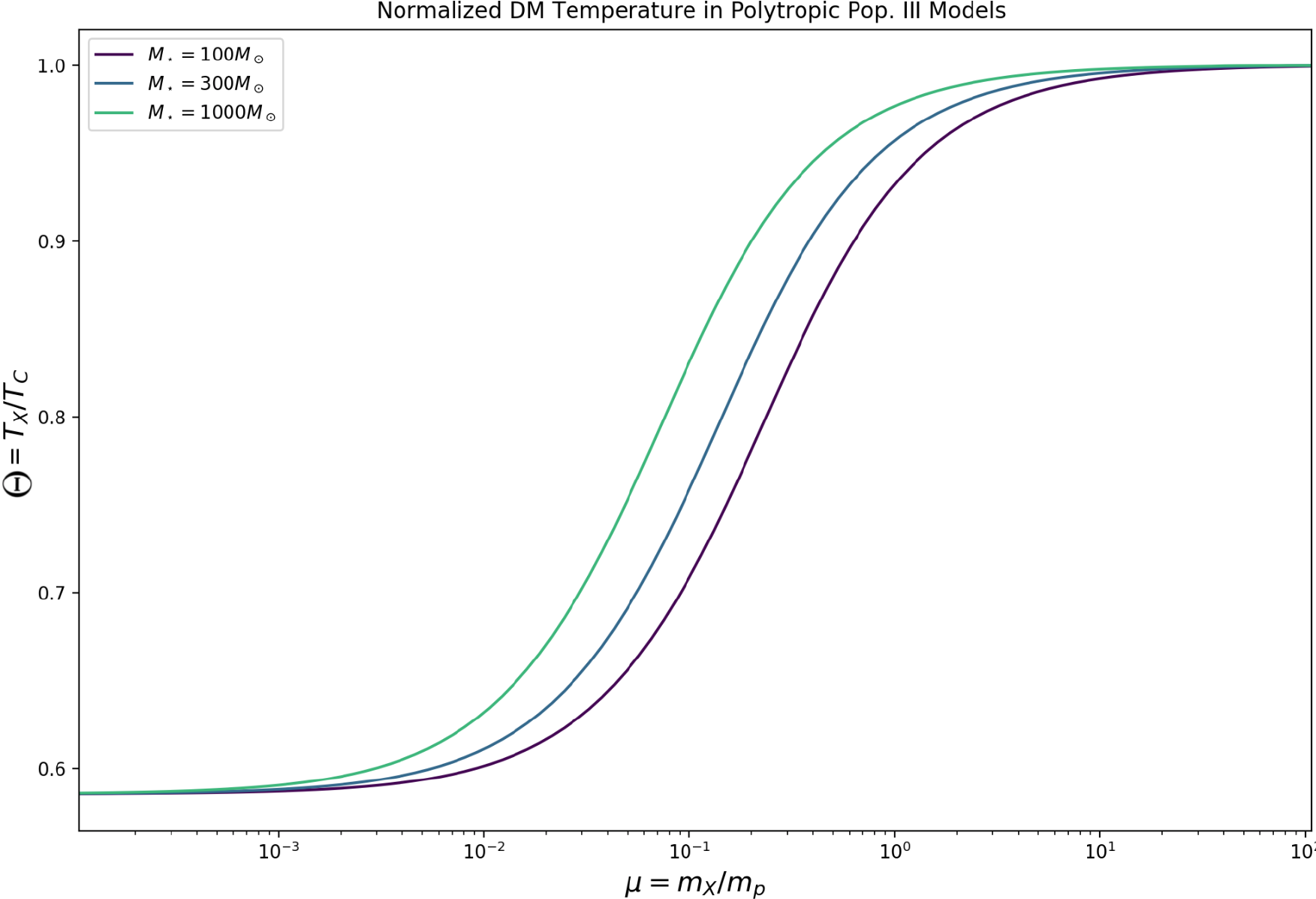}
    \caption{Numerical solution to equation \ref{eq:numsol} for a $100$, $300$, and $1000$ M$_\odot$ star. Note that when $\mu\gtrsim1, \Theta \approx 1$; when $\mu \ll 1$, $\Theta \approx 0.59$. }
    \label{fig:DMTemp}
\end{figure}

Standard numerical techniques, such as \texttt{fsolve} and \texttt{quad} in the Python \texttt{SciPy} package, can solve this equation easily, giving the dark matter temperature inside a star $T_X$ as a function of dark matter mass $m_X$. The results of this calculation are presented in Fig.~\ref{fig:DMTemp}. They key takeaway is that, to order-of-magnitude, the dark matter temperature is the core temperature of the star. When the dark matter mass is much larger than the proton mass, the dark matter temperature \textit{is} the exactly equal to core temperature of baryons inside the star; in the other limit when the dark matter mass is much less than the proton mass, the dark matter temperature is roughly half of the core temperature, 

\section{DM Evaporation rates}\label{sec:Evaporation}
For DM with $m_X\lesssim 1~\GeV$ one needs to consider the effects of evaporation~\cite{Gould:1987evap}, i.e. the process via which DM particles can be upscattered to velocities above the escape velocity via collision with nuclei. In this appendix we derive and validate an analytic approximation for the evaporation rate of Dark Matter from Pop~III stars, by assuming they are well described by $n=3$ polytropic models.

We start by estimating the $m_X$ below which evaporation becomes relevant. In order to obtain an order of magnitude approximation, we compare the average thermal velocity of DM particles at the core of a Pop~III star of a given mass to the escape velocity. Whenever the thermal velocity is higher than the escape velocity, evaporation becomes relevant. Technically one should use the escape velocity at the core, however, for the purpose of this order of magnitude analysis we will use the escape velocity at the surface, which is lower than the escape velocity at the core. This, in turn, means that we are over estimating the $m_X$ below which DM evaporation becomes relevant. As we have seen in Appendix~\ref{sec:CapDMTemp}, $T_X$ becomes constant throughout the star, and a very good order of magnitude estimate is $T_X\sim T_c$, with $T_c\sim 10^8$~K, the central temperature of the Pop~III star. The condition $v_X\gtrsim v_{esc}$, i.e. evaporation being efficient, can be recast into:
\be\label{eq:EvapmX}
m_X\lesssim 1~\GeV\left(\frac{\Rstar/\Rsun}{\Mstar/\Msun}\right).
\ee
For the $100\Msun$ Pop~III stars this becomes: $m_X\lesssim 4\times 10^{-2}~\GeV$, whereas for the heaviest Pop~III stars considered ($1000\Msun$), evaporation becomes relevant at $m_X\lesssim 1.4\times 10^{-2}~\GeV$. Therefore, at masses below $\sim 10^{-2}~\GeV$ we will need to include the effects of DM evaporation. Below we derive an analytic approximation for the evaporation rate, $E$.

In~\cite{Gould:1987evap} Gould derives analytic closed form evaporation rates from a stellar shell, assuming captured DM particles follow a truncated Maxwell Boltzmann distribution:
\be\label{eq:TruncMB}
f_X(w)=\frac{e^{-w^2/v_X^2}\Theta(v_c-w)}{\sqrt{\pi^3}v_X^3\left[\mathrm{Erf}(v_c/v_X)-\frac{2}{\sqrt{\pi}}\frac{v_c}{v_X}e^{-v_c^2/v_X^2}\right]}.
\ee
Here $w$ is the DM particle speed, $v_X\equiv\sqrt{2T_X/m_X}$ is the thermal average DM speed, $v_c$ represents the cutoff in the DM distribution, and henceforth we will assume it to be equal to the escape velocity from a given shell: $v_e$. We point out that compared to Eq.~(\ref{eq:TruncMB}), Gould does not include the appropriate normalization factor for a truncated DM distribution (i.e.  $1/\left[\mathrm{Erf}(v_c/v_X)-\frac{2}{\sqrt{\pi}}\frac{v_c}{v_X}e^{-v_c^2/v_X^2}\right]$) in his Eq.~(3.8) of~\cite{Gould:1987evap}. In what follows we will account for this factor. Below we briefly describe the steps of the calculation that will lead to our approximation of the evaporation rate used throughout this paper: Eqn.~(\ref{eq:EvapRate}). 

We start with the rate with which a DM particle of velocity $w$ will scatter to velocity $v$, as a result of collisions with nuclei inside the star. This is derived by Gould in~\cite{Gould:1987evap}, his Eq.~(3.1), which we reproduce here for clarity: 
\be\label{eq:Ratewtov}
    R^{\pm}(w \rightarrow v)= \frac{2}{\sqrt{\pi}} \frac{\mu_{+}^{2}}{\mu} \frac{v}{w} n(r) \sigma\left[\chi\left(\pm \alpha_{-}, \alpha_{+}\right)+\chi\left(\pm \beta_{-}, \beta_{+}\right) e^{\mu\left(w^{2}-v^{2}\right) / u^{2}(r)}\right].
\ee
The upper/lower sign corresponds to up-scattering ($v>w$) /down-scattering ($v<w$). The most important mechanism relevant for DM evaporation is the former. Here $n(r)$ represents the number density of target baryons inside the shell, and $u(r)\equiv\sqrt{2T(r)/m}$ is the thermal average velocity of a target nuclei of mass $m$. The following notations are used: $\chi(a, b) \equiv \int_{a}^{b} d y e^{-y^{2}}$, $\alpha_{\pm} \equiv(m / 2 T(r))^{1 / 2}\left(\mu_{+} v \pm \mu_{-} w\right)$,  $\beta_{\pm} \equiv(m / 2 T(r))^{1 / 2}\left(\mu_{-} v \pm \mu_{+} w\right)$, $\mu_{\pm} \equiv \frac{\mu \pm 1}{2}, \quad \mu \equiv \frac{m_X}{m}$.
Next, we consider the rate at which a DM particle of a fixed velocity $w$ escapes, i.e. up-scatters to any velocity $v$ greater than the escape velocity at the shell $v_e(r)$:
\be\label{eq:Evapw}
 \Omega_{v_{e}}^{+}(w) \equiv \int_{v_{e}}^{\infty} R(w \rightarrow v) d v.
\ee
Again, this has been calculated analytically by Gould~\cite{Gould:1987evap}:
\begin{align*}
    \Omega_{v_{e}}^{+}(w)=\frac{1}{2 \pi^{1 / 2}} \frac{2 T(r)}{m} \frac{1}{\mu^{2}} \frac{\sigma n(r)}{w}\left[\mu\left(\alpha_{+} e^{-\alpha_{-}^2}-\alpha_{-} e^{-\alpha_+^2}\right)+\left(\mu-2 \mu \alpha_{+} \alpha_{-}-2 \mu_{+} \mu_{-}\right) \chi\left(\alpha_{-}, \alpha_{+}\right)\right.
    \\ \left. +2 \mu_{+}^{2} \chi\left(\beta_{-}, \beta_{+}\right) e^{(-m_X/ 2 T)\left(v^{2}-w^{2}\right)}\right],
\end{align*}
where $\alpha_{\pm}$ and $\beta_{\pm}$ are evaluated for $v=v_e$. 
Next we can calculate the total evaporation rate from the shell by integrating over the velocity distribution of DM particles:
\begin{align*}
    R\left(v_{c} \mid v_{e}\right) \equiv \int_{0}^{\infty} f_{X}(w) \Omega_{v_{e}}^{+}(w) d w. 
\end{align*}
 Assuming $v_c=v_e$ [i.e. the Maxwell-Boltzmann (MB) distribution is truncated to the escape velocity], and $\mu\ll 1$, which is valid for Pop~III stars, in view of our discussion at the beginning of this section, we obtain the following estimate for the total evaporation rate from the shell:
 \be\label{eq:EvapShellApprox}
 R\left(v_{c}=v_{e} \mid v_{e}\right)\approx\frac{2}{\sqrt{\pi}}n(r)\sigma u(r)e^{-v_e^2/v_X^2}
 \ee
 The evaporation coefficient ($E$) is defined in the following way:
 \be\label{eq:DefERate}
 E=\frac{\int dV n_X R\left(v_{c}=v_{e} \mid v_{e}\right)}{\int dV n_X},
 \ee
with the integrals being done over the volume of the star, and $n_X$ representing the number density of DM particles inside the star. In Appendix~\ref{sec:CapDMTemp} we have shown that DM particles attain an isothermal sphere distribution:
\be\label{eq:nxIso}
n_X(r)=n_{X,c}e^{-m_X\Phi(r)/T_X}.
\ee
The gravitational potential is defined by: $\Phi(r)\equiv\int_0^r \,dr'G M(r')/r'^2$ and we calculated it using the $n=3$ polytropic model approximation in Appendix~\ref{sec:CapDMTemp}: Eq.~(\ref{eq:Phin3}) . Next we use this potential, in combination with $n_X$ from Eq.~(\ref{eq:nxIso}) to evaluate the integral in the definition of the evaporation coefficient from Eq.~(\ref{eq:EvapRate}). For the escape velocity from a shell at at radius $\xi$ we get: $v_e^2(\xi)=v_{esc}^2(1+\xi_1/2\theta(\xi))$. Remarkably, we find that for the case of $n=3$, the exponential term that that comes from multiplying $n_X$ and $ R\left(v_{c}=v_{e} \mid v_{e}\right)$ is now $\xi$ independent: $e^{-\Phi (r)m_X/T_X}e^{-v_e^2/v_X^2}=e^{-v_{esc}^2/v_X^2(1+\xi_1/2)}$. Therefore the integral at the numerator can be performed, if we know the radial dependence of $n(r)$ (the number density of protons) and $u(r)$ (their average thermal velocity). Using the $n=3$ polytrope, and the assumption of $P\sim P_{gas}\sim P_{rad}$ (i.e. a constant ratio between the gas and radiation pressure throughout the star) we have: $n(\xi)=n_c\theta^3(\xi)$, and $T(\xi)=T_c\theta(\xi)$, which implies $u(\xi)=u_c\theta^{1/2}(\xi)$. Lastly, We find that for an $n=3$ polytrope, the central proton density $n_c$ can be related to the average proton density: $n_c\approx \bar{n}_p\left(\frac{\xi_1^3}{6}\right)$. Putting everything together, the radial integral at the numerator of Eq.~(\ref{eq:EvapRate}) becomes, up to parameters that are $\xi$ independent, and therefore can be factored out: $\int_0^{\xi_1} d\xi\xi^2\theta^{7/2}(\xi)$. For $n=3$ we can approximate numerically this integral to $3/2$. Finally combining everything we have so far, we get the result quoted in Eq.~(\ref{eq:EvapRate}), which we reproduce here:
\be\label{eq:EvapRateAppendix}
E\approx\frac{3V_{\star}\bar{n}_pu_c\sigma}{2V_1\sqrt{\pi}}e^{-\frac{v_{esc}^2\mu}{u_c^2\Theta}(1+\xi_1/2)}.
\ee
 Throughout we denote by $V_1\equiv\int dV e^{-m_X\Phi(r)/T_X}$. This is an integral that can be performed numerically, but for which we will also find an analytic approximation. Defining the general case of the effective volume of index $j$ as: $V_j\equiv\int dV e^{-j\Phi(r)m_X/T_X}$, and using the standard second order approximation for the Lane-Emden functions of arbitrary index: $\theta(\xi)\approx 1-1/6\xi^2$, we find:
\be\label{eq:Vjapprox}
V_j\approx \frac{4\pi}{3j^{3/2}}r_X^3\left[\sqrt{\frac{\pi}{6}}\mathrm{Erf}\left(\sqrt{\frac{3j}{2}}\frac{\Rstar}{r_X}\right)-\sqrt{j}\frac{\Rstar}{r_X}\exp{\left(-\frac{3}{2}\frac{\Rstar^2}{r_X^2}j\right)}\right],
\ee
with $r_X^2\equiv\frac{9T_X}{4\pi G\rho_c m_X}$. We want to point out that when using Eq.~(\ref{eq:EvapRateAppendix}) in order to place bounds on $\sigma$ vs. $m_X$ for sub-GeV DM models, we always calculate numerically $V_1$, and do not rely on the approximation of Eq.~(\ref{eq:Vjapprox}). 

\begin{figure} [H]
    \centering
    \includegraphics[width=.8\linewidth]{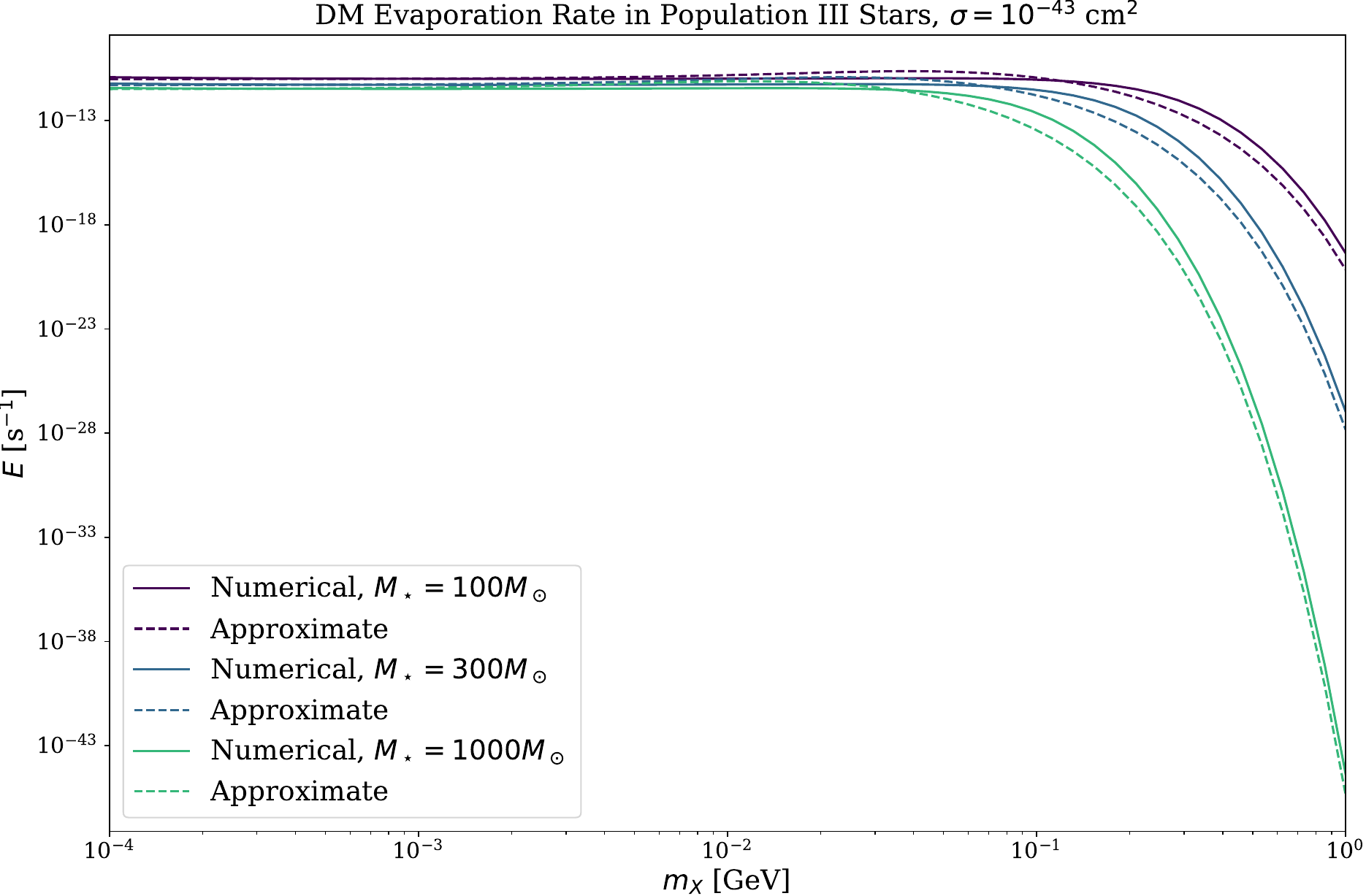}
    \caption{Comparison of the evaporation rate coefficient obtained numerically from Eq.~(\ref{eq:DefERate}) vs. our analytic approximation of Eq.~(\ref{eq:EvapRateAppendix}). Note the excellent agreement between the two, for Pop~III stars with masses ranging from $100-1000~\Msun$. Note that the rates are almost insensitive to the stellar mass, until the exponential decay factor kicks in. For higher mass stars this cutoff comes at lower $m_X$, as expected.}
    \label{fig:EvapApproxVsNumerical}
\end{figure}

We end this section with Fig.~\ref{fig:EvapApproxVsNumerical}, a plot that validates our analytic approximation for the evaporation rate coefficient. 


\section{Equilibration timescale and lower bounds on annihilation cross section}\label{sec:Equilibration}
Our formalism relies on the assumption of an efficient equilibration between capture and annihilations/evaporation of dark matter. This leads to a time independent number of DM particles inside the star, and to a simple form of the heating injected by DM annihilations, as presented in Eq.~(\ref{eq:LDMeqCtot}). In this section, we investigate the conditions under which the timescale for this equilibration is much shorter than the lifetime of the star. 

We start by briefly reviewing the DM models considered. First, for the WIMP window, which is bound at the lower end of $m_X$ (the Lee-Weinberg bound~\cite{Lee:1977}) by $m_X\gtrsim 10~\GeV$, and a the higher end of $m_X$ (the Griest-Kamionkowski bound~\cite{Griest:1990}) by $m_X\lesssim 120~\unit{TeV}$. We point out that the so called Lee-Weinberg limit has actually been found, independently, by several groups~\cite{Lee:1977,Dicus:1977nn,Hut:1977zn,Sato:1977ye,Vysotsky:1977pe} and that it's value is actually model-dependent. For instance, for Majorana fermions, where the annihilation cross section is p-wave suppressed, the Lee-Weinberg limit is enhanced by roughly one order of magnitude~\cite{Kolb:1986}. On the other hand, for scalar DM,~\cite{Boehm:2003} finds that the corresponding bound can be lowered to $\mathcal{O}(\unit{MeV})$. In fact, one of the most stringent bounds of WIMP DM lower mass limits comes from BBN, and it is roughly $\mathcal{O}(\unit{10 MeV})$~\citep{Sabti:2019}. Within this window of parameter space DM can be produced thermally, via the standard freezeout mechanism, without violating the unitarity limit, while still interacting only weakly. The thermal average DM annihilation cross section ($\sigmav$) can be expanded around $v\lesssim 1$:
\be\label{eq:sigmaVexpansion}
\sigmav\approx a+b\langle v^2 \rangle+\mathcal{O}(\avg{v^4})
\ee
Two distinct scenarios are commonly considered in the literature: the s-wave annihilations, for which $b=0$, so the thermal average cross section is a constant, independent of the DM velocity $v$. Remarkably, if the thermal average cross section is at the weak-scale, i.e. $\sigmav\approx a\sim 10^{-26}\ccpers$, one recovers, via the freeze-out mechanism, a value of the relic abundance that matches observations, i.e. $\Omega_X\sim0.3$. This is commonly known as the WIMP miracle, and was one of the main reasons WIMP DM models were theoretically favoured in the past decades, before LHC data and direct detection experiments placed severe constraints on such models. Alternatively, one can consider the p-wave annihilation, when $a=0$, and the thermal average cross section depends on the DM thermal velocity. In this case it us useful to recast Eq.~(\ref{eq:sigmaVexpansion}) as: $\sigmav=b\avg{v^2}=b'/x$, with $x\equiv m_X/T_X$, the commonly defined dimensionless decoupling parameter. In order to match the observed relic abundance the parameter $b'$ has to have a value of $b'\sim 10^{-24}\ccpers$~\cite{Lopes:2016}.

Outside of the WIMP regime, we have, at the higher mass end, what is commonly know as the Superheavy Dark Matter. Reproducing the correct thermal relic abundance via the freeze-out mechanism would violate unitarity at those high masses, if DM is considered to be a point particle. One of the most well known non-thermal production mechanisms for superheavy DM is the gravitational production during inflation, which leads to what is commonly known as WIMPZILLAs~\cite{Kolb:1999}. Those particles can be their own antipartners, and therefore annihilate, with a cross section that is not fixed by the relic abundance. So, in principle, they could annihilate with cross sections as high as the unitarity limit:
\be\label{eq:sigmavUnitarity}
\sigmav_{U.L.}=\frac{4\pi}{m_X^2v}(2J+1),
\ee
with $v\equiv\sqrt{2T_X/m_X}$ and $J=0$~(s-wave) or $J=1$~(p-wave). 
At the other end of the mass spectrum, for sub-GeV DM, in this paper we will only consider two such models: Strongly Interacting Particles (SIMP) dark matter~\cite{Hochberg:2014} and the Co-SIMP model~\cite{Smirnov:2020}. In both of those models DM is thermally produced, and the Lee-Weinberg limit is bypassed by allowing interactions with a coupling stronger than the weak scale. As opposed to the usual s/p-wave annihilations, which are $2\to2$ annihilations, those processes are $3\to2$. Namely: $DM+DM+DM\to DM+DM$(SIMP) or $DM+DM+SM\to DM+SM$(Co-SIMP). 

We discuss next in some detail first of those two models: SIMP DM. The DM number changing rate for this process is controlled by the thermal averaged cross section $\avg{\sigma_{SIMP}v^2}$, which, one usually assumes, based on dimensional grounds to be proportional to some effective, dimensionless coupling constant controlling the annihilation process: 
\be\label{eq:sigmavSIMPscaling}
\avg{\sigma_{SIMP}v^2}\sim\frac{\alpha^3_{SIMP}}{m_X^5}.
\ee
Using the standard thermal relic abundance calculation one can show that if $\alpha_{SIMP}\sim1$ and $m_X\sim0.3\GeV$, this model can produce sub-GeV DM efficiently~\cite{Profumo:2017book}:
\be\label{eq:SIMPThermalRelic}
\left(\frac{\Omega_X}{0.2}\right)\sim\left(\frac{m_X}{35~\unit{MeV}}\right)^{3/2}\left(\frac{x_{f.o.}}{20}\right)^2\left(\frac{1}{\alpha_{SIMP}}\right)^{3/2},
\ee
with $x_{f.o.}$ the value of the decoupling parameter when the freezeout condition is met: $\Gamma_{annih}=H(T)$, i.e. when the annihilation rate per DM particle is equal to the Hubble rate. Assuming $x_{f.o.}\sim20$ we can combine Eqns.~(\ref{eq:sigmavSIMPscaling})-(\ref{eq:SIMPThermalRelic}) to obtain the following mass dependence of the annihilation rate coefficient: 
\be
\avg{\sigma_{SIMP}v^2}\sim 2.7\times 10^4\left(\frac{1~\GeV}{m_X}\right)^2~\GeV^{-5}
\ee
This rough estimate, which we will use in our calculations, can be confirmed by fully solving numerically the corresponding Boltzmann equation, as done by~\cite{Hochberg:2014} (see their Fig.~2).

We point out here that one could consider $4\to2$ processes as well, and show that thermal relics with masses at the $m_X\sim 100~\unit{keV}$ scale can be produced thermally, if DM interacts strongly: i.e. $\avg{\sigma_{4\to2}v^3}\sim\alpha_{4\to2}^4/m_X^8$, with $\alpha_{4\to2}\sim1$.

For the Co-SIMP model, the $3\to2$ process of interest is: $DM+DM+SM\to DM+SM$. In order to produce $\Omega_X\sim0.3$ the thermal averaged annihilation factor must be~\cite{Smirnov:2020}:
\be\label{eq:CoSIMPThermalRelic}
\avg{\sigma_{CoSIMP}v^2}\sim 10^{12}\left(\frac{\unit{MeV}}{m_X}\right)^3\left(\frac{0.12}{\Omega_Xh^2}\right)^2~\GeV^{-5}
\ee

Next, we proceed to calculate the equilibration time scale between the capture and annihilation/evaporation processes, and compare it to the lifetime of the star, for each of the DM models described above. As discussed in Sec.~\ref{sec:MSCapture}, the number of DM particles inside a star will reach a constant, equilibrium, value, whenever at times larger than $t_{eq}\equiv\tau_{eq}/\kappa$. The usual equilibration time scale, between capture and annihilation is defined by: $\tau_{eq}\equiv 1/\sqrt{C_{tot}C_A}$, with $C_{tot}$ being the total capture rate, and $C_A$ the $N_X$ independent annihilation coefficient defined in terms of the total annihilation rate $\Gamma_A$ as: $\Gamma_A=C_A N_X^j$, with $j$ being the number of DM particles entering the annihilation process. Evaporation leads to a shortening of the equilibration timescale by a factor of $\kappa\equiv\sqrt{1+E^2\tau_{eq}^2/4}$. Imposing equilibration in a time less than a fraction of the typical lifetime of the Pop~III star, which, in view of their high masses becomes independent of $\Mstar$, and with a value approximately equal to $10^6$~yrs. The annihilation coefficient $C_A$ for $2\to2$ (s/p-wave) annihilations takes the following form:
\be\label{eq:CA2to2}
C_A^{2\to2}=\frac{\int dV n_X^2\sigmav}{\left(\int dV n_X\right)^2},
\ee
whereas for the SIMP model we have:
\be\label{eq:CASIMP}
C_A^{SIMP}=\frac{\int dV n_X^3\avg{\sigma_{SIMP}v^2}}{\left(\int dV n_X\right)^3},
\ee
\be\label{eq:CACoSIMP}
C_A^{CoSIMP}=\frac{\int dV n_X^2n_{SM}\avg{\sigma_{CoSIMP}v^2}}{\left(\int dV n_X\right)^2},
\ee
with $n_{SM}$ the numer density of the relevant SM particles entering the process. For us this will be the same as the number density of protons inside the star, since we approximate the star as being made of fully ionized $H$. 

In general, both the SIMP and the CoSIMP DM can be realized in nature, simultaneously. However, for our purposes, the DM heating due to CoSIMP DM annihilations inside the star is many orders of magnitude higher than the heating due to SIMP DM annihilations. This can be traced to the much lower $n_X$ when compared to $n_{SM}$, inside the star. Therefore, inside the star it is the CoSIMP DM that has the dominant effect. Converselly, outside the star $n_{SM}$ becomes much lower than the ambient $n_X$. So, if those two models coexist (SIMP/CoSIMP), then, outside the star we need to take into account the efects of DM annihilations on the DM densities, i.e. the ``annihilation plateau.'' We do this analysis in Appendix~\ref{sec:DMHalos}. 

\begin{figure} [!ht]
    \centering
    \includegraphics[width=.8\linewidth]{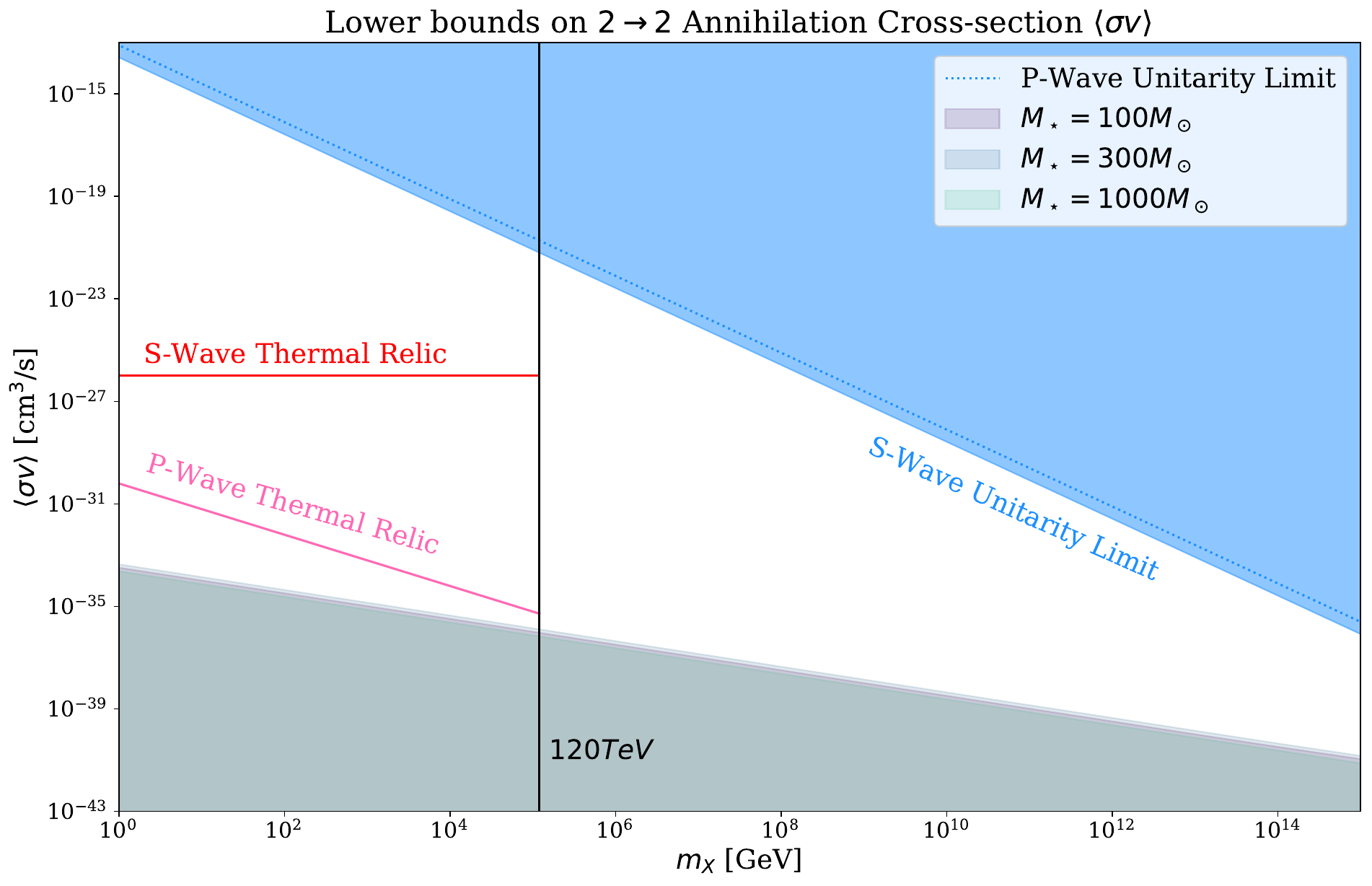}
    \caption{Lower bounds on $\sigmav$ for $2\to2$ annihilations, such that the capture and annihilation processes equilibrate inside the star in $t_{eq}\sim 10^4$~yrs, corresponding to about $1\%$ of the typical lifetime of the Pop~III stars considered here. The gray-out region at the bottom is excluded, since the equilibrium is attained in more than $10^4$~yrs. Note that this excluded region is almost the same for Pop~III stars with $\Mstar$ in the $100-1000~\Msun$ range. Additionally, we compare those lower bounds with the unitarity limit, and with the $\sigmav$ required by the freezeout mechanism for thermal relics, in the WIMP mass window.}
    \label{fig:SigmaVLBSupGeVDM}
\end{figure}
In Fig.~\ref{fig:SigmaVLBSupGeVDM} we plot the upper bound on $\sigmav$ obtained by requiring $t_{eq}\lesssim 10^4$~yrs. Note that for the case considered in that figure, $m_X\gtrsim 10~\GeV$, DM evaporation can be safely neglected, so $t_{eq}\approx\tau_{eq}$. The fact that the upper bound is always below the thermal relic $\sigmav$ demonstrates that for WIMPs equilibrium is attained well within the lifetime of the star. For Superheavy DM particles that annihilate, the same conclusion holds, as can be seen from comparing the unitarity limit to our lower bounds on $\sigmav$.
\begin{figure} [!ht]
    \centering
    \includegraphics[width=.75\linewidth]{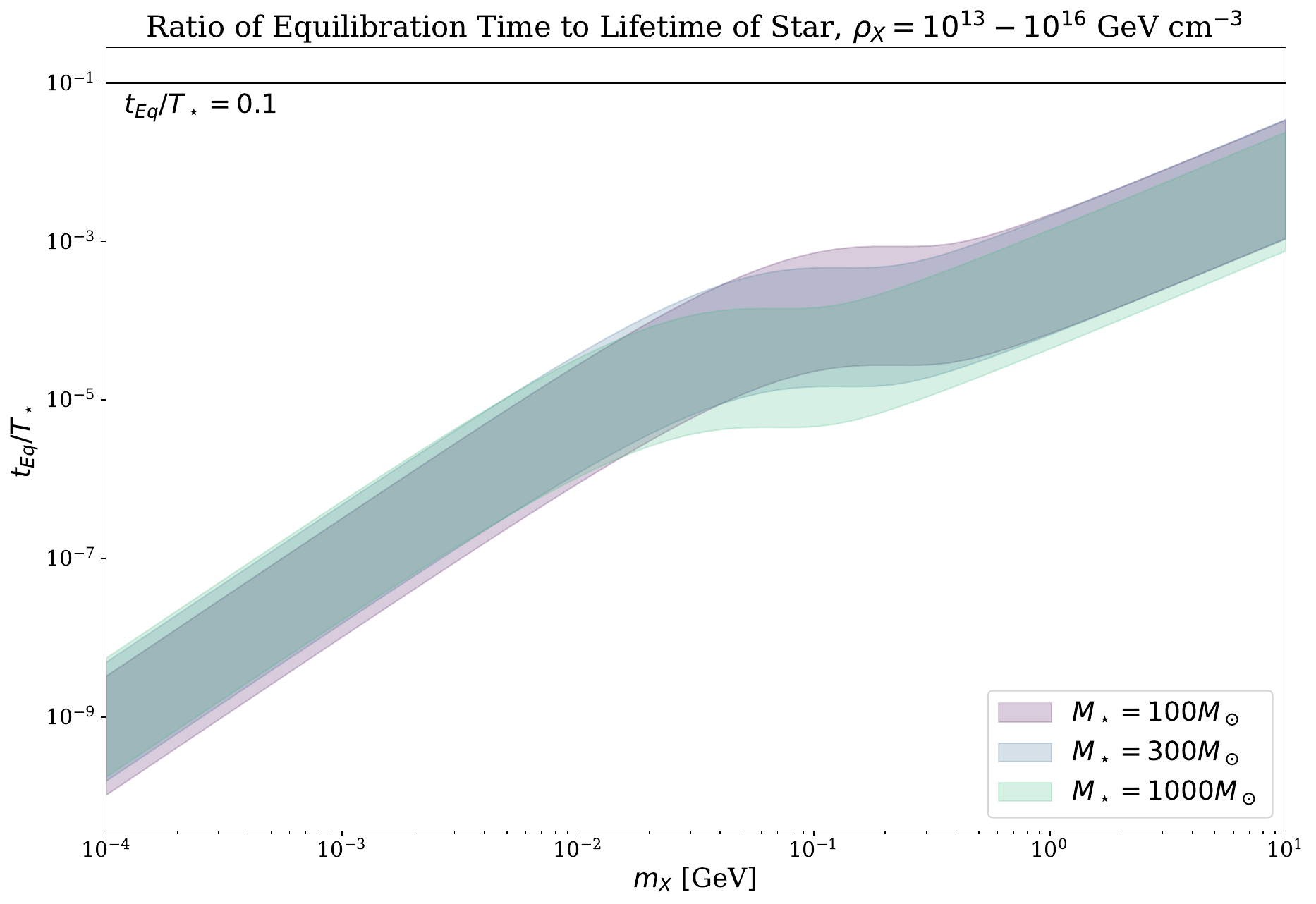}
    \caption{Ratio between equilibration timescale ($t_{eq}\equiv\tau_{eq}/\kappa$) and the lifetime of the star ($T_{\star}\sim 10^6$~yrs), as a function of $m_X$ for CoSIMP DM. Each band corresponds to ambient DM densities $10^{13}~\GeV\percc\lesssim\rho_X\lesssim10^{16}~\GeV\percc$. The trhee different bands correspond to three different values of $\Mstar$ labeled in the legend.}
    \label{fig:SigmaVLBSubGeVDM}
\end{figure}
In Fig.~\ref{fig:SigmaVLBSubGeVDM} we plot the equilibration timescale normalized to the lifetime of the star, for CoSIMP DM. For the entire mass range considered we took $\sigma$ at the deepest edge of our excluded regions for each star. In order to estimate the number density of the relevant SM particles, we assumed we used the $n=3$ polytrope approximation: $n_{SM}\sim\theta^3(\xi)$. Note that the equilibration timescale decreases rapidly with $m_X$. More importantly, at the highest $m_X$ considered here for the CoSIMP DM model, equilibrium is still attained well within the lifetime of the star.
\begin{figure} [H]
    \centering
    \includegraphics[width=0.75\linewidth]{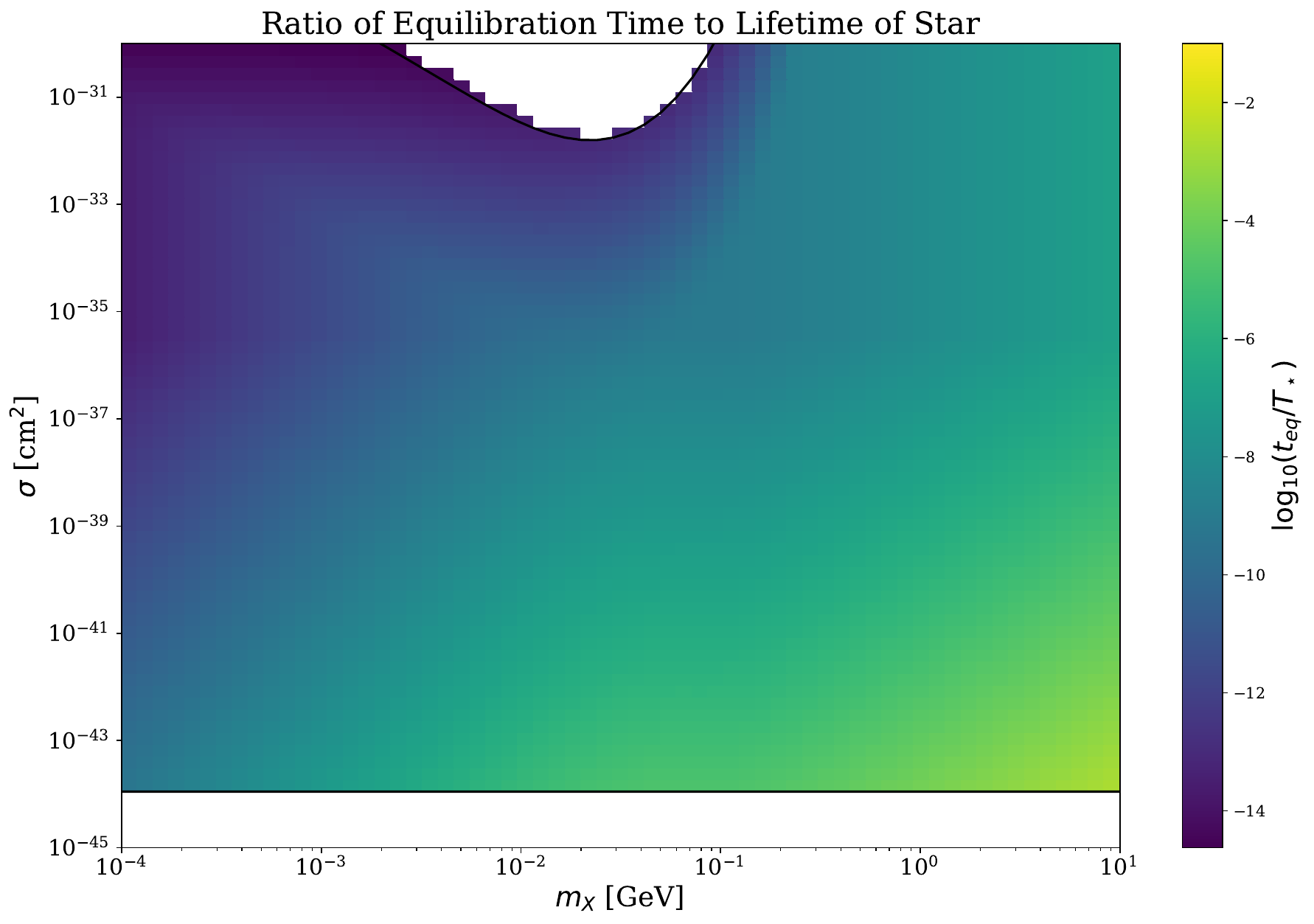}
    \caption{Ratio between equilibration timescale ($t_{eq}\equiv\tau_{eq}/\kappa$) and the lifetime of the star ($T_{\star}\sim 10^6$~yrs), as a function of $m_X$ for CoSIMP DM. At each point within the region we exclude in view CoSIMP DM we calculate the corresponding ratio, and color code it, according to the color bar on the right. Note that throughout this region $t_{eq}\ll T_{\star}$, i.e. equilibrium between capture and annihilations/evaporation is attained in a timescale that is much shorter than the lifetime of the star, for the entire parameter space considered.}
    \label{fig:teqToTstar}
\end{figure}
We can see the Same effect, for the entire swath of parameter space excluded by Pop~III stars for CoSIMP DM-proton cross section, in Fig.~\ref{fig:teqToTstar}.

To sum up, in this Appendix we investigated whereas our assumption of equilibration between capture and annihilations/evaporation is reached within a small fraction of the lifetime of the star. We find that this is certainly the case for two of the thermal DM models considered: WIMPs and for CoSIMP DM. If we allow for the possibility of SIMP interactions as well, we find that their annihilations, inside the SM rich environment of a star, are negligeable, when compared with CoSIMP DM. Therefore, equilibrium is reached mostly due to efficient CoSIMP annihilations (at lower $\sigma$), and aided by the effects of evaporation (at higher $\sigma$). Regarding non-thermal DM we considered superheavy dark matter mdels, such as WIMPZILLAs. We find that an equilibrium can be attained well within the lifetime of the star if WIMPZILLAs self-annihilate. Moreover, we find the lower bound on $\sigmav$ for such models, for which equilibrium is reached within $1\%$ of the lifetime of the star. 

In the next section we investigate the role of DM annihilations in the environment surrounding the star, and check the robustness of our results when when including this effects on the ambient DM density.


\section{DM Mini-halos}\label{sec:DMHalos}
The first stars in the universe are believed to have been formed in DM mini-halos of mass $M_{halo} = 10^5 - 10^6 M_\odot$, at typical redshifts of $z = 10 - 50$ \cite{Abel:2001}. The DM profiles formed at these redshifts hosting Pop~III stars have been studied extensively, particularly within the context of DM's effect on the stellar formation~\cite{Spolyar:2008dark,Freese:2008dmdens}. These works have also discussed the effects of baryonic in-fall to the halo's core on its density profile. The process of adiabatic contraction has been used to approximate this through calculations involving the conservation of adiabatic invariants, assuming an adiabatic process~\cite{Young:1980,Blumenthal:1985}. Recent work has also demonstrated that massive Pop~III star formation can persist up to redshifts of $z \sim 6$ in extreme cases~\cite{Mebane:2018}. This supports the recent claim of detection of a Pop~III stellar complex at $z \sim 7$  by~\cite{Vanzella:2020}.

In this section we will mainly utilize the methods in \cite{Freese:2008dmdens} and \cite{Blumenthal:1985} to calculate the density profiles of DM mini-halos at redshifts of $z \sim 7$ and $z \sim 10 - 50$ to find the DM density at the edge of the baryonic core, $\rho_X$. This parameter is necessary for accurately calculating the DM capture rates in Pop~III stars and thus calculating constraints on the DM-nucleon scattering cross section. We also calculate the  DM dispersion velocity, $\bar{v}$, for these different halos.

We start by describing the initial DM halo profile before contraction using the standard NFW profile~\cite{Navarro:1997}~\footnote{Our results regarding the adiabatic compression of DM densities are largely insensitive to the initial profile, as shown in~\citet{Freese:2008dmdens}, who demonstrates that even for the most extreme case of a purely cored profile, there is significant enhancement of DM densities, and that this enhancement is largely insensitive to the choice of the initial profile, as seen in their Figs.~2 and~3. See also Fig.~1 of~\cite{Freese:2008ds}.}:
\be
\rho_{halo} = \frac{\rho_0}{\frac{r}{r_s}\left(1+\frac{r}{r_s}\right)^2}
\label{eq:NFWDensityProf}
\ee
where $\rho_{halo}$ is the DM density at a point $r$ from the center, $r_s$ is the scale radius and $\rho_0$ is a normalization called the central density~\cite{Freese:2008cap, Ilie:2019}. The virial raidus $r_{vir}$ is related to the virial mass of the halo ($M_{halo}$) in the following way: 
\be
\frac{M_{halo}}{\frac{4\pi}{3}r_{vir}^3}=200\rho_{crit}(z),
\ee
where $\rho_{crit}$ is the critical density of the universe at redshift $z$. The central density is calculated as a function of the halo's concentration parameter, $c \equiv \frac{r_{vir}}{r_s}$, and the redshift, z, via the following equation:
\be
\rho_0 = \rho_{crit}\left(z\right) \frac{200}{3} \frac{c^3}{\text{ln}\left(1+c\right) - c/\left(c+1\right)}.
\label{eq:CentralDensity}
\ee
 From the virial theorem, we can calculate the dispersion velocity of DM, $\bar{v}$:
\be
\langle \bar{v}^2 \rangle = \frac{\bar{W}}{M_{halo}}
\label{eq:vbar}
\ee
where 
\be
W = -4 \pi G \int \rho_{halo} M_{halo}\left(r\right) r dr
\ee
is the gravitational potential of the DM halo. The typical pop III star forming at $z \sim 10 - 50$, in halos of mass $M_{halo} = 10^5 - 10^6 M_\odot$, with concentration parameters from $c = 1 - 10$, will have dispersion velocities of $\bar{v} = 1 - 15$ km/s, corroborated by \cite{Freese:2008cap, Ilie:2019}. For the case of a Pop III star forming at $z \sim 7$, \cite{Mebane:2018} showed a minimum halo mass for formation of $M_{halo} \sim 10^8 M_\odot$. Assuming concentration parameters ranging from $c = 1 - 10$ as well, the dispersion velocities range from $\bar{v} = 22 - 55$ km/s.

Under the assumption of adiabaticity for the collapse of a protostellar cloud, we can assume adiabatic invariants are well conserved and use this fact to calculate the effect of baryonic in-fall on the DM profiles, following~\cite{Freese:2008dmdens}. We utilize the Blumenthal method~\cite{Blumenthal:1985}, where conservation of angular momentum is assumed as the halo is compressed, to solve for the final mass profile given by the following equation: $M_f (r_f) r_f = M_i (r_i) r_i$. This equation essentially says that a particle at an initial radius $r_i$, is pulled into a final radius $r_f$, where $M(r)$ is the total enclosed mass at r. Note that~\cite{Freese:2008dmdens} has shown that, within factors of a few, this method reproduces results from the more elaborate Young~\cite{Young:1980} and Gnedin~\cite{Gendin:2004} methods, which allow for non circular DM orbits. 

\begin{figure} [bht]
    \centering
    \includegraphics[width=0.8\linewidth]{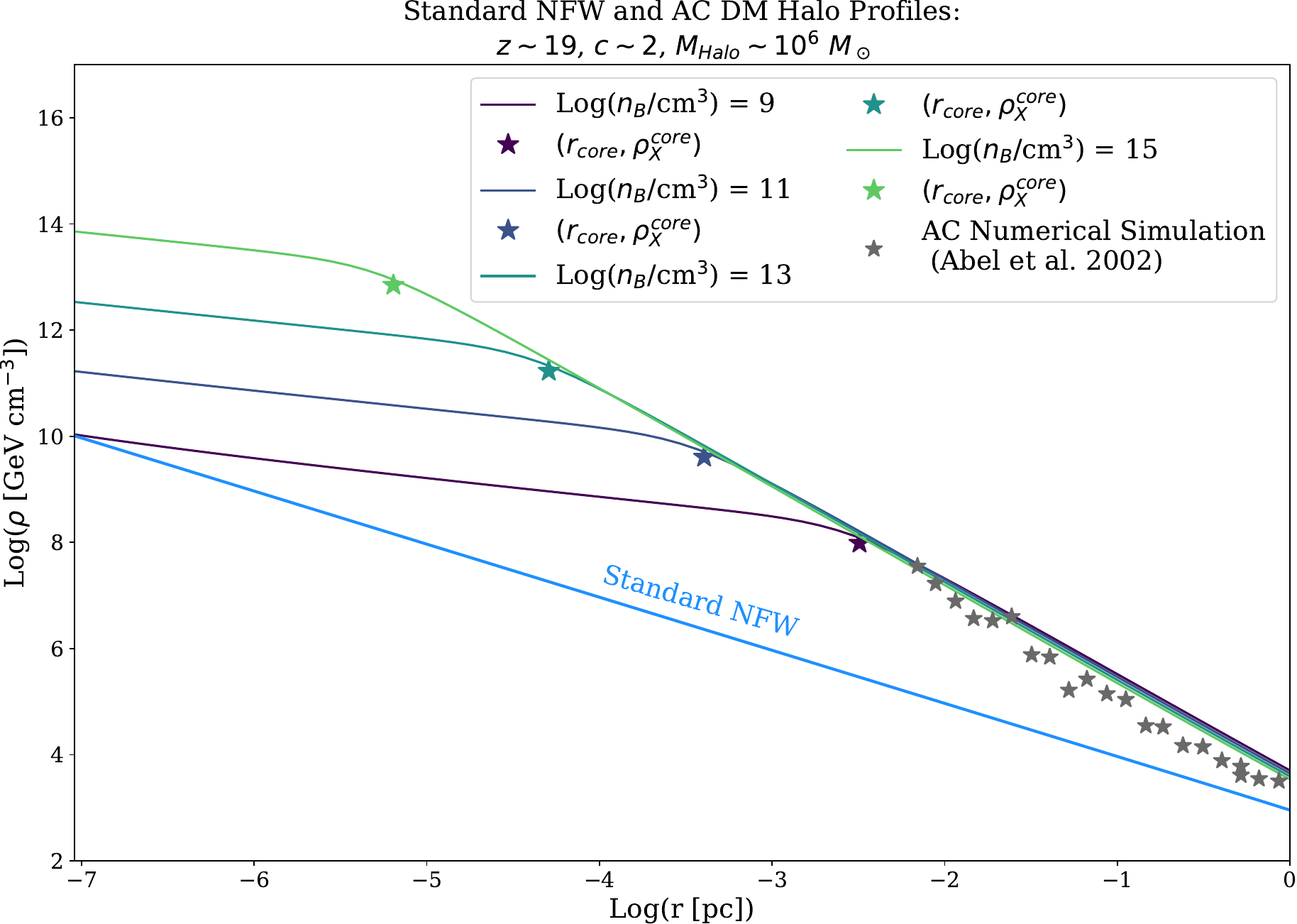}
    \caption{Adiabatically contracted NFW profiles vs. numerical simulation of DM densities during the runaway collapse of a pre-Pop III star molecular gas cloud. Each profile corresponds to a different value for the protostellar core density ($n_{core}$), labeled in the legend. The simulation data points are taken from Fig.2 of~\cite{Abel:2001}, which corresponds to $n_{core}\sim 10^{13}\unit{cm}^{-3}$. Resolution limits the simulation from probing the DM densities to scales smaller than $\sim 10^{-2}$~pc. Note the excellent agreement with the AC contracted profile for the same $n_{core}=10^{13}\unit{cm}^{-3}$. We additionally plot the predicted DM density at the edge of the baryonic core from Eq.~(\ref{eq:rhoXCore}), at the predicted radius of the core given by Eq.~(\ref{eq:rCore}).} 
    \label{fig:DM_AC}
\end{figure}

It is worth mentioning that most numerical N-body simulations do not have the required resolution to follow the DM profiles directly, especially in the inner miliparsec of the microhalo. However, the numerical results of ~\cite{Abel:2001} support an adiabatically contracted DM density profile, for a baryon gas density up to $n_{core}\sim 10^{13}\percc$, and as far inward as the resolution limit of the simulation, $\sim 10^{-2}$~pc, as one can see in Fig.~\ref{fig:DM_AC}. 

\begin{figure} [!ht]
    \centering
    \includegraphics[width=.8\linewidth]{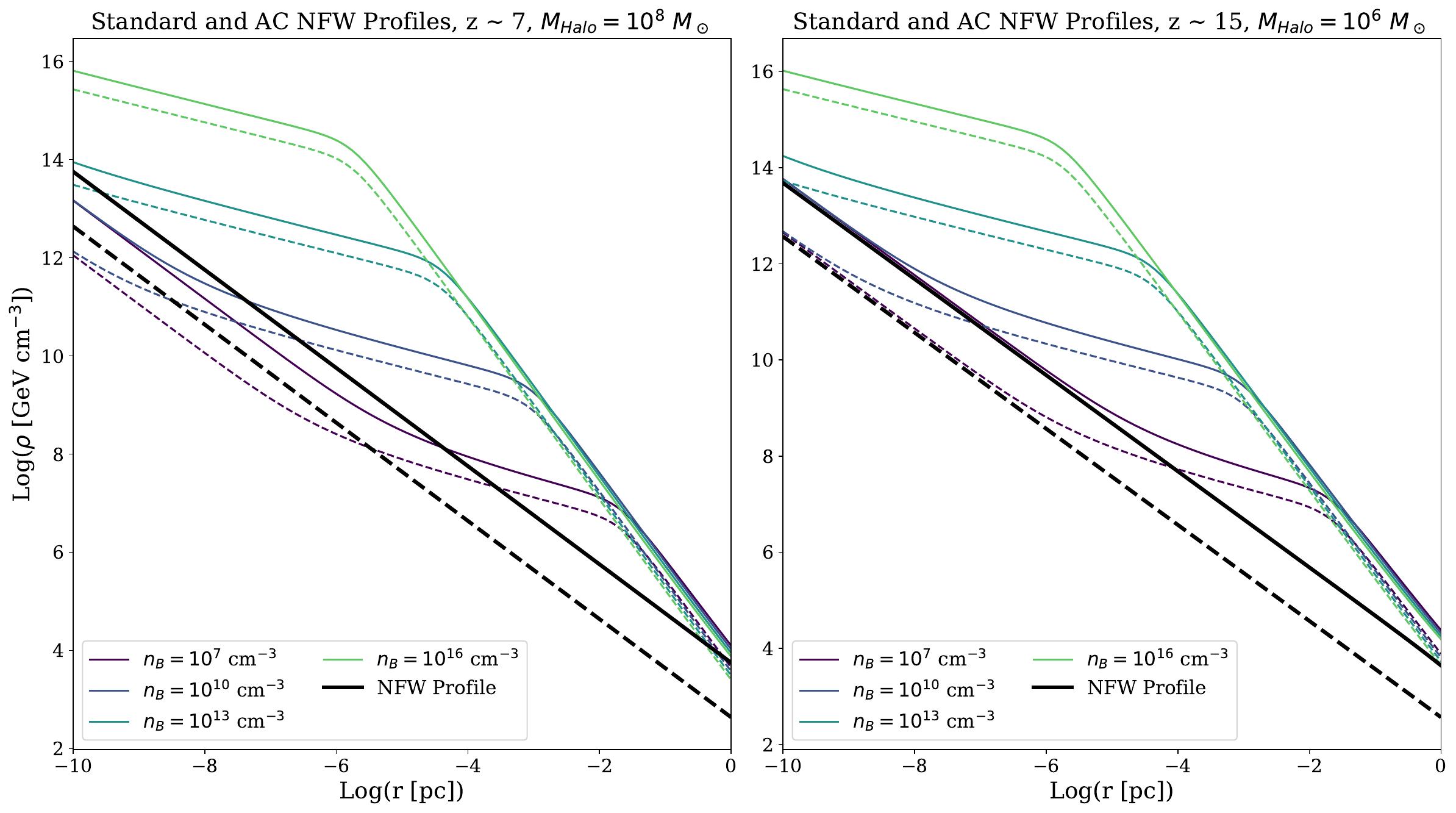}
    \caption{Adiabatically contracted NFW DM profiles for redshifts $z \sim 7$ and $z \sim 15$. The solid lines represent the profiles for $c = 10$ while the dotted lines are those for $c = 1$. The different colors of the lines represent varying the densities of the core baryonic gas cloud collapsing at the center of the DM halo, on its way to becoming a proto Pop~III star ($n_{core}\sim 10^{22}~\unit{cm}^{-3}$). Despite the different redshifts, the DM profiles are very similar and demonstrate that the effect of concentration parameter is mostly insignificant in the ranges discussed. Both cases show a significant enhancement and lead to DM densities at the edge of the core as high as $\rho_X = 10^{16}$ GeV cm$^{-3}$, assuming adiabatic contraction operates until the formation of a protostellar core.}
    \label{fig:DMDensityProfiles}
\end{figure}

As evidenced in Fig.~\ref{fig:DM_AC}, the adiabatically enhanced DM densitiy profiles have a broken power law behavior. This is due to the sharp decrease of the baryonic density outside of the baryonic core. As found in~\cite{Freese:2008ds,Smith:2012} the value of the adiabatically contracted dm density at the edge of the baryonic core can be estimated in terms of the number density of the protons inside the core:
\be\label{eq:rhoXCore}
\rho_X\approx 5 \left(\frac{n_{core}}{\cc}\right)^{0.81}\GeV\percc.
\ee
Moreover, the profile outside of the baryonic core scales as:
\be\label{eq:rhoXout}
\rho_X(r)\approx\rho_X(1~\unit{pc}) \left(\frac{r}{1~\unit{pc}}\right)^{-1.8},
\ee
as found in~\cite{Freese:2008ds,Smith:2012}, and additionally verified by us in this work. Remarkably, both the numerical simulations of~\citet{Abel:2001} and the adiabatic contraction formalism predict very similar values for the DM densities outside of the baryonic core (see Fig.~\ref{fig:DM_AC}). Moreover, the DM density at 1 pc can be estimated with $\rho_X(1~\unit{pc})\sim 10^{4}~\GeV\percc$. Note that this is only mildly sensitive to the concentration parameter or the redshift where the Pop~III star forms, as shown in Fig.~\ref{fig:DMDensityProfiles}. Equating the values of the DM density from Eqns.~(\ref{eq:rhoXCore}) and (\ref{eq:rhoXout}) one gets the following estimate for the radius of the baryonic core~\cite{Smith:2012}:
\be\label{eq:rCore}
r_{c}\approx 16.7 \left(\frac{n_{core}}{10^{14}~\percc}\right)^{-0.81/1.8}~\unit{AU}
\ee

We can now estimate the DM density at the edge of baryonic core from the simulation of ~\citet{Abel:2001}, which corresponds to an $n_{core}\sim 10^{13}~\percc$, and for which the numerical resolution limits the computation of DM densities in the inner milliparsec, as seen in Fig.~\ref{fig:DM_AC}. However, in view of the agreement between simulation data and the adiabatic contraction (AC) profile, we expect this trend to continue at least up to the edge of the baryonic core. This means that in fact the numerical simulations of~\citet{Abel:2001} support a DM density at the edge of the baryonic core of $\rho_X\approx 5 \times 10^{13\cdot 0.81}~\GeV\percc\sim 10^{11}~\GeV\percc$. If adiabatic compression operates up to higher $n_{core}$, then this value will be correspondingly enhanced by $(n_{core}/10^{13})^{0.81}$. Specifically, assuming that AC ceases to operate at $n_{core}\sim 10^{16}~\percc$, we estimate $\rho_X$ at the edge of the core to be $~5\times 10^{13}~\GeV\percc$, whereas for an $n_{core}\sim 10^{19}~\percc$ the corresponding DM density is $\rho_X\sim 10^{16}~\GeV\percc$.  As one can see from Fig.~\ref{fig:DM_AC}, DM densities continue to increase, albeit at a milder rate, at scales smaller than the baryonic core. Conservatively, we will always set the ambient DM density to be equal to the DM density at the edge of the baryonic core, corresponding to the $n_{core}$ where AC is assumed to cease to operate. For the later assume a value between $n_{core}\sim 10^{16}-10^{19}~\percc$, leading to ambient DM densities ranging between $10^{13}-10^{16}~\GeV\percc$, which are the values we used in this work to place bounds on DM-proton interaction cross section. We want to additionally emphasise that results from other numerical simulations~\citep[for example][]{Sellwood:2005,Gendin:2011,Davis:2013mha}, in addition to the aforementioned~\cite{Abel:2001}, are in good agreement with those obtained via the adiabatic contraction formalism, especially for high redshift halos, such as those where Pop~III stars form, since baryonic feedback effects are not important in this case.    

We summarize our results from contracting the initial NFW profiles given by Eq.~(\ref{eq:NFWDensityProf}) in Fig.~\ref{fig:DMDensityProfiles} for Pop~III star-forming halos at $z \sim 7$ and $z \sim 15$. We find that, at both redshifts, the DM densities at the edge of the baryonic cores are greatly enhanced by the process of adiabatic contraction, largely irrespective of the concentration parameter. It is also evident that there is little variation in the densities at the edge of the baryonic core when considering the different redshifts. As discussed in Section \ref{sec:Constrain}, these enhanced DM densities allow for competitive constraints on the DM-nucleon scattering cross-section.

In contrast to the enhancement of the ambient DM density due to adiabatic contraction, we also consider the effect of DM annihilation on the density profile, which reduces the ambient density over time. We start by considering what effect this may have on the initial DM profile, i.e. before star formation. To estimate this, we first take an initial NFW profile that evolves from annihilations as baryons fall inwards and collapse to form a proto-stellar core. In doing so, we assume that the collapse is rapid enough that the DM profile does not respond gravitationaly, but rather only through annihilations. For a more conservative result, we also start with an initial AC profile with a baryon core density made artificially high. This is not physically realistic, as the DM profile would, in reality, take time to become enhanced, but we consider it to show that even for initially higher density profiles, the effects of annihilation are not relevant at the distances that the star would capture dark matter.

For DM annihilating via a $2\to 2$ process, the differential equation governing the rate at which DM particles are annihilated out of the halo is given by:
\be\label{eq:dNxdt_WIMP}
\frac{dN_X}{dt} = -\Gamma_{ann}^{2\rightarrow 2} = -\int dV n_X^2 \langle \sigma v\rangle.
\ee
Solving for the DM density at a given time and position gives:
\be
\rho_X^{2 \rightarrow 2}(r, t) = \frac{\rho_0(r) \rho_{AP}^{2 \rightarrow 2}(t)}{\rho_0(r) + \rho_{AP}^{2\rightarrow 2}(t)},
\label{eq:rhot_22}
\ee
where $\rho_0(r) = \rho_0(0, r)$ is the initial DM density, and $\rho_{AP}^{2\rightarrow2}$, the so-called ``Annihilation Plateau," is given by $\rho_{AP}^{2\rightarrow 2}(t) = \frac{m_X}{\langle \sigma v \rangle t}$. Since the $2\to 2$ process considered does not require baryonic matter for annihilation, there is no functional dependence on the baryon content, and so it is straightforward to calculate the radius at which the annihilation plateau begins to be relevant as a function of time. This radius can be found by finding where the initial profile $\rho_0(r)$ equals the annihilation plateau density $\rho_{AP}^{2\to 2}(t)$. Doing so for an initial NFW profile gives the following scaling relation:
\be
r_{AP} \approx 2\times 10^{-9} \text{ pc} \left(\frac{r_s}{190\text{ pc}}\right)\left(\frac{\rho_0}{30\text{ GeV cm}^{-3}}\right)\left(\frac{\langle\sigma v\rangle}{10^{-26}\text{ cm}^{3}\text{ s}^{-1}}\right) \left(\frac{t}{10^6\text{ yrs}}\right)\left(\frac{1\text{ GeV}}{m_X}\right).
\ee
Since DM particles are generally captured outside the $10$ A.\ U.\ region ($\approx 5\times 10^{-5}$ pc), it is safe to say that the initial NFW profile will not be affected at the distance scales relevant for capture during the time it takes for star formation ($t\sim 10^6$ yrs). For the SIMP model, an equivalent analysis can be done, except with the following differential equation governing the rate of particle loss in the halo:
\be\label{eq:dNxdt_SIMP}
\frac{dN_X}{dt} = -\Gamma_{ann}^{3\rightarrow 2} = -\int dV n_X^3 \langle \sigma v^2\rangle.
\ee
Solving for the ambient DM density gives:
\be
\rho_X^{3\rightarrow2}(r, t) = \frac{\rho_0(r) \rho_{AP}^{3 \rightarrow 2}(t)}{\sqrt{\rho_0(r)^2 + (\rho_{AP}^{3 \rightarrow 2}(t))^2}}.
\label{eq:rhot_simp}
\ee
where $\rho_{AP}^{3 \rightarrow 2} = \frac{m_X}{\sqrt{2 \langle \sigma v^2 \rangle t}}$. Similar to the $2 \to 2$ process, one can estimate the radius where the annihilation plateau begins to be relevant by solving $\rho_{AP}^{3 \rightarrow 2}(t)=\rho_0(r)$. Doing so for a NFW and SIMP annihilation leads to the following scaling relation:
\be
r_{AP} \approx 7\times 10^{-9}\text{ pc} \left(\frac{r_s}{190\text{ pc}}\right)\left(\frac{\rho_0}{30\text{ GeV~cm}^{-3}}\right)\left(\frac{t}{10^6\text{ yrs}}\right)^{1/2}\left(\frac{10^{-4}\text{ GeV}}{m_X}\right)^2.
\ee
Again, we find that the profile is not affected at timescales and distance scales relevant for captured DM and so the initial profile can be well approximated by a NFW profile for both the SIMP and WIMP models.

\begin{figure}
    \centering
    \includegraphics[width=0.8\linewidth]{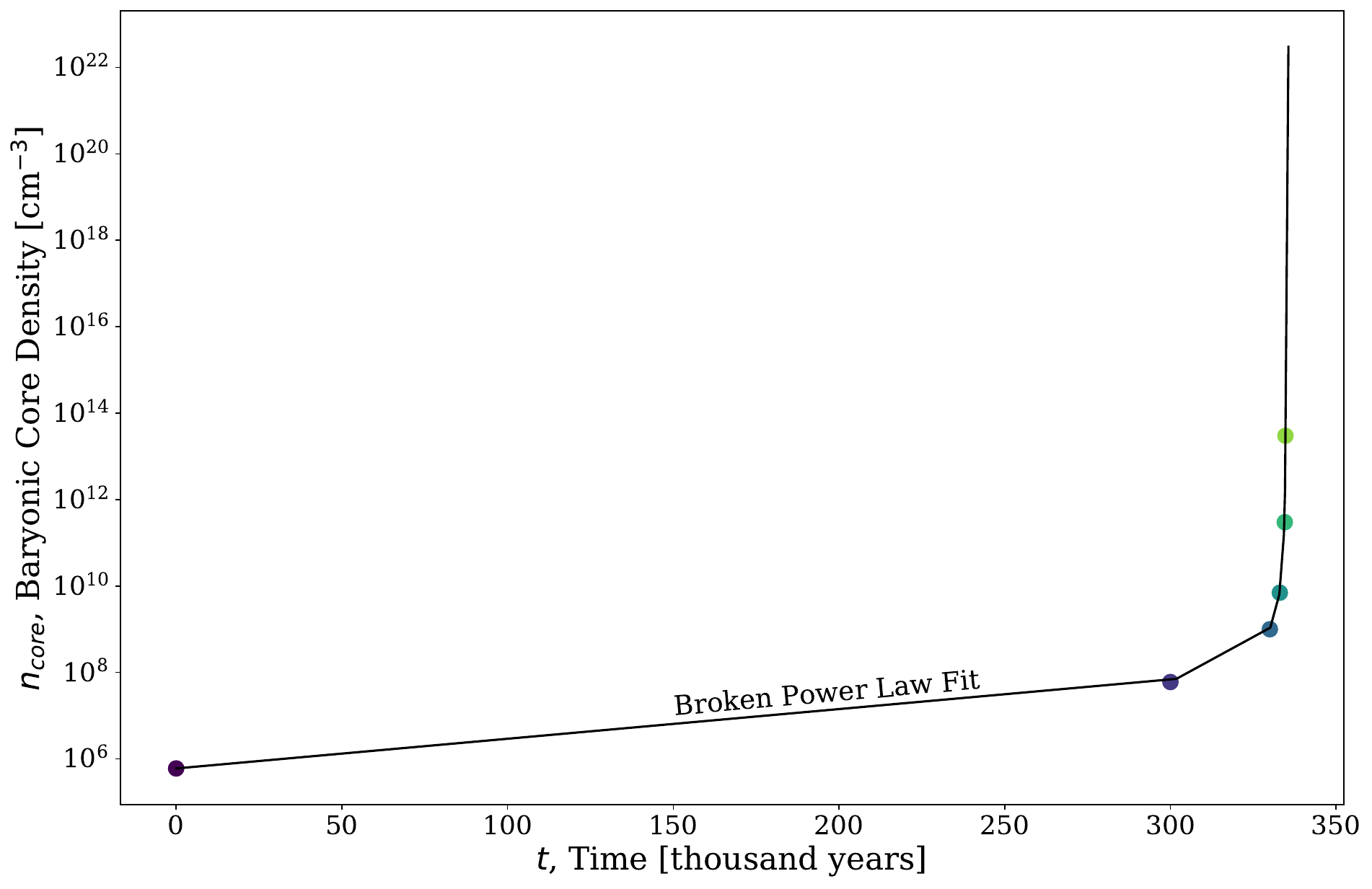}
    \caption{Core density of baryon profile $n_{core}$ as a function of time based on a power-law fit of simulation data from \cite{Abel:2001}. Here, the points represent data from\cite{Abel:2001}, while the black line is a broken power law fit. Here we have defined $t=0$ to be the time at which the core density reaches $n_{core} \sim 10^6$ cm$^{-3}$, and have extrapolated past the last point in the simulation data ($n_{core}\sim 10^{13}$ cm$^{-3}$, $t\sim 334,700$ years) by continuing with the same power-law behavior fitted between the final two points. In reality, one would expect a larger power for the final region, making this estimate slightly conservative. We have extrapolated up to the point at which a proto Pop~III star forms, $n_{core}\sim 10^{22}$ cm$^{-3}$. Due to the rapid contraction of the baryons for $t\gtrsim 300,000$ years, we conservatively estimate this to be at time $t\sim 335,600$ years, only $900$ years after reaching $n_{core}\sim 10^{13}$ cm$^{-3}$.}
    \label{fig:nbcore_time}
\end{figure}

For the CoSIMP model, a more detailed calculation is required since a standard model particle is required for annihilation in this process. Therefore, to find how much DM is annihilated away when baryons began falling into the DM Halo, we must know the baryonic profile at each point in time. The baryon profile can be well approximated by the following function \cite{Freese:2008dmdens}:
\be\label{eq:nBaryon_r}
n_B(r) = \frac{n_{core}}{1+(r/r_c)^{2.3}},
\ee
where $n_{core}$ is the baryonic core density and $r_c$ the core radius ($r_c$ ultimately depends on $n_{core}$ via Eq.~(\ref{eq:rCore})). This function was obtained from fitting the data in the simulations of \cite{Abel:2001}. The core densities at different times can be found in \cite{Abel:2001}, and the core radii is then given exactly by Eq.~(\ref{eq:rCore}). Since the core densities from the simulations in \cite{Abel:2001} are given at discrete times, to approximate the profile at any given time, we have fitted between data points a power function of the form: $n_{core}(t)=\alpha t^{\beta}$, where $\alpha$ and $\beta$ are found from fitting a line in logarithmic space. The result of fitting this data can be seen in Fig.~\ref{fig:nbcore_time}, where we have extrapolated to the time at which the core density is high enough that a proto Pop~III star is formed, i.e. $n_{core}\sim 10^{22}$ cm$^{-3}$. Annihilations past this point will still occur, but will be subdominant in the region of capture. This is because after the runaway collapse of baryons into a small, dense core, most of the baryons outside the core will be in an accretion disk within $~10$ A.\ U.\ of the core, as shown by simulations \cite{Abel:2001}. This is much farther inward from where most DM will be captured, and so this effect will cease to be relevant for the CoSIMP model past this point.

\begin{figure}
    \centering
    \includegraphics[width=0.8\linewidth]{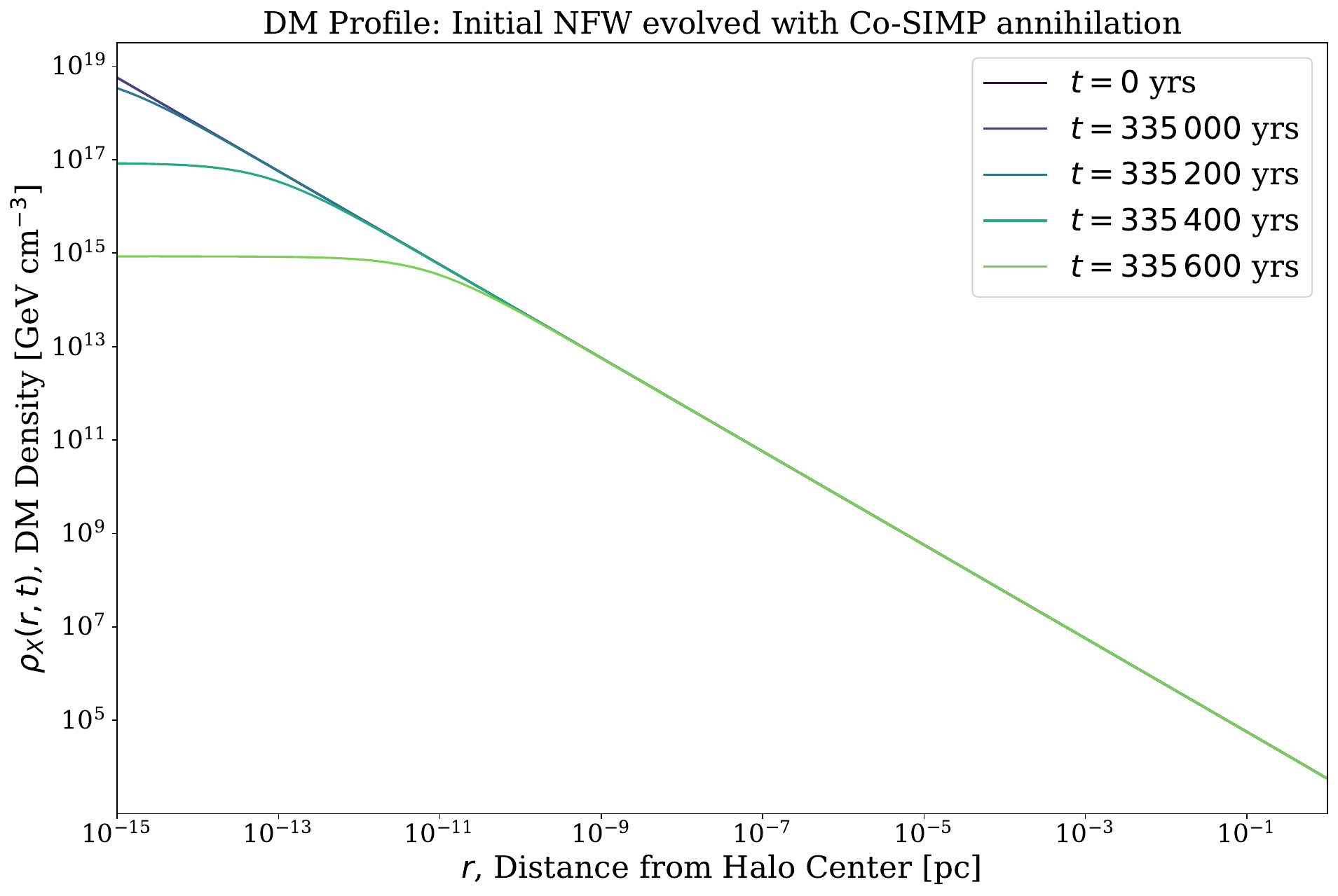}
    \caption{Time-evolution of an initial NFW DM profile due to DM annihilations in the CoSIMP model during the collapse of baryons up to the formation of a stellar proto-core when $n_{core}\sim 10^{22}$ cm$^{-3}$. This supports the fact that the initial profile can be taken as a NFW as it is unchanged in the regions relevant to capture, $\gtrsim 10$ AU.}
    \label{fig:rhort_initialNFW}
\end{figure}

Equipped with the baryon profile as a function of time, we are able to estimate how much DM is annihilated away during the baryon cloud collapse. However, to do this, we make the assumption that the initial halo changes only due to annihilations and not through adiabatic contraction. In order to verify our claim that annihilations will not affect the initial profile, we will examine two extreme cases for the profile before baryon collapse: an initial profile that is NFW and one that is artificially enhanced by adiabatic contraction. The AC profile is used to show that even steeper profiles are not significantly altered at the distance scales relevant to capture. To solve for the DM density at a given time and radius, we point to the following equation:
\be\label{eq:dNxdt_CoSIMP}
\frac{dN_X}{dt} = -\Gamma_{ann}^{3\to2} = -\int dV n_X^2 n_B \langle\sigma v^2\rangle,
\ee
which has the following solution for the DM density:
\be\label{eq:rhoxrt_CoSIMP}
\rho_X(r,t) =  \frac{\rho_0(r) m_X}{m_X + \rho_0(r)\langle\sigma v^2\rangle\int_0^t n_B(r,t') dt'},
\ee
with $\rho_0(r)$ being the initial DM profile. One can then solve this numerically using the broken power law function used to fit the baryon core density as a function of time. The results of this when taking $\rho_0(r)$ to be a NFW profile is shown in Fig.~\ref{fig:rhort_initialNFW}. Here we see that the annihilation of DM and baryons during collapse does affect the initial NFW profile, but only at radii much smaller than the edge of the accretion disk, which is at $\sim 10$ A.\ U.\ ($\sim 5\times 10^{-5}$ pc). It is therefore safe to make the assumption that a NFW profile would not change at the distance scales relevant to capture from annihilations during collapse.
However, as previously mentioned, this is an underestimate of these effects since, in actuality, the DM profile would respond to the infall of baryons through adiabatic contraction by becoming steeper, which would naturally lead to larger annihilation rates. Thus, to be the most conservative, we have also taken the initial profile $\rho_0(r)$ to be an AC profile with a baryon core density of $n_{core} \sim 10^{22}$ cm$^{-3}$, which is approximately the core density for the formation of a proto Pop~III star. We would like to strongly emphasize here that we are not suggesting that this would be the physically-motivated initial profile, and thus the resulting profiles do not represent the distribution of dark matter at star formation. However, the high-density nature of such a profile provides the most conservative estimate for the effects of annihilation on the DM profile at star formation, and is thus useful to demonstrate that even for an nonphysical, steep profile, these effects can be ignored. A complete treatment of this question would require coupling the time-dependence of adiabatic contraction to Eq.~(\ref{eq:dNxdt_CoSIMP}). 

\begin{figure}
    \centering
    \includegraphics[width=0.8\linewidth]{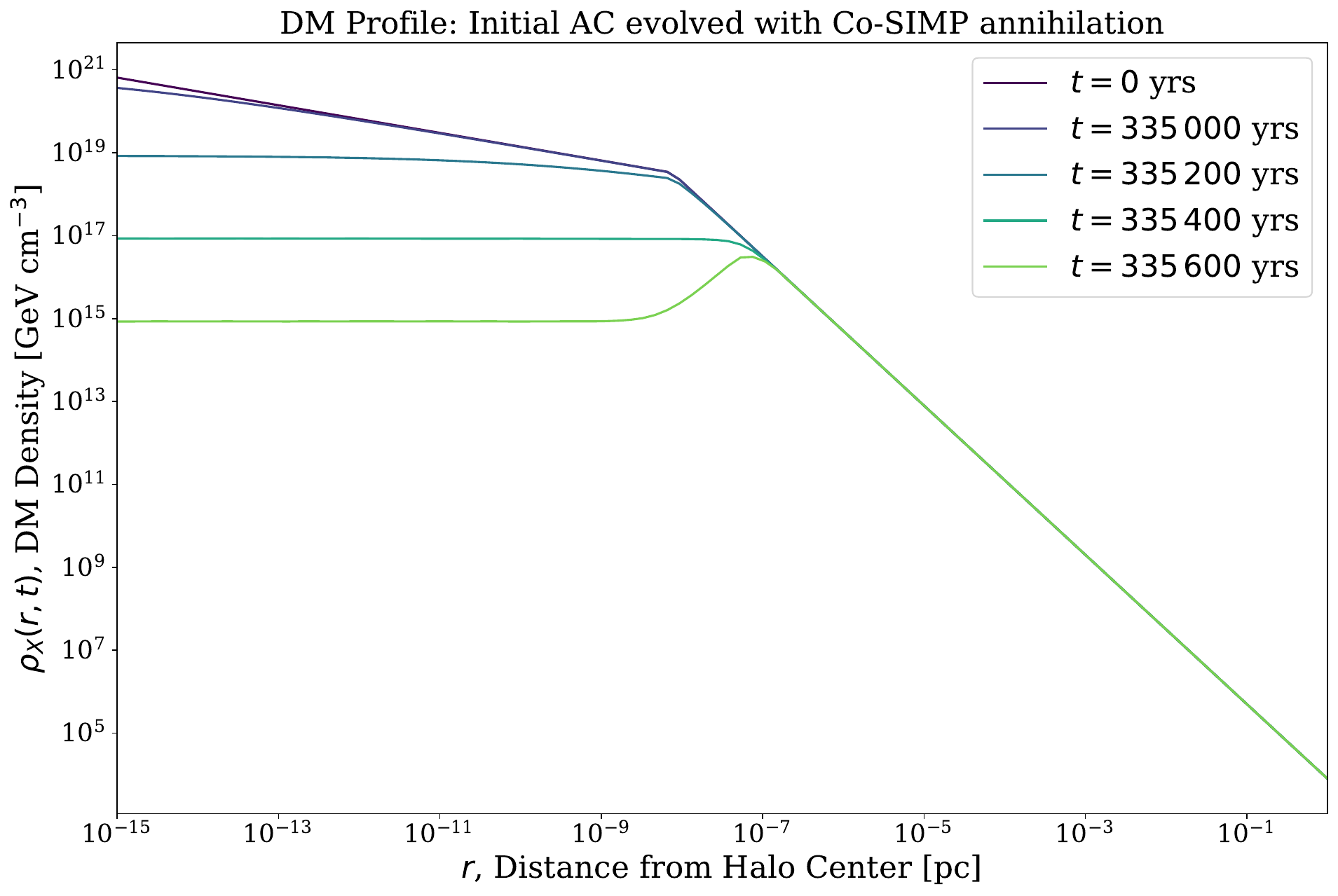}
    \caption{Time-evolution of an initial AC DM profile ($n_{core}\sim 10^{22}$ cm$^{-3}$) due to DM annihilatins in the CoSIMP model during the collapse of baryons up to the formation of a stellar proto-core when $n_{core}\sim 10^{22}$ cm$^{-3}$. This represents an extreme case where the initial profile is made much steeper than it would be to show that even in the most conservative case, the densities in the region where capture is relevant, $r\sim 5\times 10^{-5}$ pc, is unaffected during the collapse of the baryons.}
    \label{fig:rhort_initialAC}
\end{figure}

In Fig.~\ref{fig:rhort_initialAC} we plot the result of Eq.~(\ref{eq:rhoxrt_CoSIMP}) with an initial AC profile with $n_{core}\sim 10^{22}$ cm$^{-3}$. Again, it is clear that annihilations during collapse do affect the initial profile, however this is limited to the region $r\lesssim 10^{-7}$ pc, which is still inwards of the distance relevant for capture, $r\sim 5\times 10^{-5}$ pc. Thus, even for the most extreme case of a steep AC profile, the DM densities in the regions relevant for capture are unaffected.
An intriguing feature of the profile at $t=335,600$ yrs is the peak in the density at $r\sim 10^{-7}$ pc, before it falls and flattens as it moves inwards. This can be explained by examining the behavior of the denominator of Eq.~(\ref{eq:rhoxrt_CoSIMP}). When $m_X \gg \rho_0(r) \langle\sigma v^2\rangle \int_0^t n_B(r,t') dt'$, i.e. when the annihilation term is sub-dominant, the profile is simply described by the initial profile, which is the case for the larger radii ($r\gtrsim 10^{-7}$) in Fig.~\ref{fig:rhort_initialAC}. However, as annihilation becomes more relevant, which depends not only on time, but on the radius (since the baryon distribution is considered), the initial profile actually vanishes from the equation and the profile is described by $\rho_X(r,t)\approx \frac{m_x}{\langle\sigma v^2\rangle \int_0^t n_B(r,t') dt'}$. Thus, inward of a given radii at a specific time, the profile scales like $\rho_X(r)\sim n_B(r)^{-1}$. Now, the baryon profile is described by Eq.~(\ref{eq:nBaryon_r}), which scales like $n_B(r)\sim 1/r$ when $r\gg r_c$, and like $n_B(r)\sim r^0$ when $r\ll r_c$. Thus, when $t=335,400$ yrs, for example, inwards of $r\sim 10^{-7}$ pc the profile completely flattens as the annihialtion term becomes dominant and the baryon profile in this region is flat. However, for the $t=335,600$ yrs case, $r_c$ has actually shifted farther inwards since $r_{c}\sim n_{core}^{-0.81/1.8}$ and $n_{core}$ depends on time via Fig.~\ref{fig:nbcore_time}. Thus, there is a small region $r\sim 10^{-8}-10^{-7}$ pc where the $n_B(r)\sim 1/r$ relation is captured, but inverted, since $\rho_X(r,t)\sim 1/n_B(r)$ there. As shown in Fig.~\ref{fig:rhort_initialAC}, going further inwards, the profile flattens again as $r \ll r_c$ and so $n_B(r)\sim r^0$.

\begin{figure} [H]
    \centering
    \includegraphics[width=.75\linewidth]{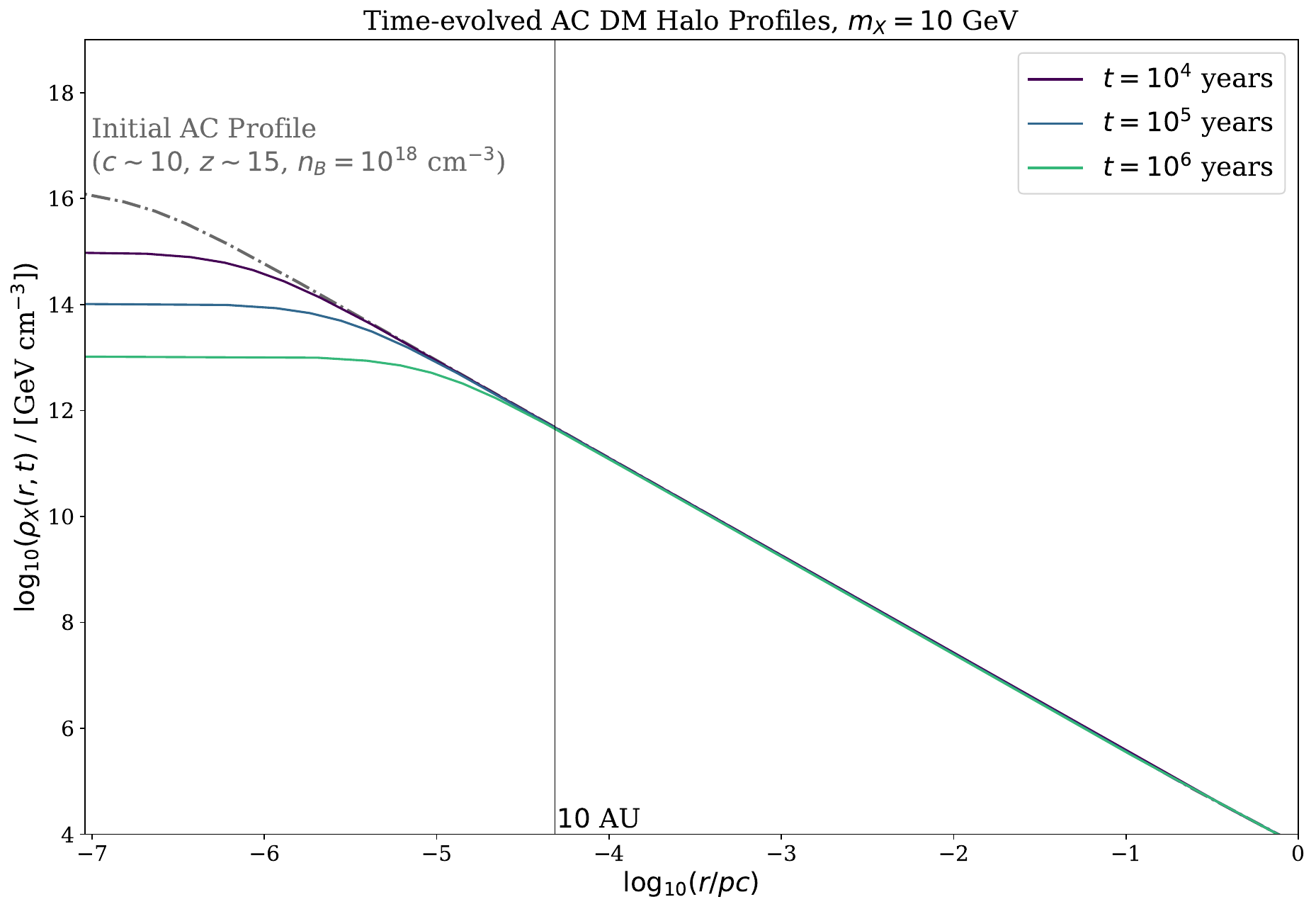}
    \caption{Time-evolved AC DM profiles under the influence of $2\rightarrow 2$ S-Wave annihilations for a $10$ GeV WIMP. The initial AC profile ($t = 0$) is represented by the grey dashed-dotted line while the varying colors represent the same profile at a later time, $t$. The annihilations of DM particles in the region surrounding the star ($\lesssim 10$ AU) lead to a flattening of the DM profile in the inner region known as the ``Annihilation Plateau." This effect becomes more pronounced over time as more particles annihilate and leads to lower ambient DM densities.}
    \label{fig:DMDensityProfiles_AnnPlat}
\end{figure}

After star formation, DM particles in the region outside of the star will self-annihilate, thus reducing the density of DM in the capturing region. The differential equation governing the number of DM particles in the region outside the star for $2 \rightarrow 2$ processes is given by Eq.~(\ref{eq:dNxdt_WIMP}) and has a solution shown in Eq.~(\ref{eq:rhot_22}). For $3 \rightarrow 2$ processes, namely the SIMP and Co-SIMP models, it can be shown that SIMP annihilation, which requires 3 DM particles, is far more efficient than Co-SIMP annihilation in the region outside the star due to the low baryonic density. Thus, the equivalent differential equation for the $3 \rightarrow 2$ models effectively reduces to that of SIMP annihilation, given in Eq.~(\ref{eq:dNxdt_SIMP}) with solution in Eq.~(\ref{eq:rhot_simp}). An important feature of Eqs.~(\ref{eq:rhot_22}) and (\ref{eq:rhot_simp}) is that in the limit of $\rho_{AP} \gg \rho_{0}$ ($\rho_0 \gg \rho_{AP}$), the DM density simply becomes $\rho_X \simeq \rho_0$ ($\rho_X \simeq \rho_{AP})$. In words, this means that, at a given position and time, the density is defined by the lower of the two. This fact is portrayed in Fig.~\ref{fig:DMDensityProfiles_AnnPlat} for $2 \rightarrow 2$ s-wave annihilations. Note that at $t = 0$, $\rho_{AP} \rightarrow \infty$, and thus the profile is unaffected by ambient DM annihilations. However, as time increases and more DM particles in the region around the star begin annihilating, the profile begins flattening around the higher densities (towards the profile's center), creating the annihilation plateau. This effect has implications for the DM capture rate, and thus the strength of our constraints on the DM scattering cross section (See Section \ref{sec:Constrain}), as lower ambient densities cause the capture rate to drop. However, the suppression of the ambient density due to this process is within our uncertainty of the ambient DM density for $2\rightarrow 2$ and SIMP annihilations. To see this, first note that the more pronounced effects of the annihilation plateau occur for lower DM masses ($\rho_{AP} \sim m_X$), larger times ($\rho_{AP} \sim 1/t$), and an efficient annihilation cross section. For $2 \rightarrow 2$ s-wave annihilation, Fig.~\ref{fig:DMDensityProfiles_AnnPlat} shows that, for the largest time considered ($t \sim T_\star = 10^6$ years) and the smallest mass in this regime ($m_X = 10$ GeV), the annihilation plateau reduces the ambient density to $\rho_X \sim 10^{13}$ GeV cm$^{-3}$, which is the lower limit we take for the ambient DM density. A similar analysis of the SIMP annihilation plateau shows a minimum density of $\rho_X \sim 10^{12}$ GeV cm$^{-3}$, which, while below our lower limit, is still within the uncertainty for the ambient DM density. Thus, while the annihilation plateau is an important process that must be factored into the constraints placed by this method, it does not reduce constraining in a drastic manner. We do, however, always include the effects of the DM annihilations on $\rho_X$ on all of our $\sigma-m_X$ bounds. 

An intriguing question that arises when considering the high DM density in the region surrounding the star is that of ambient annihilations producing diffuse emissions, which could potentially provide a signal that is distinct from the star's luminosity. To determine whether this effect is negligible relative to the star's luminosity, one must calculate the diffuse emissions from DM annihilations in the halo. This effect is relevant only for the $2 \rightarrow 2$ processes we consider, as SIMP DM produces no SM particles upon annihilation, and CoSIMP DM requires a SM particle for annihilation and, as mentioned previously in the discussion of the annihilation plateau, the baryonic density outside the stellar region is too small for any considerable effects. For $2 \rightarrow 2$ processes, the luminosity from ambient annihilations within a given volume around the star is given by:
\be
L_{amb} = m_X \Gamma_{ann}^{2\rightarrow 2} = m_X \int dV n_X^2 \langle \sigma v\rangle.
\ee
It is straightforward to compute this integral analytically by taking the outer profile of the halo from Eq.~(\ref{eq:rhoXout}). The final result is given by:
\be\label{eq:LDM_ambient}
L_{amb} = 5 V_\star \frac{\langle \sigma v\rangle \rho_0^2}{m_X} \left[1 - \left(\frac{R_\star}{r_{cutoff}}\right)^{0.6}\right],
\ee
where $\rho_0$ is the DM density at the edge of the core, and $r_{cutoff}$ is the radius at which the integral is truncated. One can thus calculate the approximate diffuse luminosity for a given DM profile out to some point around the star. For the WIMP regime, the annihilation cross section can be taken from the standard cross section giving the correct relic abundance. For more massive DM particles, a very conservative approach would be to take the annihilation cross section at the unitarity limit. For the study of Pop~III stars as considered in this paper, the annihilation cross section could also be taken by the lower bounds placed in Fig.~\ref{fig:SigmaVLBSupGeVDM}. For DM to equilibrate in the star, the annihilation cross section would necessarily have to fall between the unitarity limit and the lower bounds placed in Fig.~\ref{fig:SigmaVLBSubGeVDM}, which thus provides a natural range of cross sections to explore this effect.

\begin{figure} [H]
    \centering
    \includegraphics[width=.75\linewidth]{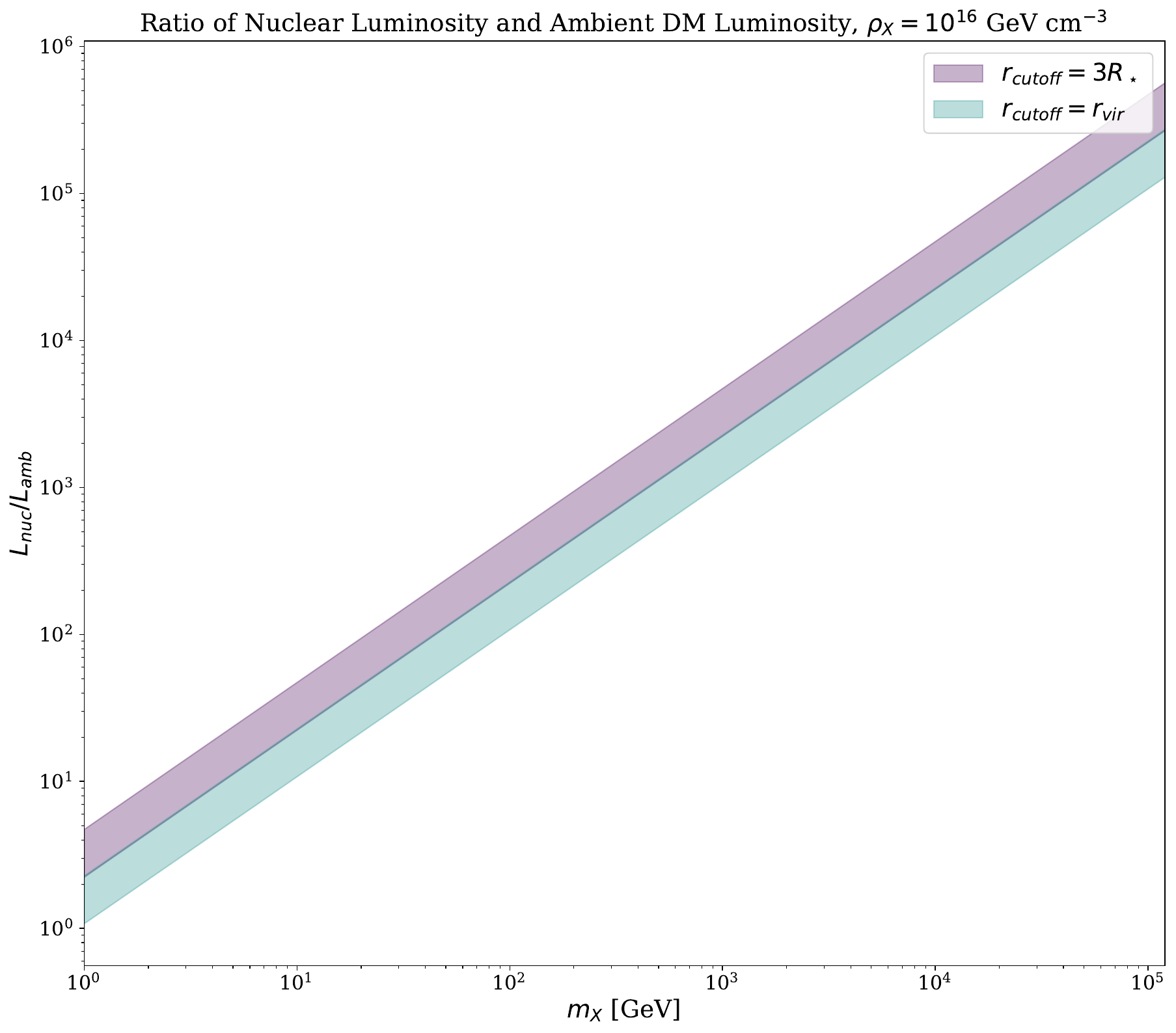}
    \caption{Ratio of a $M_\star = 100 M_\odot - 1000 M_\odot$ star's nuclear luminosity (represented by a band of a given color) to the diffuse emissions due to WIMP annihilations in the region directly surrounding the star (out to 3 stellar radii, purple line) and from the entire halo (out to the virial radius, blue line). The annihilation cross section is taken from the standard WIMP miracle cross section. This plot demonstrates that, in the WIMP regime and for the highest ambient DM density considered, the star's nuclear luminosity is always dominant relative to the diffuse emissions.}
    \label{fig:Lstar_LambRatioWimp}
\end{figure}

\begin{figure} [H]
    \centering
    \includegraphics[width=.75\linewidth]{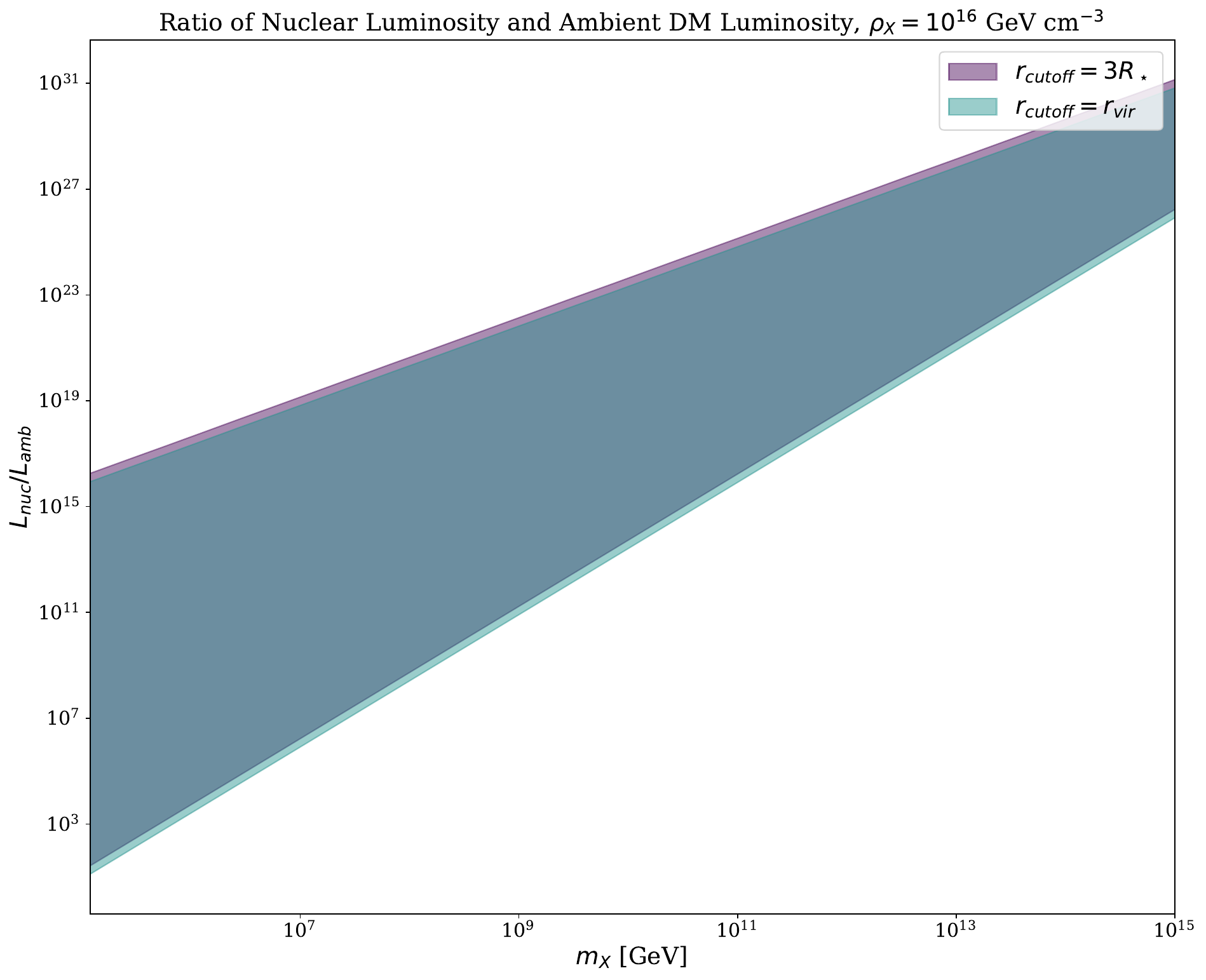}
    \caption{Ratio of a $M_\star = 100 M_\odot$ star's nuclear luminosity to the diffuse emissions due to heavy DM annihilations in the region directly surrounding the star (out to 3 stellar radii, purple band) and from the entire halo (out to the virial radius, blue band). The annihilation cross section is taken in a range between the bounds placed in Fig.~\ref{fig:SigmaVLBSupGeVDM} arising from the equilibrium condition and the unitarity limit (this is represented by the band of a given color). In this figure, it is evident that across all annihilation cross sections allowable for heavy DM that equilibrates in the star, the diffuse emissions are sub-dominant relative to the star's nuclear luminosity.}
    \label{fig:Lstar_LambRatioNonWimp}
\end{figure}

We estimate the upper bound on the effect of the ambient annihilations to the star's total luminosity by considering the ratio $L_{nuc}/L_{amb}$. For the most conservative approach, where $L_{amb}$ is maximized, we consider the highest density we take in this paper, $\rho_0 = 10^{16}$ GeV cm$^{-3}$. The results of this calculation are presented in Figs.~\ref{fig:Lstar_LambRatioWimp} and \ref{fig:Lstar_LambRatioNonWimp}, which show in the respective mass regimes that the diffuse emissions are always subdominant to the star's total luminosity. 

Although, as Figs.~\ref{fig:Lstar_LambRatioWimp} and~\ref{fig:Lstar_LambRatioNonWimp} show, the ratio between the nuclear luminosity ($L_{nuc}$) and the diffuse DM halo emission from DM annihilations inside the halo ($L_{amb}$) is always greater than one, and typically much greater than one, we point out the intriguing possibility of estimating the diffuse emission, by removing from the total spectrum the expected stellar spectra. If the remaining residuals are statistically significant, one could infer DM annihilations are the cause, and as such infer properties of the DM particle, in a very similar fashion to the DM explanation of center of the galaxy excess gamma-ray excess in the FERMI data~\cite{Goodenough:2009,hooper2011dark,Cholis:2019}. However, in the latter case, other more mundane astrophysical sources, such as unresolved pulsars, could explain away the excess~\cite{Gordon:2013,YUAN:2014}. This is in contrast to the situation of a possible excess from DM microhalos at high redshifts, where pulsars are not expected to be present. 

\newpage

\bibliography{RefsDM}
\end{document}